\documentclass[structabstract]{aa}
\usepackage{txfonts}

\usepackage{natbib}
\bibpunct{(}{)}{;}{a}{}{,} 

\usepackage{graphicx}

\newcommand{\trick}[1]{#1}
\begin{document}
\input epsf
\title{p, He, and C to Fe cosmic-ray primary fluxes in diffusion models}
\subtitle{Source and transport signatures on fluxes and ratios.}
\author{A. Putze\inst{1}
   \and D. Maurin\inst{2,3,4,5}
   \and F. Donato\inst{6}
       }

\offprints{A.~Putze, {\tt antje@fysik.su.se}}

\institute{The Oskar Klein Centre for Cosmoparticle Physics,
   Department of Physics, Stockholm University,
   AlbaNova, SE-10691 Stockholm, Sweden
   \and Laboratoire de Physique Subatomique et de
	Cosmologie ({\sc lpsc}),
	Universit\'e Joseph Fourier Grenoble 1, CNRS/IN2P3, Institut Polytechnique de Grenoble,
	53 avenue des Martyrs,
	Grenoble, 38026, France
   \and Laboratoire de Physique Nucl\'eaire et des Hautes Energies,
    Universit\'es Paris VI et Paris VII, CNRS/IN2P3,
	  Tour 33, Jussieu, Paris, 75005, France
   \and
   Dept. of Physics and Astronomy, University of Leicester,
   Leicester, LE17RH, UK
   \and
   Institut d'Astrophysique de Paris, UMR7095 CNRS,
   Universit\'e Pierre et Marie Curie, 98 bis bd Arago,
   75014 Paris, France
    \and Dept. of Theoretical Physics and INFN, 
		via Giuria 1, 10125 Torino, Italy
}

\date{Received / Accepted}

\abstract
{The source spectrum of cosmic rays is not well determined by diffusive
shock acceleration models. The propagated fluxes of proton, helium, and
heavier primary cosmic-ray species (up to Fe) are a means to indirectly
access it. But how robust are the constraints, and how degenerate are the
source and transport parameters?
}
{We check the compatibility of the primary fluxes with the transport
parameters derived from the B/C analysis, but also ask whether they add
further constraints. We study whether the spectral shapes of these fluxes
and their ratios are mostly driven by source or propagation effects. We then
derive the source parameters (slope, abundance, and low-energy shape).
}
{Simple analytical formulae are used to address the issue of degeneracies
between source/transport parameters, and to understand the shape of the p/He
and C/O to Fe/O data. The full analysis relies on the USINE propagation
package, the MINUIT minimisation routines ($\chi^2$ analysis) and a Markov
Chain Monte Carlo (MCMC) technique.
}
{Proton data are well described in the simplest model defined by a power-law
source spectrum and plain diffusion. They can also be accommodated by models
with, e.g., convection and/or reacceleration. There is no need for breaks in
the source spectral indices below $\sim 1$~TeV/n. Fits to the primary fluxes
alone do not provide physical constraints on the transport parameters. If we
leave the source spectrum free, parametrised by the form $dQ/dE = q
\beta^{\eta_S} {\cal R}^{-\alpha}$, and fix the diffusion coefficient $K(R)=
K_0\beta^{\eta_T} {\cal R}^{\delta}$ so as to reproduce the B/C ratio, the
MCMC analysis constrains the source spectral index $\alpha$ to be in the
range $2.2-2.5$ for all primary species up to Fe, regardless of the value of
the diffusion slope $\delta$. The values of the parameter $\eta_S$
describing the low-energy shape of the source spectrum are degenerate with
the parameter $\eta_T$ describing the low-energy shape of the diffusion
coefficient: we find $\eta_S-\eta_T\approx 0$ for p and He data, but
$\eta_S-\eta_T\approx 1$ for C to Fe primary species. This is consistent
with the toy-model calculation in which the shape of the p/He and C/O to
Fe/O data is reproduced if $\eta_S-\eta_T\approx 0-1$ (no need for different
slopes $\alpha$). When plotted as a function of the kinetic energy per
nucleon, the low-energy p/He ratio is determined mostly by the modulation
effect, whereas primary/O ratios are mostly determined by their destruction
rate.
}
{\trick{Models based on fitting B/C are compatible with primary fluxes. The
different spectral indices for the propagated primary fluxes up to a few
TeV/n can be naturally ascribed to transport effects only, implying
universality of elemental source spectra}.
}


\authorrunning{A. Putze, D. Maurin, \& F. Donato}
\titlerunning{p, He, and C to Fe cosmic-ray fluxes in diffusion models}
\maketitle


\section{Introduction}
The measured Galactic cosmic-ray (GCR) fluxes at Earth result from a
three-step journey: i) the diffusive shock acceleration (DSA) mechanism
provides a source spectrum; ii) these particles are then transported (by diffusion,
but also convection and reacceleration) and also interact in the interstellar medium (ISM) until
they reach the Solar neighbourhood; iii) they enter the solar cavity where
they decelerate due the effect of solar modulation (active for GCRs below
a few tens of GeV/n).

The last step prevents direct measurements of low-energy (beyond a few
hundreds of GeV/n) interstellar fluxes (IS). 
The first step can be investigated by means of semi-analytical or numerical studies
of the DSA mechanism. However, due to the variety of  possible sources for
the GCRs, and the intrinsic complexity of this mechanism, the source
spectral index and especially its low-energy shape is not very well predicted
\citep[e.g.,][]{2010APh....33..160C}. Awaiting further progress along this line, an
indirect route to access the source spectrum is to start with the
top-of-atmosphere (TOA) fluxes
and go back to the source spectra. To do so, it is generally assumed that the
steady state holds for the propagation, and that the first and second step
are independent. The first assumption is known to fail at high enough energies,
whereas the second one may only be approximate. In this study, given the
success of simple steady-state diffusion models for the nuclear component, we
follow the same route in order to draw some constraints on the source
parameters. Note that this paper does not consider primary electrons,
whose spectrum may depend on local sources above a few tens of GeV
\citep[e.g.,][]{1989ApJ...342..807B, 2010arXiv1002.1910D}.

The propagation step is the focal point of many phenomenological studies 
addressing the flux of secondary species (created by 
interactions of the primary species in the ISM and radiation fields of the
Galaxy) including light nuclei, antiprotons, positrons, radioactive
isotopes and also gamma rays. The transport parameters are usually determined
by fitting data on secondary-to-primary ratios of nuclei, as for example the B/C
(boron-to-carbon) ratio. However, present data on B/C ratio,  even when
combined with CR radioactive isotope measurements, lead to a constrained but 
large range of allowed values for these parameters
\citep{2001ApJ...555..585M,2010A&A...516A..67M,2010A&A...516A..66P}. The
study of these secondary-to-primary ratios is |to first order|insensitive to
the details of the source spectra \citep[e.g.,][]{2002A&A...394.1039M}. 
Therefore, a common phenomenological approach is to 
first  extract the transport parameters, then to fit the source spectra; however 
the source and transport parameters may be correlated
\citep{2009A&A...497..991P}. 

The importance of the primary fluxes and their ratios was recognised a long time
ago \citep{1974Ap&SS..30..361W}. In this paper we reconsider their study,
trying to answer the following questions: what phenomena shape the TOA and IS fluxes?
Is it the source spectrum or the propagation step, and are there degeneracies
between the two effects? To accuracy accuracy can we determine the low-energy
source spectra, the spectral indices, and the source abundances? Are the
source spectra universal or species-dependent?

On the experimental side, accurate data are available up to few hundreds of
GeV, whilst at higher energies data are less abundant and have large error
bars and scatter between experiments. On the modelling side, we have a semi-analytical propagation
model proved to work well with many GCR observables
\citep{2001ApJ...555..585M,2001ApJ...563..172D,2002A&A...381..539D,2009PhRvL.102g1301D}
and an implementation of the Markov Chain Monte Carlo technique (MCMC) to derive
the probability density functions of the analysed model parameters
\citep{2009A&A...497..991P,2010A&A...516A..66P}. We take advantage of
these to address the above questions. Our results are also
supported by toy-model calculations, especially for the shape of ratio p/He.
As abundant and accurate data are expected in the near future by the orbiting
PAMELA experiment and forthcoming AMS-02 detector (to be installed on the
International Space Station), such a study also aims at providing some
guidelines on how to tackle the information contained in the primary flux
propagated spectra. We finally note that the recent ATIC-2 and CREAM-I
measurements for proton and Helium data hint at a spectral change $\gtrsim$~TeV/n.
This has consequences for the secondary production of $\gamma$-rays, antiprotons,
and positrons \citep{2010arXiv1010.5679D,2010arXiv1011.3063L}. However, this
occurs only for the high-energy part of these spectra. In particular, as
our previous antiproton calculations \citep{2001ApJ...563..172D} relied on a fit
to the data, the conclusions obtained in \citet{2008PhRvD..78d3506D} remain unchanged
even with the new source spectra provided in this study.

The paper is organised as follows:  Sect. \ref{sec:prop} contains a short
overview of the propagation scheme employed in the present study. In
Sect.~\ref{sec:pHe_alone}, we discuss the p and He data and whether they
can provide any constraints on the transport parameters or if they can be
fitted in any propagation configuration (e.g. with or without convection, with or without
reacceleration). In Sect.~\ref{sec:pHe_propagparamfixed}, we seek for generic
constraints on the source spectra (p, He, and C to Fe) comparing  the
values obtained in different configurations of propagation models. In 
Sect.~\ref{sec:ratios}, the origin for the observed shape for the
ratio of primary species is outlined. Our conclusions and perspectives
are given in Sect.~\ref{sec:conclusion}.

\section{The propagation model}
\label{sec:prop}

The framework employed to calculate the fluxes is the diffusion model 
with convection and reacceleration discussed in \citet{2001ApJ...555..585M,2002A&A...394.1039M},
updated and fully detailed in \citet{2010A&A...516A..66P}. Here we only summarise the main features of the model. 

The Galaxy is shaped as a gaseous thin disk with half-thickness $h=0.1$~kpc
and an infinite radial extension (1D model, as also used in \citealt{2001ApJ...547..264J}),
hosting the interstellar medium and the stars, and surrounded by a thick halo for cosmic-ray transport whose
half-height is $L$.

Assuming a steady-state situation, the transport equation for the CR nucleus $j$ can be written as 
\begin{equation}
{\cal L}^j N^j + \frac{\partial}{\partial E}\left( b^j N^j - c^j 
\frac{\partial N^j}{\partial E} \right) = {\cal S}^j\;.
\nonumber
\end{equation}
where the differential density  $N^j\equiv N^j(E,\vec{r})$
depends on the position $\vec{r}$ in the Galaxy and on the energy
(throughout the paper, $E$ is the total energy, $E_k$ is the kinetic energy,
$E_{k/n}$ the kinetic energy per nucleon, and $E_{/n}$ the total energy per
nucleon). 
${\cal L}^j$ sums up the physics of the transport in the Galaxy, while 
${\cal S}^j$ contains the source term. 

   \subsection{Transport parameters}
The operator ${\cal L}$ (we omit the superscript $j$) describes the diffusion
$K(\vec{r},E)$ and convection $\vec{V_C}(\vec{r})$ in the Galaxy, the  decay rate
$\Gamma_{\rm rad}(E)= 1/(\gamma\tau_0)$ for radioactive species, and the destruction
rate $\Gamma_{\rm inel}(\vec{r},E)=\sum_{ISM} n_{\rm ISM}(\vec{r}) v \sigma_{\rm inel}(E)$
on the interstellar matter (ISM). It reads
\begin{equation}
{\cal L}(\vec{r},E) =  -\vec{\nabla} \cdot (K\vec{\nabla}) + \vec{\nabla}\cdot\vec{V_C} +
     \Gamma_{\rm rad} + \Gamma_{\rm inel}.
 \nonumber
\end{equation}
The spatial diffusion coefficient is parametrised as
\begin{equation}
K(E)= \beta^{\eta_T} \cdot K_0 {\cal R}^\delta \;.
\label{eq:eta_T}
\end{equation}
where ${\cal R}=pc/Ze$ is the rigidity of the particle, $\beta$ is the velocity
of the particle in units of c and $\eta_T$ parameterizes the low-energy
behaviour of diffusion.  The nominal form is given by $\eta_T=1$, which is just
the inevitable effect of particle velocity on the diffusion rate. Hence this
$\beta$ term is always present, and other values of $\eta_T$ are relative to
this. \citet{2006ApJ...642..902P} argued that the form of the spatial diffusion
coefficient could be modified at low energy, due to the possibility that the
nonlinear MHD cascade sets the power-law spectrum of turbulence. Indeed,
\citet{2010A&A...516A..67M} found that the value of this parameter was crucial
for the determination of $\delta$ given the current B/C data.
The convective wind acts in the whole diffusive volume with a  constant
velocity $\vec{V_C}=\pm V_C\vec{e_Z}$ pointing perpendicularly to the Galactic disk. 
The coefficients $b$ and $c$ account for the first
and second order energy changes
\begin{eqnarray}
\label{eq:b}
b\,(\vec{r},E)&=& \big\langle\frac{dE}{dt}\big\rangle_{\rm ion,\,coul.} 
   - \frac{\vec{\nabla}.\vec{V_C}}{3} E_k\left(\frac{2m+E_k}{m+E_k}\right)
	 \nonumber\\
   &  & + \;\; \frac{(1+\beta^2)}{E} \times K_{pp},\nonumber\\
c\,(\vec{r},E)&=&  \beta^2 \times K_{pp}.\nonumber
\end{eqnarray}
Coulomb and ionisation losses add to possible energy gains due to reacceleration, described 
by the coefficient $K_{pp}$ in momentum space. 
The parameterisation for  $K_{pp}$ is taken from the model of minimal reacceleration by the
interstellar turbulence \citep{1988SvAL...14..132O,1994ApJ...431..705S}:
\begin{equation}
K_{pp}\times K= \frac{4}{3}\;V_a^2\;\frac{p^2}{\delta\,(4-\delta^2)\,(4-\delta)}.
\label{eq:Va}
\end{equation}
where $V_a$ is the Alfv\'enic speed.

\subsection{Source parameters}

The source term ${\cal S}$ includes the initial spectrum at the sources and the  
secondary contributions (spallations of heavier nuclei). 
Acceleration models typically predict $dQ/dp\propto p^{-\alpha}$ 
(e.g., \citealt{1994ApJS...90..561J}),
which leads to $dQ/dE\propto p^{-\alpha}/\beta$,
where the low-energy behaviour is unknown. 
For future reference, we note that
\begin{equation}
\frac{dQ}{dE} =  \frac{dQ}{dE_k} 
              = \frac{1}{A} \cdot \frac{dQ}{dE_{k/n}} 
              = \frac{1}{\beta}\cdot \frac{dQ}{dp} 
              = \frac{1}{Z\beta} \cdot \frac{dQ}{d{\cal R}}\;.
\label{eq:E_p_R}
\end{equation}

In this paper, we model the low-energy shape by adding one free
parameter, $\eta_S$, active at low energy:
\begin{equation}
        Q_{E_{k/n}}(E) \equiv \frac{dQ}{dE_{k/n}}  
        = q \cdot \beta^{\eta_S} \cdot {\cal R}^{- \alpha},
       \label{eq:source_spec}
    \end{equation}
where $q$ is taken to be the normalisation for a differential energy
per nucleon source spectrum. The reference low-energy shape corresponds
to $\eta_S=-1$ (to have $dQ/dp\propto p^{-\alpha}$, i.e. a pure power-law).

\subsection{Free parameters of the model}
\label{sec:free_par}
The present model contains of necessity several free parameters: the parameters
in the transport sector $\{K_0,\, \delta,\, V_a,\, V_c,\, \eta_T\}$,  the
ones in the source term $\{q,\,\eta_S,\, \alpha\}$, and the halo size of the
Galaxy $L$ in the geometry sector.
We will see in the following that not all these parameters have
the same relevance to the physics of primary cosmic nuclei, and we will
therefore operate within a critical sub-range of parameters.   In diffusion models, $L$ 
cannot be solely determined from the B/C ratio because of the well-known
degeneracy between $K_0$ and $L$  when only stable species are considered. If
not differently stated, we will work with the default values $\eta_S=-1$ and
$\eta_T=1$, and the reference value $L=4$~kpc. In most of the analyses we
let free the source normalisation $q_i$ for each primary species.

The low energy ($\lesssim 10$ GeV/n) charged particles are braked in the 
heliosphere by the solar wind, modulated with variations of the scale of the 11-year cycle.  We
adopt  the force-field approximation, which  provides a simple analytical
one-to-one correspondence between the modulated top-of-the atmosphere (TOA)
and the demodulated interstellar (IS) fluxes, and whose only effective parameter is the modulation 
potential $\phi$ (GV). For a species $j$, the IS and
TOA energies per nucleon are related by $E_{/n}^{\rm IS} = E_{/n}^{\rm TOA} +
\Phi$ ($\Phi=Z/A\times \phi$ is the modulation parameter), and the
fluxes by ($p$ is the momentum)
\begin{equation}
\psi^{\rm IS} \left(E^{IS}\right) 
  = \left( \frac{p^{\rm IS}}{p^{\rm TOA}}\right)^2 
   \psi^{\rm TOA} \left(E^{\rm TOA}\right).
 \label{eq:modul}
\end{equation}

The force-field method is an approximated solution to the diffusion equation
for charged particles in the heliosphere. A more accurate treatment is far
beyond the scope of our paper, which is centred on the source and
diffusive processes active in much a wider energy range and responsible for strong
effects of the primary spectra.

In principle, the solar modulation potential could also be a free
parameter of the study. Since the low-energy spectrum of the primary fluxes is
determined mainly by the low-energy injection spectrum (see
Eq.~\ref{eq:source_spec}), the low-energy diffusion scheme (see
Eq.~(\ref{eq:eta_T})), and the solar modulation (see Eq.~\ref{eq:modul}),  a
non-trivial correlation is expected between these parameters. For instance, we
find using the {\sc{minuit}} minimisation routines (not showed) that there is
a negative correlation between $\phi$ and $\eta_S$. This holds for all primary
cosmic-ray fluxes studied in this work. Given that there is already a great
number of degeneracies between the different source parameters, which cannot be
lifted with the current available data, we choose to fix the modulation
potential $\phi$ to the values given by the experiments. A more detailed study,
in the spirit of the one done in \citet{2010arXiv1011.0037T}, will be undertaken
in a future work.

\section{Analysis with free source and transport parameters}
\label{sec:pHe_alone}

In this Section, we use different data sets for proton and helium
fluxes\footnote{The isotopic separation is not always achieved, so that many
data actually correspond to $^1$H+$^2$H for the proton flux and $^3$He+$^4$He
for the helium flux, where $^2$H and $^3$He come from secondary contributions
only. This amounts to a $\lesssim 10\%$ error at low energy, which is
contained in the error bars.} in order to determine which propagation models
describe data and to try to set constraints on the free parameters of our model. 
The fitting procedure is based on the {\sc minuit} routine, which minimises a
$\chi^2$ function.

\subsection{Data\label{sec:data}}
\begin{figure}[!t] 
\centering
\includegraphics[width=\columnwidth]{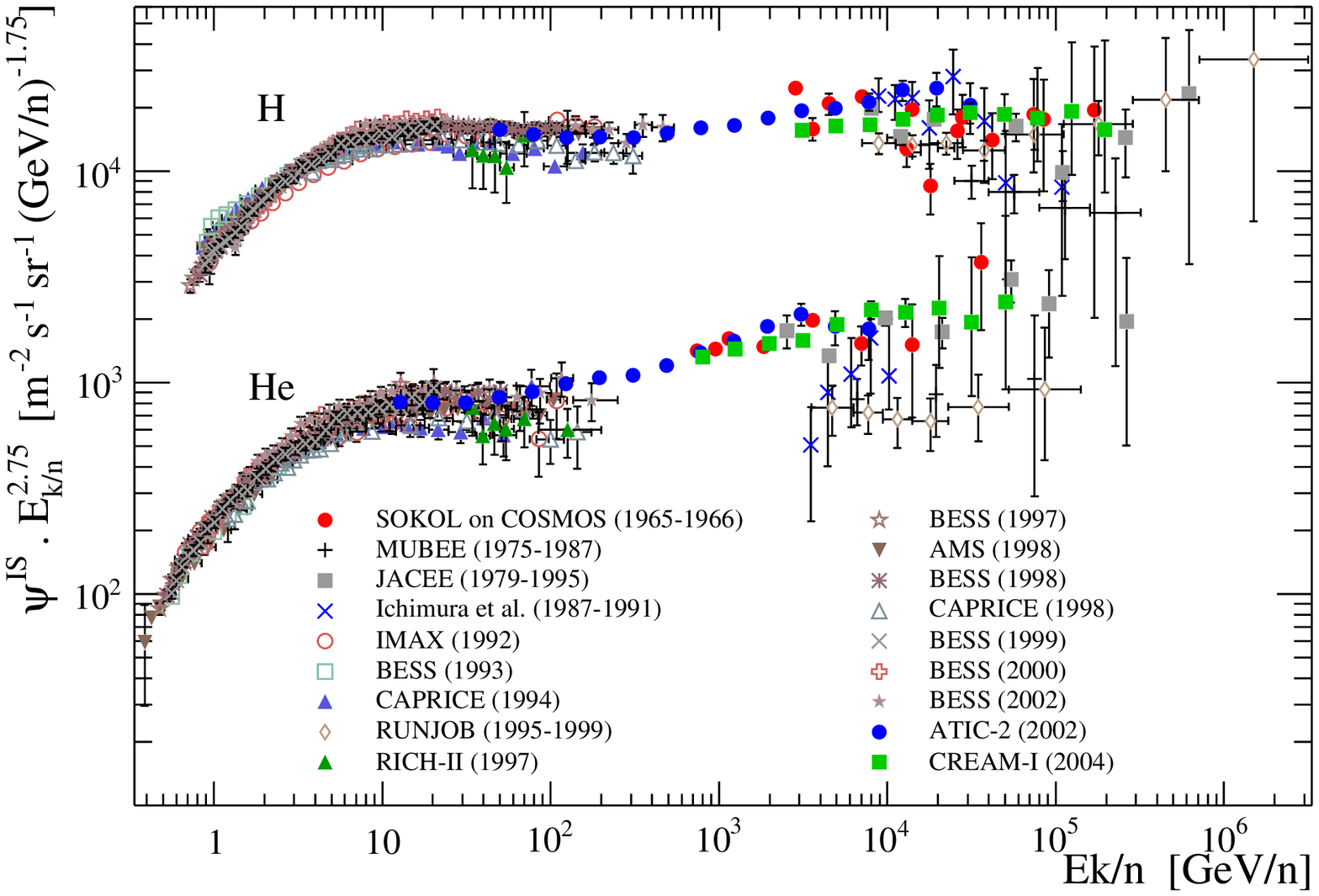}
\caption{Demodulated p and He data ($\times E_{k/n}^{2.75}$) as a function of
$E_{k/n}$. The low-energy data and the associated modulation parameter are:
AMS-01 (\citealt{2000PhLB..490...27A} for H and \citealt{2000PhLB..494..193A}
for He) with $\phi=600$~MV; CAPRICE94 \citep{1999ApJ...518..457B} and
CAPRICE98 \citep{2003APh....19..583B} with $\phi=650$~MV and $600$~MV
respectively; IMAX \citep{2000ApJ...533..281M} with $\phi=750$~MV; and the
series of BESS balloon flights BESS93 \citep{2002ApJ...564..244W} BESS97
\citep{2007APh....28..154S}, BESS98
\citep{2000ApJ...545.1135S,2007APh....28..154S}, BESS99
\citep{2007APh....28..154S}, BESS00 \citep{2007APh....28..154S}, and BESS02
(BESS-TeV, \citealt{2004PhLB..594...35H,2007APh....28..154S}), for which
$\phi=700$~MV, $491$~MV, $591$~MV, $658$~MV, $1300$~MV, $1109$~MV
respectively. The intermediate and high-energy data are:
ATIC-2 \citep{2009APh....30..133A},
CREAM-I \citep{2010ApJ...714L..89A},
JACEE \citep{1998ApJ...502..278A},
MUBEE \citep{1993ICRC....2...13Z},
RICH-II \citep{2003APh....18..487D},
RUNJOB \citep{2005ApJ...628L..41D}, 
SOKOL \citep{1993ICRC....2...17I},
and \citet{1993PhRvD..48.1949I}. 
}
\label{fig:data_all}
\end{figure}

A first important consideration is the choice of the data used to constrain the models. 
The top panel in Fig.~\ref{fig:data_all} shows the available data for p and He fluxes.
The abscissa is the kinetic energy per nucleon ($E_{k/n}$) and the ordinate
$\psi^{\rm IS}\times E_{k/n}^{2.75}$, where $\psi^{\rm IS}$ is the IS, 
demodulated using the force-field approximation.
The low-energy region (below 100 GeV/n) has been covered by many 
balloon--borne, Space-Shuttle based and satellite experiments, whereas above a
few TeV/n the data come from several balloon long-exposure flights
(accumulated over several flights in a decade). ATIC data cover the gap at
a few TeV/n energy. The overall agreement between the data is fair, the
scatter between the data being higher at high energy.

\begin{figure}[!t] 
\centering
\includegraphics[width=\columnwidth]{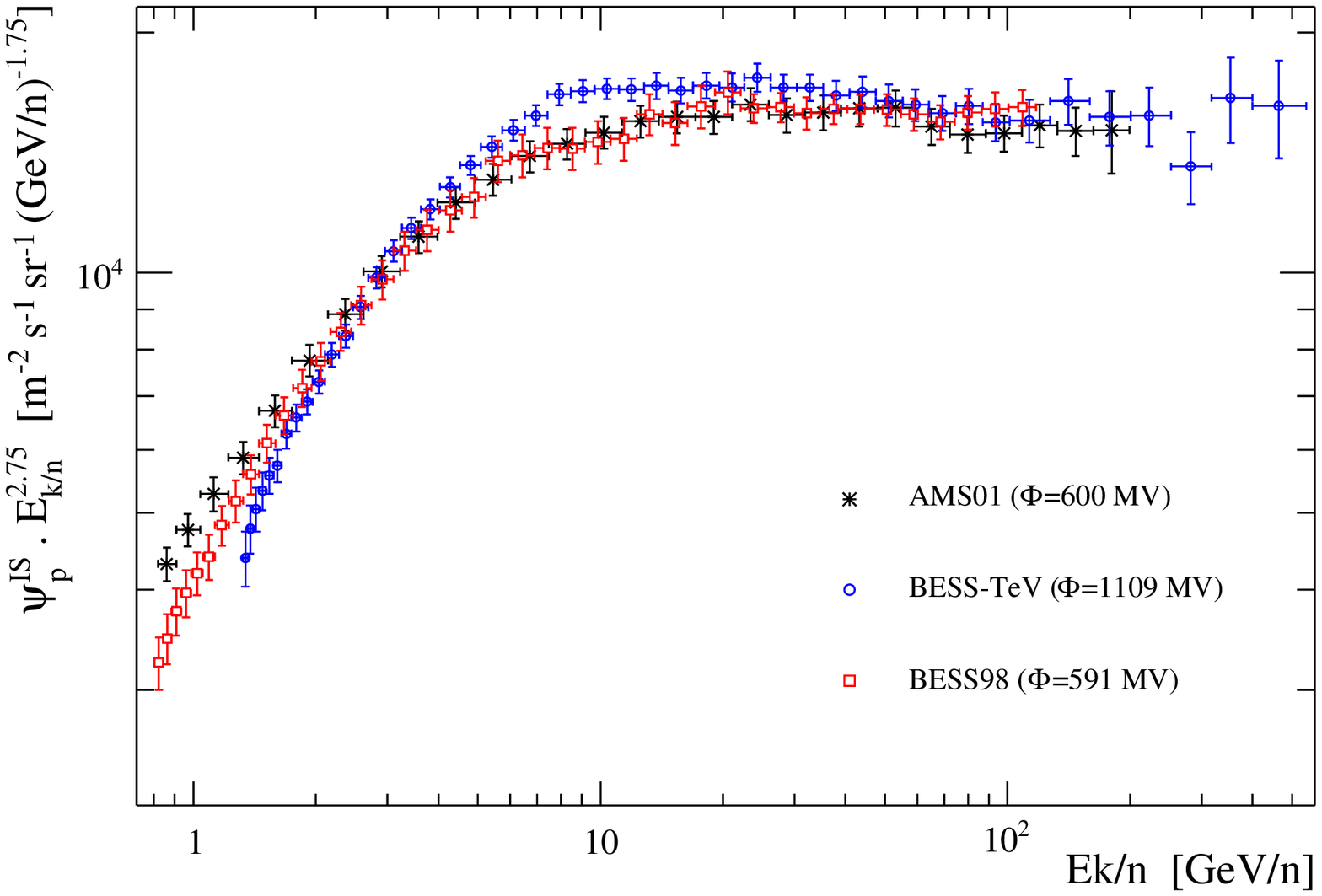}
\includegraphics[width=\columnwidth]{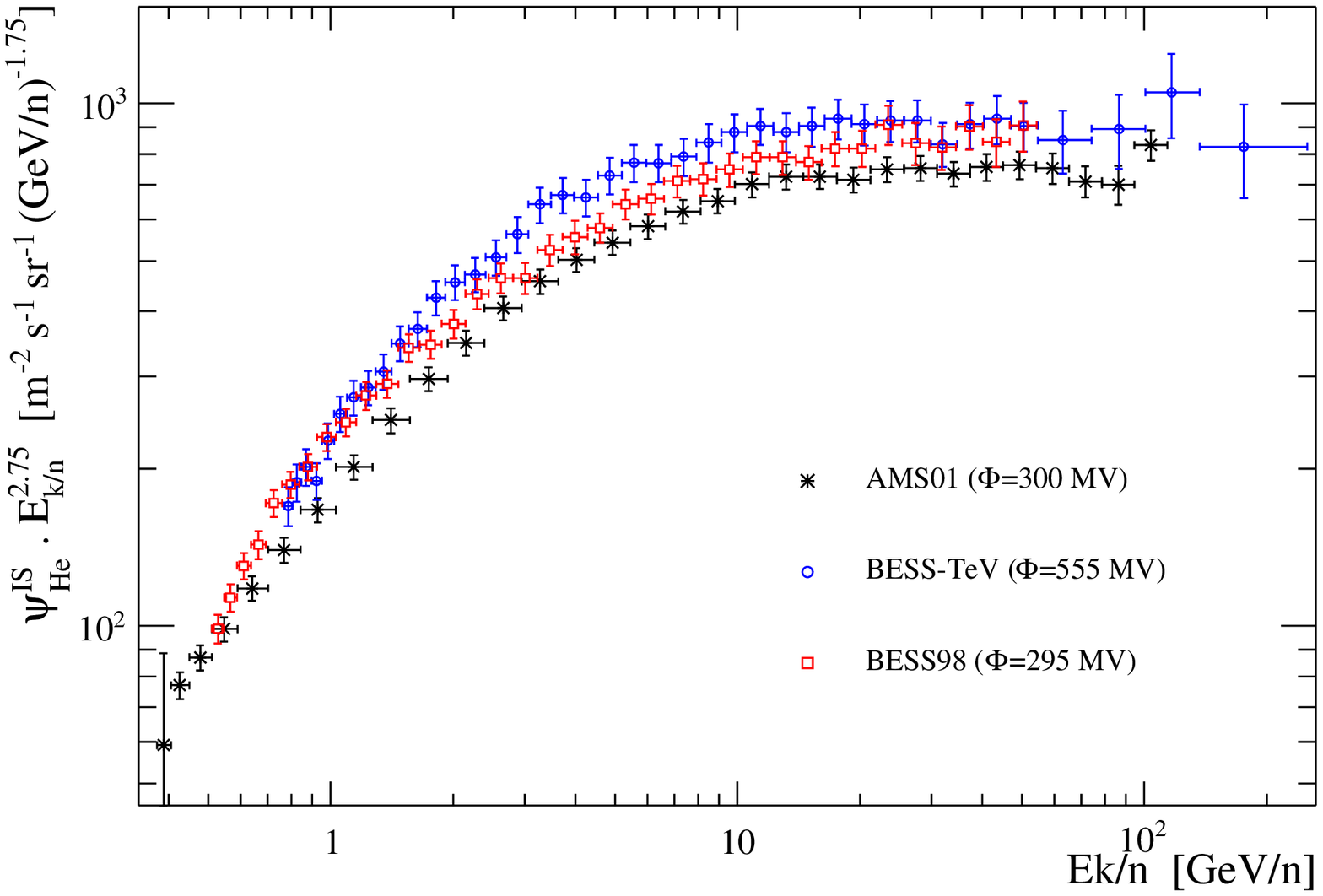}
\caption{Demodulated p (top panel) and He (bottom panel) flux $\times E_{k/n}^{2.75}$ as a
function of $E_{k/n}$ for AMS-01 (black stars), BESS98 (open red squares) and BESS-TeV
(open blue circles).}
\label{fig:data}
\end{figure}

For our analysis, the criterion is to select data samples covering a broad
energy range and consistent between experiments. We show a subset of
demodulated data in Fig.~\ref{fig:data} to illustrate the error bars and
 differences between the most recent and
consistent sets of p and He data, namely AMS-01
\citep{2000PhLB..490...27A,2000PhLB..494..193A}, BESS98
\citep{2000ApJ...545.1135S,2007APh....28..154S} and BESS-TeV
\citep[a.k.a. BESS02,][]{2004PhLB..594...35H,2007APh....28..154S}.
For the proton flux, the AMS-01 and BESS98 data, both taken in 1998 in the same
solar period, are consistent except at low energy. BESS-TeV data taken in 2002
during a high level of solar activity show a different behaviour at low and intermediate
energies. Note that usually the solar modulation level is obtained
by fitting $\phi$ and  a simple two-parameter proton spectrum to the data
\citep{2007APh....28..154S}. This is the standard lore in the field, although it
is expected to give a biased modulation level (for example, we do not know the true
interstellar proton spectrum, there is the problem of polarity in the solar magnetic field, etc). 
The use of demodulation has the same physical basis as modulating in the
same scheme (the force-field in our case, see Eq.~\ref{eq:modul}). In Figs. 1 and 2 we have chosen to
demodulate data to compare all existing data on the same foot. Since the solar
modulation is effective up to few GeV/n, only the very low side of the figures
are affected by the demodulation procedure.
The goal of this paper is not
to deal with these issues, but simply the fact that the proton data are already inconsistent
among themselves implies that we may expect inconsistencies in the fitted models.

\subsection{Pure diffusive transport}
The first step is to test a minimal model containing only
acceleration and plain diffusion, as well as nuclear reactions and
electromagnetic energy losses, but without convection and reacceleration
($V_a=V_c=0$). 

\paragraph{Main degeneracies}
In the pure diffusion model, the flux of any primary species at high energy
can be approximated by\footnote{Throughout the paper, the quantity
$\psi$ denotes the differential flux in kinetic energy per nucleon,
i.e. $d\psi/dE_{k/n}$. The notation $d\psi/d{\cal R}$ used in
Sect.~\ref{sec:ratios} is the only place where it will refer to the
differential flux in rigidity.}
\begin{equation}
\psi(E)\propto \frac{Q(E)}{K(E)} \propto \frac{q}{K_0} \cdot E^{-(\alpha+\delta)}.
\label{eq:approx}
\end{equation}
This formula shows two degeneracies between the source and transport
parameters: the first one is in the normalisation $q/K_0$,  the second one
is in the total spectral index $\alpha+\delta$. 

We start with a minimisation procedure setting the free parameters 
$\delta, \alpha$ and $q_{\rm p, He}$ (in order to break the  degeneracy
between $q$ and $K_0$ the latter is set to 0.0048 kpc$^2$~Myr$^{-1}$).
The results for protons and helium are presented in two different columns in
Table~\ref{tab:chi2-pure-diffusion}.
\begin{table}[!t]
\caption{Best-fit to p and He data for pure diffusive transport
(experiment, number of data points, best $\chi^2/{\rm d.o.f.}$, and best-fit
$\alpha+\delta$ value).}
\label{tab:chi2-pure-diffusion}
\centering
\begin{tabular}{ccccccc} \hline\hline
    Data        &          \multicolumn{3}{c}{p}             &          \multicolumn{3}{c}{He}  \\  
                &  \# & $\frac{\chi^2}{\rm d.o.f.}$& $\alpha+\delta$\,& \# &$\frac{\chi^2}{\rm d.o.f.}$&$\alpha+\delta$     \vspace{0.05cm} \\\hline
& \multicolumn{1}{c}{} \vspace{-0.20cm} \\ 
  AMS-01 (1)     &    28   &  0.19  &  2.99   \,    &   31   &  0.72  & 2.81 \vspace{0.cm}\\ 
  BESS98 (2)    &    41   &  0.39  &  2.99   \,    &   36   &  0.36  & 2.77 \vspace{0.cm}\\ 
  BESS-TeV (3)  &    47   &  0.72  &  2.89   \,    &   40   &  0.80   & 2.80     \vspace{0.1cm}\\ 
  $(1+2)$       &    69   &  0.66  &  3.00   \,    &   67   &  1.75  & 2.81 \vspace{0.cm}\\ 
  $(1+2+3)$     &   116   &  1.95  &  2.99   \,    &  107   &  2.52  & 2.81 \vspace{0.1cm}\\ 
All LE$^\dagger$&   304   &  6.22  &  2.96   \,    &  287   &  2.77  & 2.79 \vspace{0.cm}\\ 
ATIC-2          &    15   &   22.46 & 2.75     \,    &   15   &  37.4  & 2.81 \vspace{0.cm}\\ 
All HE$^\star$  &    59   &  2.58  & 2.87    \,    &   39   &   4.18  & 2.85 \vspace{0.1cm}\\ 
\hline
\end{tabular}
{\scriptsize 
$^\dagger$ IMAX92, BESS93, CAPRICE94, BESS97, AMS-01, BESS98, CAPRICE98, BESS99, BESS00, BESS02.\\
$^\star$ CREAM, SOKOL, MUBEE, JACEE, Ichimura et al., RUNJOB.}
\end{table}
The $\chi^2_{\rm min/d.o.f.}$ values are very small for a number of cases,
indicating a possible over-fitting of the data. The values of the best-fit
parameters are not reported since they are not relevant at this stage of
the analysis. For the different sets of data, the values of both $\alpha$
and $\delta$ vary from almost any value between 0 and 2.8 (not shown in the
Table), but the sum of them is close to $3.0$ for p and $2.8$ for He.  The
first three lines show that a fit to p and He on the AMS or BESS data  is
always possible in a simple diffusion scheme (although it provides
unphysical values for $\alpha$ and $\delta$). The fourth line shows the
combined analysis of AMS-01 \citep{2000PhLB..490...27A} and BESS98
\citep{2000ApJ...545.1135S} data. They were collected in the same year
(1998)|which may help reduce the systematics due to solar wind
modelling|and span nearly the same energy range. The $\chi^2_{\rm
min/d.o.f.}=1.75$ for the fit of combined He data,  compared to the
corresponding $\chi^2_{\rm min/d.o.f.}$ values of 0.72 and 0.36 for the separate
data shows inconsistencies among the data sets, as underlined in
Sect.~\ref{sec:data}. When combining the three experiments (fifth line),
this is even more visible, also for protons (the best-fit slope is
unaffected). We have fitted also all the available data in the low energy 
(sixth line) and  in the highest energy range (last line).  The agreement
among the data sets is poor, both for protons and helium, as already
visible in Fig. \ref{fig:data_all}.  ATIC data, which connect the low and
high energy sectors, are badly fit by any  pure diffusive model.

We may naively interpret the results of Table~\ref{tab:chi2-pure-diffusion}
as showing that p and He data can be well accommodated in any purely
diffusive transport models, in the low energy range ($\lesssim 100$ GeV/n),
for each experimental data set taken separately. But such models lead to
unphysical values for $\delta$ and $\alpha$. So it could also mean that
the hypothesis of a standard source spectrum (i.e., $dQ/dp\propto {\cal
R}^{-\alpha}$ when setting $\eta_S=-1$) and a standard propagation scheme
($\eta_T=1$) is unsupported by the data, or that additional effects (e.g.,
convection and/or reacceleration) are required to match the data. Before
resolving this issue, we go further with the comparison of the approximate formulae
and the full calculation.

\paragraph{Inelastic interaction: a link between $\alpha$, $\delta$ and $K_0$.}
Explicitly, the effect of the catastrophic losses at low energy, Eq. 
(\ref{eq:approx}) gives, for a 1D model,
\begin{equation}
\psi(E)=\frac{v}{4\pi} \cdot \frac{Q(E)}{\frac{K_0\beta {\cal R}^\delta}{hL} + n_{\rm ISM} \sigma v}.
\label{eq:flux_spal}
\end{equation}
For $\delta$ fixed, Eq.~(\ref{eq:flux_spal}) implies a correlation between
$K_0$, $\alpha$ and $\delta$, given some primary data. Indeed, as $K_0$
decreases, the inelastic interaction term $n_{\rm ISM} \sigma v$ becomes more efficient
in the denominator of Eq.~(\ref{eq:flux_spal}). The effect of
species destruction is therefore more pronounced at low energy. Fixing $\delta$ and
going to small $K_0$ we expect that, in order to  balance the increased
destruction rate, the numerator compensates but increasing $\alpha$. If $K_0$
is fixed and small, so that inelastic interactions can dominate, the same flux of
protons (or helium) can be obtained with a larger $\alpha+\delta$.

\begin{figure}[!t] 
\centering
\includegraphics[width=\columnwidth]{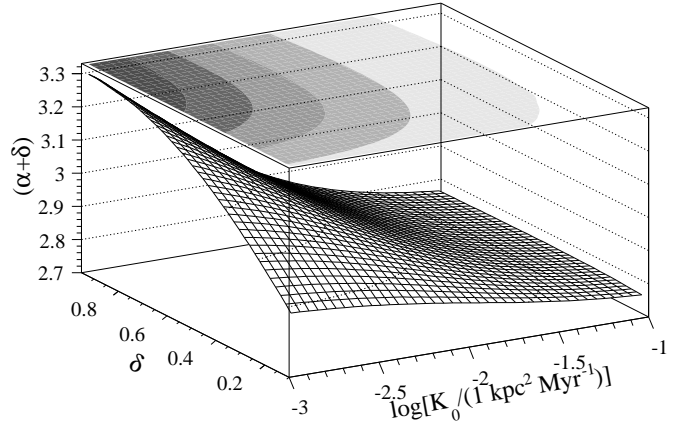}
\caption{Surfaces of $\alpha+\gamma$ for the best-fit models in the plane $K_0-\delta$
(free parameters are $\alpha$ and $q_{\rm p}$) on AMS-01 proton data. The colour code
(from light to darker shades) for the contours superimposed on top of each graph correspond to
 $(\alpha+\gamma)$ = \{2.8, 2.9, 3.0, 3.1, 3.2\}.}
\label{fig:gamma_contours}
\end{figure}

This effect is confirmed by the numerical results, as seen in
Fig.~\ref{fig:gamma_contours}. For each point in the $K_0-\delta$ plane, we
plot $\alpha+\delta$ for the best-fit model on AMS-01 proton data (the free
parameters are $\alpha$ and $q_{\rm p}$). We checked (not shown) that similar
values for $\alpha+\delta$ are obtained when the fits is performed on other
proton fluxes (BESS98 or BESS-TeV), or for other species (He or a combined
fit p+He).  We can see from the figure that, for any fixed $\delta$ value,
the data require higher  $\alpha+\delta$ while $K_0$ decreases, due to the increasing importance of the destruction rate.  This effect is
less pronounced for small values of the diffusion coefficient slope,
namely  when $\delta$ is close to 0.2-0.3.

\paragraph{Asymptotic behaviour (or why $\alpha+\delta\gtrsim \gamma_{\rm
data}$)} Even for light primary species such as protons, which suffer
less from destruction in the ISM, the asymptotic purely diffusive regime is
not reached. If we fit a primary flux with  (as is usually done in the
literature)
\[
   \psi(E) \propto  E^{-\gamma_{\rm data}}\,,
\]
then we are bound to have 
\[
\alpha+\delta \equiv \gamma_{\rm asympt}\gtrsim \gamma_{\rm data}\,.
\]
\noindent
This implies that caution is in order whenever we wish to compare the
result of studies fitting the propagated fluxes with a power-law function
\citep[e.g.,][]{2007APh....28..154S} to those (such as this one) fitting
directly the source spectrum. Catastrophic, but also continuous losses
flatten the propagated spectrum below $\lesssim $ few tens of GeV/n
energies. The inequality $\gamma_{\rm asympt}\gtrsim\gamma_{\rm data}$ is
also valid for convection and/or reacceleration, as derived from an analysis of B/C \citep{2010A&A...516A..66P,2010A&A...516A..67M}.

\paragraph{Simultaneous fit of p and He to lift the $\alpha+\delta$
degeneracy?} As the residence time|hence the destruction rate of
any species|depends on the energy through the transport parameter $\delta$
(and not on $\alpha+\delta$), the different inelastic cross-sections for
each species ($\sigma_{\rm inel}^p\sim 30$~mb and $\sigma_{\rm inel}^{\rm
He}\sim 90$~mb) leave different imprints on the corresponding p and He
low-energy spectra. This is expected, to some degree, to lift the
degeneracy on $\alpha+\delta$ when using a combined fit to various primary
species.

\begin{figure*}[!t] 
\centering
\includegraphics[width=\textwidth]{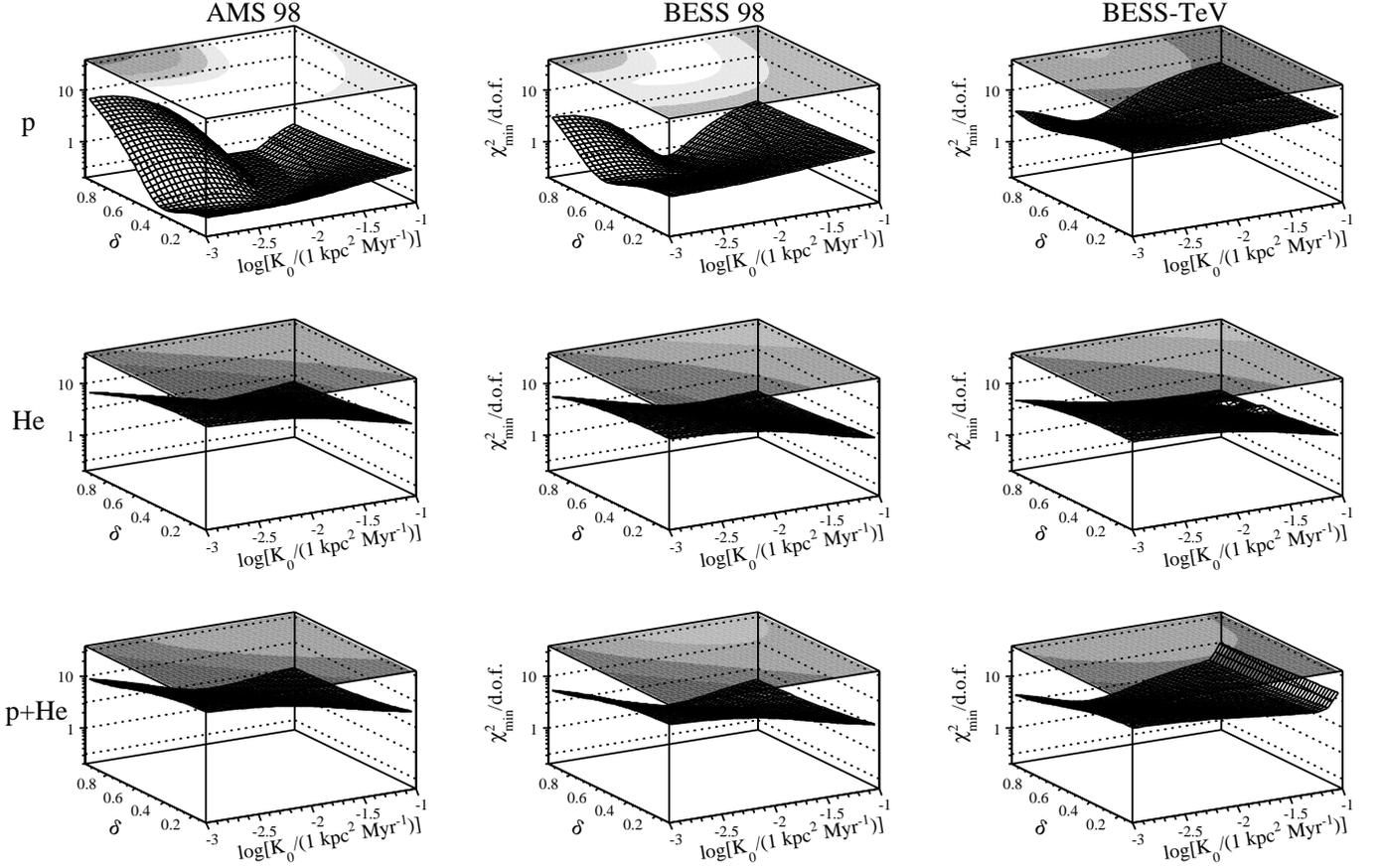}
\caption{Surfaces of $\chi^2_{\rm min/d.o.f.}$ (best-fit models) in the
plane $K_0-\delta$. First row: fit performed on p data (free parameters
$\alpha$ and $q_{\rm p}$). Second row fit performed on He data (free parameters
$\alpha$ and $q_{\rm He}$). Third row: fit performed simultaneously on p
and He data (free parameters $\alpha$, $q_{\rm p}$ and $q_{\rm He}$). The columns
from left to right correspond respectively to AMS-01, BESS88 and BESS-TeV
data. The colour code for the contour plots (iso-$\chi^2_{\rm min/d.o.f.}$)
superimposed on each graph is (from white to darker shades) $\chi^2_{\rm
min/d.o.f.}$ = \{0.5, 1, 2, 5, 10\}.}
\label{fig:chi2_contours}
\end{figure*} 
Figure~\ref{fig:chi2_contours} shows the $\chi^2_{\rm min/d.o.f.}$ contours
in the $K_0-\delta$ plane for the separate fits of p (first row), He (second
row) and for the combined fit p+He (last row), for the different sets of
data  used before. The ranges chosen for $K_0$ and $\delta$ correspond to
extreme but not impossible values of these parameters that can accommodate
the secondary-to-primary B/C ratio \citep{2010A&A...516A..67M}. The
top-left plot shows the strong degeneracy of $\alpha$ and $\delta$  (for
AMS-01 data), as almost any configuration is acceptable. The top-right plot
shows that for other data (BESS-TeV) no good fit can be achieved
($\chi^2_{\rm min/d.o.f.}>2$) in the selected $K_0-\delta$ region: the good
fits occur for unrealistic $\delta$ only. We note here that there is
no  inconsistency with the results shown in Table
\ref{tab:chi2-pure-diffusion},  which have been obtained by spanning
larger ranges for the free parameters including unphysical values.  The second row of Fig.~\ref{fig:chi2_contours}
shows the fits to He data. All the experiments tend to prefer large $K_0$
and large $\delta$ (both unrealistic if we demand $\chi^2_{\rm
min/d.o.f.}$$\sim 1$), but the $\chi^2$ does not change significantly, meaning that no
particular class of models is selected by the helium data.  

The third row of Fig.~\ref{fig:chi2_contours} is the combined p+He fit. The
resulting $\chi^2_{\rm min/d.o.f.}$ surfaces corresponds to a trade-off
between the best-fit for p and He. The best-fit $\delta$
(not shown in the figure) still falls in region of $\delta\gtrsim 1$, and
the degeneracy $\alpha+\delta$ is not lifted as no specific value for
$\delta$ is preferred.

\paragraph{The role of $\eta_S$ and $\eta_T$.}
We have introduced the possibility to have a non-standard low-energy diffusion
coefficient by means of the parameter $\eta_T$, as given by Eq.~(\ref{eq:eta_T}).
The low-energy shape of the source spectrum is driven by the parameter
$\eta_S$, see Eq.~(\ref{eq:source_spec}). If for illustration we neglect the nuclear
interactions, we obtain
\begin{equation}
\psi(E)\propto \frac{Q(E)}{K(E)}\propto \beta^{\eta_S+1-\eta_T}
{\cal R}^{-\alpha-\delta},
 \label{eq:eta_S-eta_T}
\end{equation} 
where the extra $\beta$ factor comes from the 
$v/4\pi$ in front of Eq.~(\ref{eq:flux_spal}).

The parameters $\eta_S$ and $\eta_T$ introduce a similar shape correction
at the lowest energies. If it were not for energy losses and inelastic reactions 
(see next section), only the quantity $\eta_S-\eta_T$ would be expected
to be constrained. We  fit AMS-01 and BESS98 data, as
well as all the proton data, with $\alpha$, $\eta_S$, $\delta$, and $q_{\rm p}$
as free parameters. The best
$\chi^2$ is slightly smaller than the one obtained with only $\alpha$, 
$\delta$, and $q_{\rm p}$ free (and $\eta_S=-1$). Similar results are achieved when the
acceleration scheme is fixed to the standard lore ($\eta_S=-1$) and the
fourth free parameter is $\eta_T$. However, the corresponding
values for $\delta$ and $\alpha$ are again unsupported.

A further degeneracy can be produced by solar modulation, whose action 
is a decrease of the flux with the increase of the solar wind strength 
(parameter $\phi$). On the other hand, the TOA flux increases with
$\phi$ when $\eta_S \leq \eta_T$ (in the standard scenario $\eta_s=-1$ and
$\eta_T=1$), so that some compensation of the solar modulation can be produced.

\subsection{Summary for the transport parameter constraints from p and He data}

The existing proton and helium data are unable to select any particular
propagation model. This is consistent with the fact that the transport and
source parameters are degenerate, as shown from simple arguments in our toy
formulae. The best present data (AMS-01, BESS98 and BESS-TeV) can be quite
well reproduced by solar modulated pure diffusive transport for instance, 
but they favour unphysical values of the source and transport parameters.
Hence other ingredients are required. This could be a modification of the
low-energy source spectrum or diffusion coefficient, or the addition of
convection and/or reacceleration (which can also accommodate the current
data), or an improvement of the calculation of the solar modulation effect. 
However, given the present data and the physics of primary spectra, such an
approach is bound to fail: the increase of the parameter space merely brings
new degeneracies. Moreover, most of the parameter space is already ruled
out by the constraints from the B/C ratio. To go further, we thus have to restrict the
parameter space to the source parameter space only (and use some prior
on the propagation parameters).

\section{Analysis with fixed transport parameters\label{sec:pHe_propagparamfixed}}

In the previous sections we have shown that present data on primary fluxes alone cannot constrain
significantly the transport parameters. The next natural strategy  is to
fit simultaneously primary and secondary species. However, as emphasised
in \citet{2009A&A...497..991P}, the large body of data for primary species
drives the fit away from the best-fit regions of the B/C ratio. Actually,
the standard lore is to fix the transport parameters to
their best-fit value, and then constrain the source parameters. But the
latter values are then biased\footnote{\citet{2009A&A...497..991P} show
that the values of the source-spectrum parameters (slope and abundances)
are positively correlated among themselves and with the reacceleration
strength, but are negatively correlated with the other propagation
parameters.}. We nevertheless follow this approach, but we repeat
the analysis on several possible transport configurations. This allows us to derive explicitly 
the systematic effects of the source parameters arising from 
this bias.

In this section, we first gather several sets of transport parameters shown
to be consistent with B/C data (Sect.~\ref{sef:goodBC}). We then fit the
source parameters for p and He, the best-measured primary fluxes to
date (Sect.~\ref{sec:BCfixed_pHe}). We repeat the analysis for other
primary species, to inspect the universality of the source slopes and
obtain their relative source abundances
(Sect.~\ref{sec:BCfixed_otherPrim}).

    \subsection{Transport parameters consistent with B/C\label{sef:goodBC}}

The transport parameters are usually constrained from secondary-to-primary
ratios (e.g. B/C). In the literature, various classes of models have been
used, leading to very different values of their respective best-fit
parameters (see, e.g., \citealt{2007ARNPS..57..285S} for a review). For instance, a model with diffusion + reacceleration is
characterised by a best-fit propagation slope $\delta\approx 0.3-0.4$
\citep[e.g.,][]{2005JCAP...09..010L}, leaky-box inspired models points to
$\delta\approx 0.5-0.6$
\citep[e.g.,][]{2003ApJS..144..153W,2009A&A...497..991P}, whereas diffusion
$+$ convection models (w/wo reacceleration) points to
$\delta\approx 0.75-0.85$ 
\citep[e.g.,][]{2001ApJ...555..585M,2010A&A...516A..66P}. 

This sensitivity to the CR transport mode (pure diffusion, w/wo convection,
w/wo reacceleration) is discussed in \citet{2010A&A...516A..67M}. Although the
best-fit model is one with both convection and reacceleration, it predicts
$\delta\sim 0.8$, a value quite high compared with theoretical
expectations (1/3 for a Kolmogorov spectrum of turbulence, and 1/2
for Kraichnain, e.g., \citealt{2007ARNPS..57..285S}). Following \citet{2010A&A...516A..67M}, we use below four
configurations of the diffusion model covering a large but plausible range for
the transport parameters. These models along with their best-fit parameters
are reproduced in Tab.~\ref{tab:BC_models}:
\begin{itemize}  
  \item Model~II is with reacceleration only;
  \item Model~III is with convection and reacceleration;
  \item Model~I/0 is with a low-energy upturn of the diffusion coefficient
  ($\eta_T<0$);
  \item Model~III/II is as I/0, but with reacceleration.
\end{itemize}
The first two models correspond to the best-fit parameters for a standard
spatial diffusion coefficient [i.e. $\eta_T$ set to 1, see
Eq.~(\ref{eq:eta_T})]. The last two lines correspond to a modified diffusion
scheme: negative values of $\eta_T$ are associated to an upturn of the
diffusion coefficient at low energy \citep{2006ApJ...642..902P}. These two
models are respectively termed I/0 and III/II because both allow some
convection, but both favour $V_c^{\rm best}=0$. As shown in Fig.~7 of
\citet{2010A&A...516A..67M}, these models fit reasonably well the B/C data.
\begin{table}[!t]
\caption{Best-fit transport parameters for various configurations of
the diffusion model (fitted on B/C data).}
\label{tab:BC_models}
\centering
\begin{tabular}{lcccccc} \hline\hline
Model$\!\!\!\!$  & $\eta_T$ & $\!\!\!K_0^{\rm best}\times 10^2\!\!\!$     & $\delta^{\rm best}$ &  $V_c^{\rm best}$  &  $V_a^{\rm best}$  & $\!\!\!\!\chi^2$/d.o.f$\!\!\!\!$   \\
       &          & $\!\!\!\!\!\!$(kpc$^2\,$Myr$^{-1}$)$\!\!\!\!\!\!$ &         &  $\!\!\!\!$(km$\;$s$^{-1}$)$\!\!$  & $\!\!$(km$\;$s$^{-1}$)$\!\!\!\!$   & \\\hline
& \multicolumn{1}{c}{} \vspace{-0.20cm} \\ 
II             &   1. & 9.76 & 0.23 & \dots& 73.2 & 4.73 \vspace{0.05cm}\\  
III            &   1. & 0.48 & 0.86 & 18.8 & 38.0 & 1.47 \vspace{0.1cm}\\  
I/0$^\ddagger$ & -2.6 & 2.05 & 0.61 &   0. & \dots& 3.29 \vspace{0.05cm}\\  
III/II         & -1.3 & 3.16 & 0.51 &   0. & 45.4 & 2.26 \vspace{0.05cm}\\  
\hline
\end{tabular}
\note{\tiny These results were obtained for input ingredients described in \citet{2010A&A...516A..67M},
and for $L=4$~kpc.\vspace{-0.2cm}}
\end{table}

    \subsection{Constraints on p and He source parameters\label{sec:BCfixed_pHe}}

\subsubsection{Generalities}
On the one hand, the low-energy shape of the source spectra are not well
known theoretically. They result from the diffusive shock acceleration
mechanisms at play in supernova \citep[e.g.][]{1983RPPh...46..973D} or
super-bubble shocks \citep[e.g.][]{2008MNRAS.383...41F,2010A&A...510A.101F}. Power-laws
close to $-2$ (in energy space) are predicted at high energy, but
there is still no agreement about the low-energy spectrum
\citep[e.g.][]{2010APh....33..160C}.

On the other hand, we have access only to
propagated spectra, where effects such as destruction on the ISM, energy
losses, Galactic winds and reacceleration change the energy spectra up
to a few tens of GeV/n. This is the route followed in this section,
where we try to constrain the source parameters from a fit to
the propagated fluxes. However, the shape of the flux at low-energy
is only an extrapolation since the low-energy interstellar spectrum 
is masked by solar modulation effects. 

Note that the low-energy (below 100~MeV) IS spectrum can be indirectly
constrained due to its interaction with the interstellar medium. In
this approach, the IS flux is based on empirical fits to the data
\citep[e.g.,][]{2010JGRA..11500I20H}, and its extrapolation at low energy
is used to calculate, e.g., the ionisation of the ISM
\citep{1987A&A...179..277W,1994MNRAS.267..447N,1998ApJ...506..329W} or and
of molecular clouds \citep{2009A&A...501..619P}, or the LiBeB Galactic
enrichment and production
\citep{1992Natur.357..379G,1994MNRAS.267..447N,1998ApJ...499..735L}. Some
of these studies find an increase in the low-energy spectrum, others a decrease
(with respect to a pure power law). 
Actually, data from the Voyager 1 \& 2 spacecraft near the heliospheric
termination shock could also be helpful for such studies, as they are
close to interstellar conditions, their level of modulation being $\approx 60$ MV 
\citep{2008JGRA..11310108W,2009JGRA..11402103W}. However, it has been
argued recently that anomalous cosmic rays could contribute to
an important fraction of the proton spectrum below 300 MeV \citep{2008ApJ...680L.105S}.
For this reason, we do not include Voyager data in our fits, and will
only compare them to the best-fit spectra (based on the other data)
at the end of this section.

\subsubsection{Results}
Due to the lack of robust information about the low-energy spectrum, we
choose to rely on a simple parametrisation allowing for an increase or
decrease at low-energy, as given by Eq.~(\ref{eq:source_spec}), i.e.
$Q_{E_{k/n}}(E) =  q \cdot \beta^{\eta_S} \cdot {\cal R}^{- \alpha}$.
For each configuration given in Table~\ref{tab:BC_models}|i.e., for a
given choice of the transport parameters $K_0$, $\delta$, $\eta_T$,
$V_a$, and $V_c$|we then use the MCMC technique to get the probability
density function (PDF)\footnote{The MCMC technique and the interface
with the propagation code is explained in details in
\citet{2009A&A...497..991P}. A shorter introduction is given in
\citet{2010A&A...516A..66P}.} of the three source parameters $q_i$,
$\eta_S^i$ and $\alpha_i$ (where $i$ is either p or He). The three
data sets on which we base the analysis are AMS-01, BESS98 and BESS-TeV
(see Sect.~\ref{sec:pHe_alone}). 

\paragraph{PDF of $\alpha_{\rm p}$, $\alpha_{\rm He}$, and $\alpha_{\rm He}-\alpha_{\rm p}$.}
\begin{figure}[t!]
\centering
\includegraphics[width = .5\textwidth]{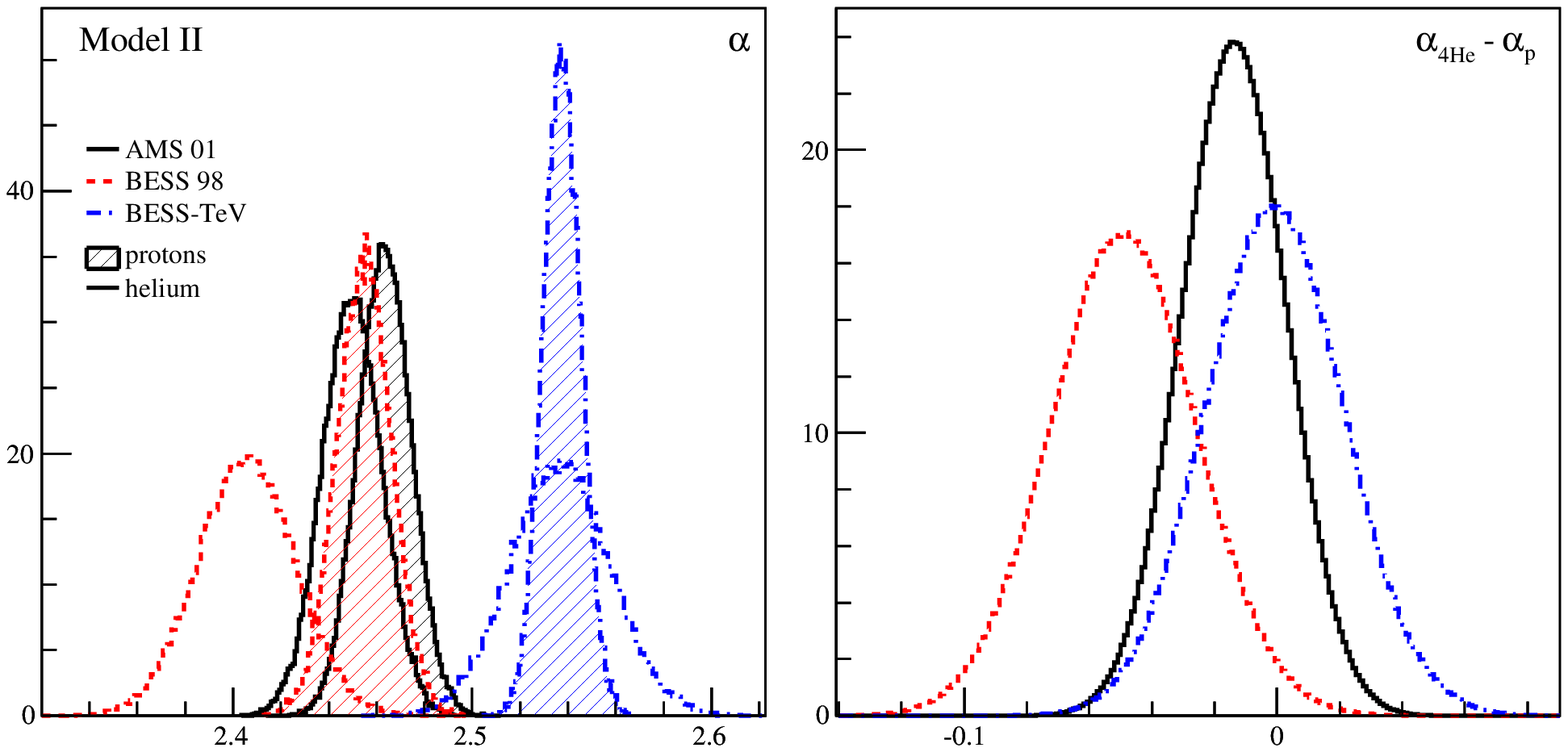}
\includegraphics[width = .5\textwidth]{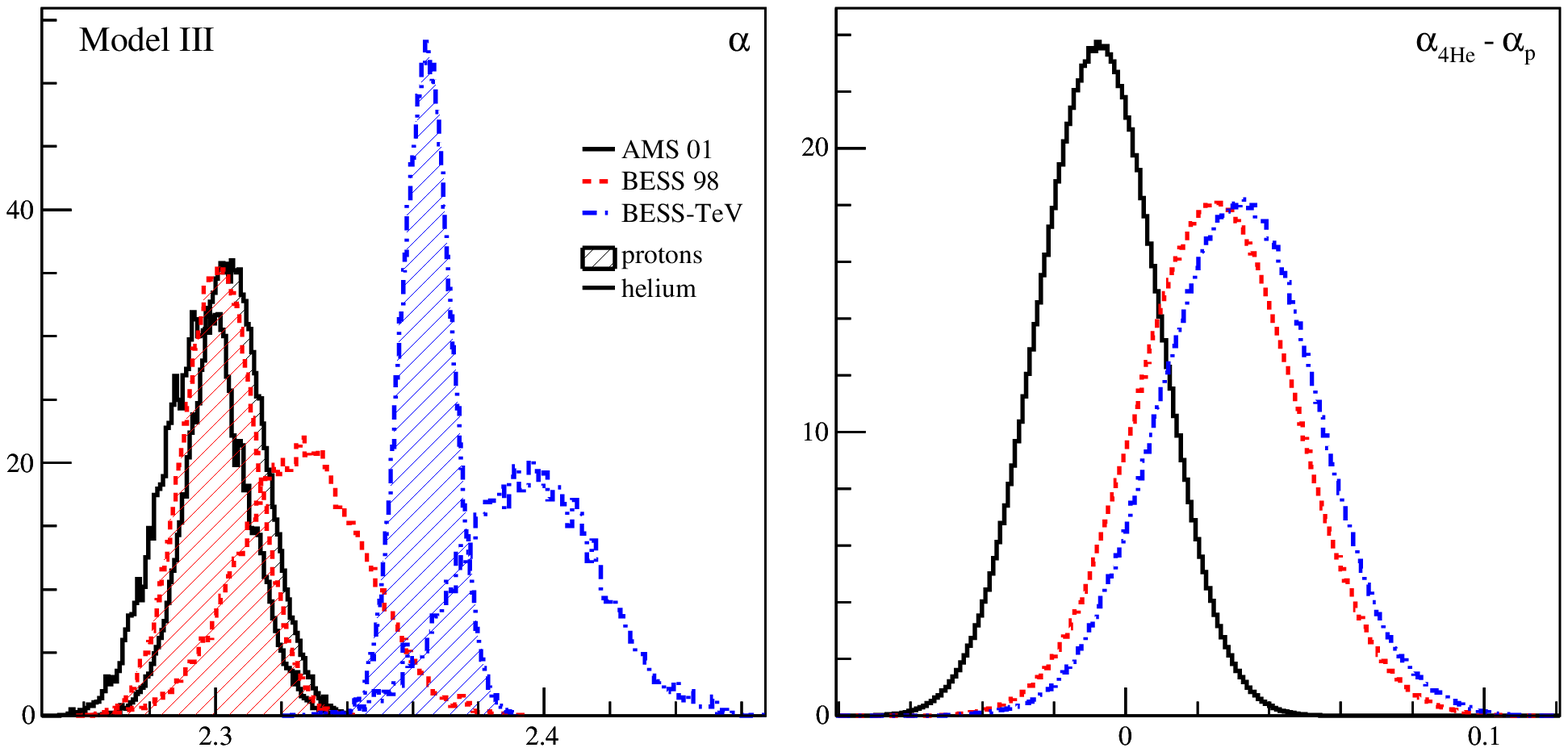}
\caption{Left panels: PDF of the source slope $\alpha$ for p (unfilled
histograms) and He (hatched histograms). Right panels: PDF for $\alpha_{\rm
He}-\alpha_{\rm p}$. The colour code corresponds to the three experimental data
used: AMS-01 (solid black line), BESS98 (dashed red lines) and BESS-TeV
(dash-dotted lines).}
\label{fig:pHe_comparison}
\end{figure}

\begin{figure}[t!]
\centering
\includegraphics[width = .5\textwidth]{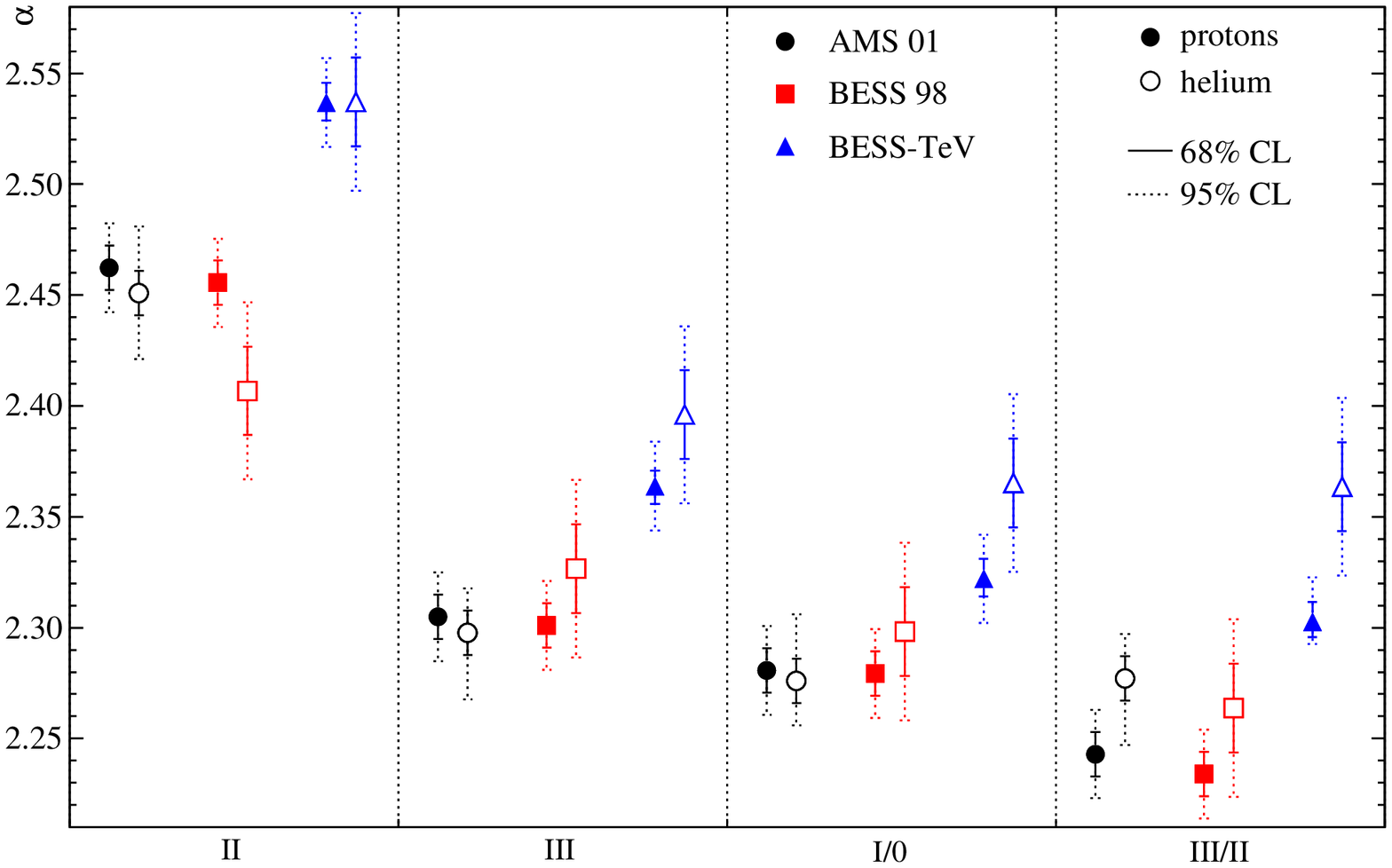}
\includegraphics[width = .5\textwidth]{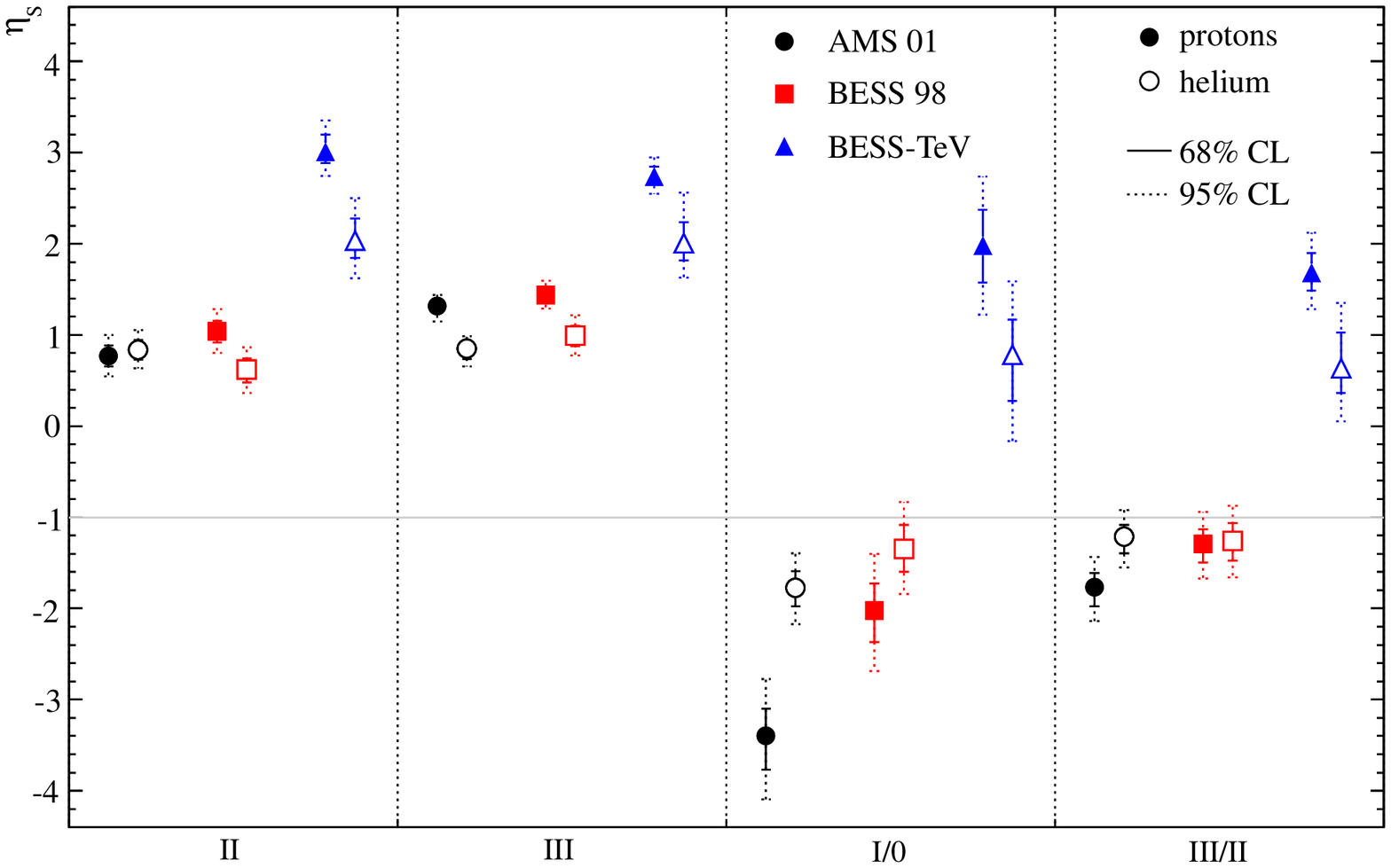}
\caption{Most-likely value (symbols), 68\% and 95\% CIs for p (filled symbols)
and He (open symbols) for the four propagation configurations gathered in
Tab.~\ref{tab:BC_models}. Top panel: spectral index $\alpha$. Bottom
panel: low-energy source parameter $\eta_S$ [see Eq.(\ref{eq:source_spec})].
The grey line $\eta_S=-1$ corresponds to the value for which the source
spectrum is a pure power-law in rigidity (i.e. $dQ/d{\cal R}\propto {\cal R^{-\alpha}}$).}
\label{fig:pHe_comparison2}
\end{figure}
The two left panels of Fig.~\ref{fig:pHe_comparison} show the PDF of
$\alpha_{\rm p}$ (hatched histograms) and $\alpha_{\rm He}$ (empty histograms),
whereas the two right panels show the PDF of  $\alpha_{\rm He}-\alpha_{\rm p}$ to
visually inspect any discrepant spectral index for the two species. For AMS-data
(solid black lines), both Model II (reacceleration, $\delta=0.23$) and
Model III (reacceleration and convection, $\delta=0.86$) show a very good
agreement between their p and He spectral index, with  respectively
$\alpha_{\rm II}\approx 2.45$ and $\alpha_{\rm III}\approx 2.3$. There are
significant differences for BESS98 (red dashed lines) and BESS-TeV (blue
dash-dotted lines) data: first, the match between $\alpha_{\rm
He}$ and $\alpha_{\rm p}$ is not as good as for AMS-01, yet
$\alpha_{\rm He}-\alpha_{\rm p}$ remains marginally
consistent with 0. The plots on the right panels and the width of the PDF
tell us that the data current precision does not allow us to separate
differences $\lesssim 0.1$ in the spectral indices. 

The 68\% and 95\% CIs on the spectral indices for the four transport
configurations of Table~\ref{tab:BC_models} and the three sets of data
are shown in the top panel of Fig.~\ref{fig:pHe_comparison2}.
From a quick visual inspection, the following trends are found:
\begin{itemize}
  \item for any given data set and species (p or He), the spread in the
  source slopes is $\alpha_p -\alpha_{\rm He}\lesssim 0.2$,
  regardless of the model (values in the same column in
  Table~\ref{tab:pHe_BCfixed});
  \item for any given model, the typical spread in  $\alpha$ when fitting
  different data sets  (values in the same row in
  Table~\ref{tab:pHe_BCfixed}) is $\approx 0.05$ for $\alpha_p$ 
  and $\approx 0.1$ for $\alpha_{\rm He}$. This is larger than
  the errors extracted from the {\sc minuit} minimisation routine, which gives
  a statistical uncertainty $\alpha_p -\alpha_{\rm He}\lesssim 0.01$ (not shown);
  \item the spectral indices obtained from BESS-TeV data are systematically
  larger and marginally incompatible with those found for AMS-01 and
  BESS98. This may be related to the systematically higher value obtained
  for $\eta_S$ (see below).
  \item model II (reacceleration only) gives larger spectral indices,
  inconsistent with the values found for the three other models. This
  is not unexpected as it has the smallest $\delta$ of all models 
  (considered in Table~\ref{tab:BC_models}).    
\end{itemize}
A scatter of $\sim0.2$ is thus attributed to the fact that
we do not know which model is best, and a scatter $\sim0.1$
because of systematics in the data. 

\paragraph{Low energy and confidence intervals (CIs) on $\eta_S$}
The 68\% and 95\% CIs on the parameter $\eta_S$ controlling the low-energy
behaviour of the source spectrum\footnote{We underline that for IS fluxes,
the lowest energy data points are at $\sim 0.8$ GeV/n for p and $\sim 0.4$
GeV/n for He, see Fig.~\ref{fig:data}. This corresponds to $\beta_{\rm p}\sim
0.8$ ($\beta_{\rm He}\sim 0.7$), so that having a pre-factor $\beta^3$
amounts to a difference of $\sim 1/2$ ($\sim 1/3$ for He) with respect to
the case $\eta_S=0$ for this low energy point.} are shown for the same
models/data in the bottom panel of Fig.~\ref{fig:pHe_comparison2}. BESS-TeV
data being at slightly higher energy than AMS-01 and BESS98, its source
spectrum low-energy parameter $\eta_S$ is less constrained. Otherwise, the
p and He $\eta_S$ point to fairly similar values for any given propagation
configuration. However, this value depends on the model chosen: the
reacceleration model (II) and  convection/reacceleration model (III) both
favour $\eta_S\approx 1$, whereas $\eta_S$ is close to -2 for Model I/0 and
-1.5 for Model III/II. The latter value is consistent with a source spectrum
being a pure power-law in rigidity, whereas the former value implies a
flattening at low energy. This is understood if we inspect the quantity
$\eta_S-\eta_T$, appearing in Eq.~(\ref{eq:eta_S-eta_T}): for models II and
III that have $\eta_T=1$, this give  $\eta_S-\eta_T\approx 0$. For models I/0
and III/II that have respectively $\eta_T=-2.6$ and -1.3, this gives
$\eta_S-\eta_T\approx 0.6$ and -0.2. So it seems that the constraint
$\eta_S-\eta_T\approx 0$ should be met for any propagation model.

\paragraph{Source abundances $q_i$}
The scatter is quite large when all the different models/data are considered.
The absolute values are not meaningful since they depends on the choice
of $L$ that is arbitrary set to 4 kpc in this analysis. We nevertheless
note that the ratio $q_{\rm He}/q_{\rm p}$ falls in the range 0.3-0.6 (not shown),
with a typical spread of $\approx 0.1-0.2$ for the PDF.

\paragraph{Spectra, goodness of fit, and high-energy asymptotic regime}
\begin{figure}[t!]
\centering
\includegraphics[width = 0.5\textwidth]{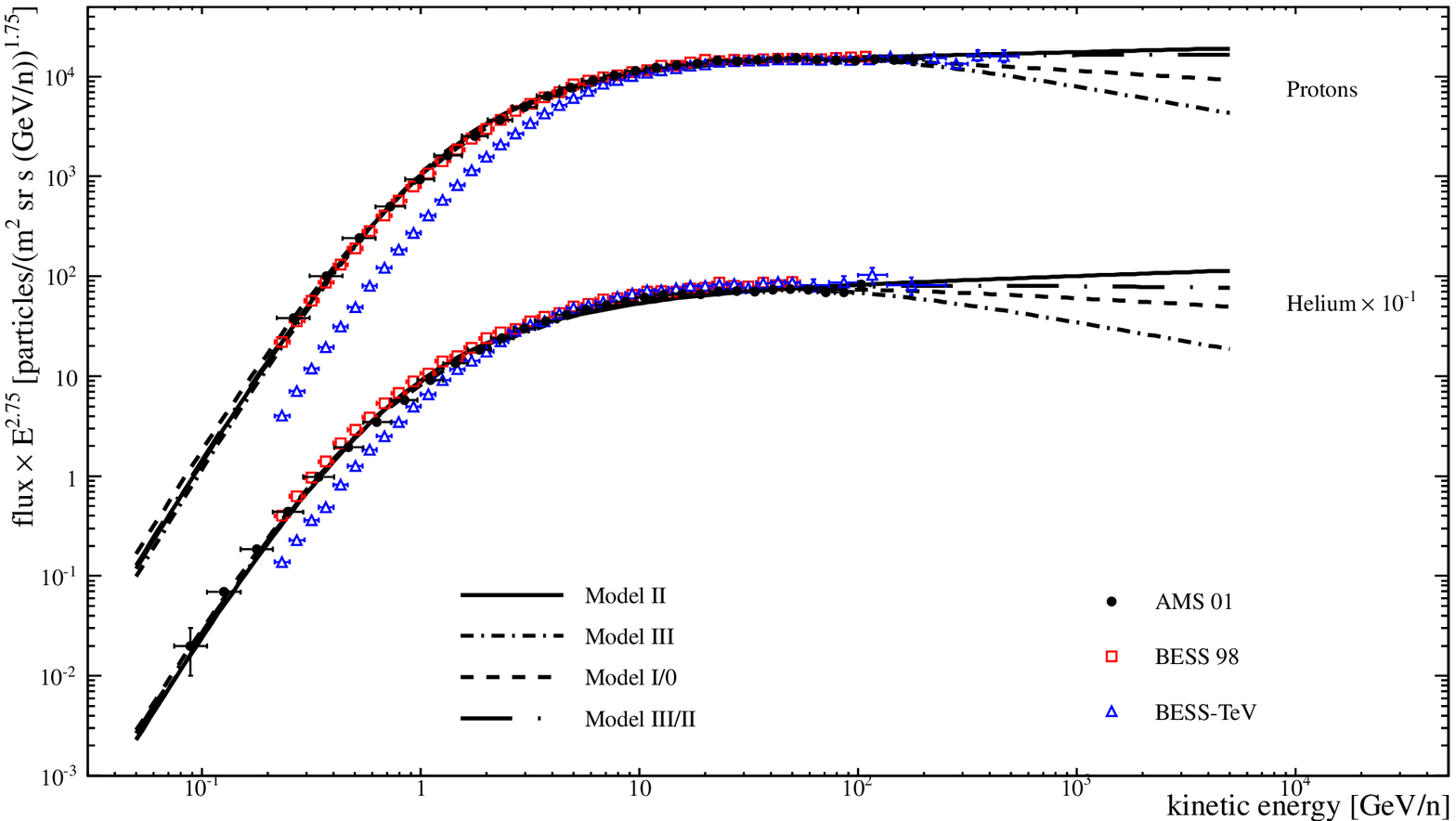}
\includegraphics[width = 0.5\textwidth]{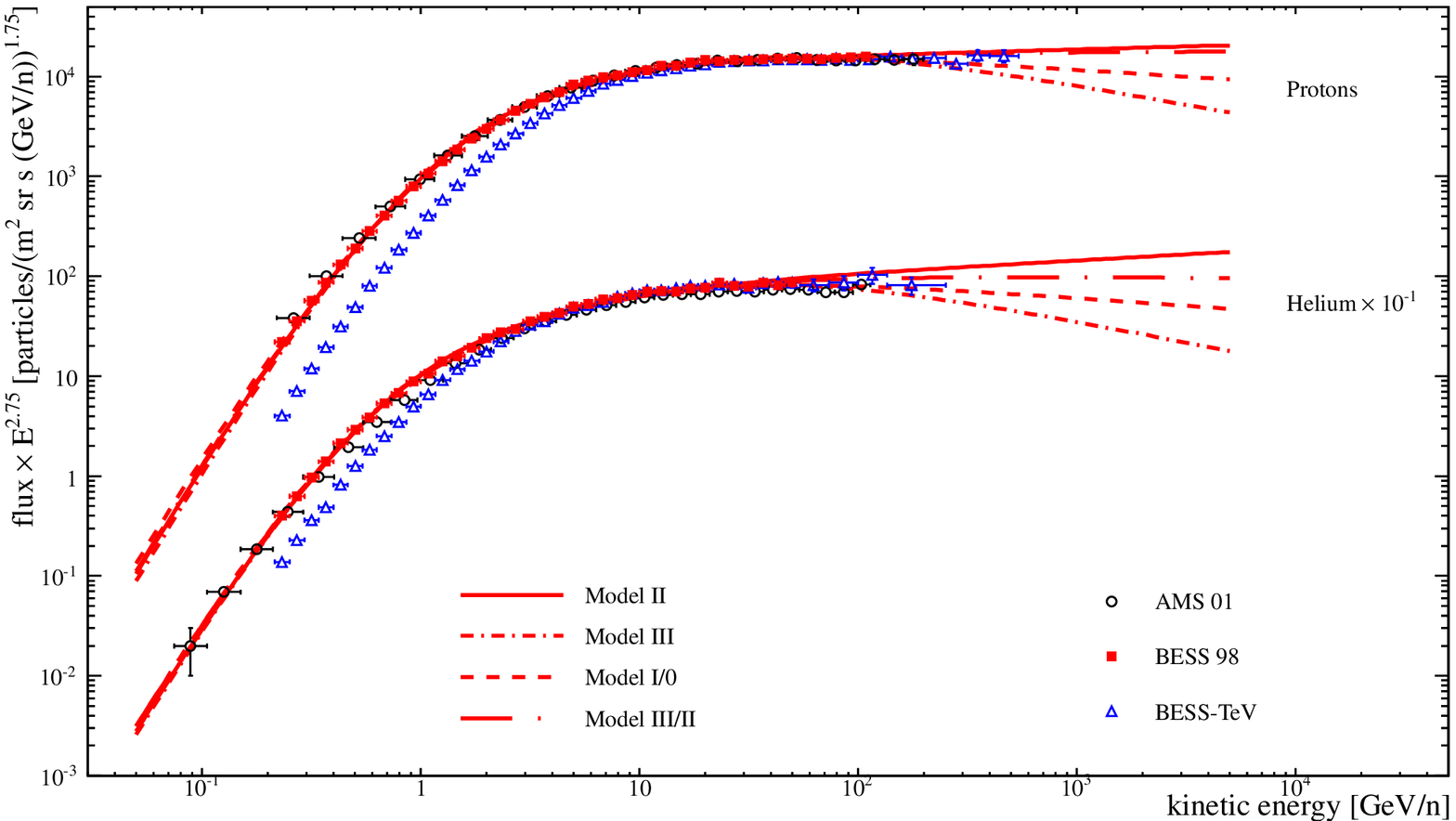}
\includegraphics[width = 0.5\textwidth]{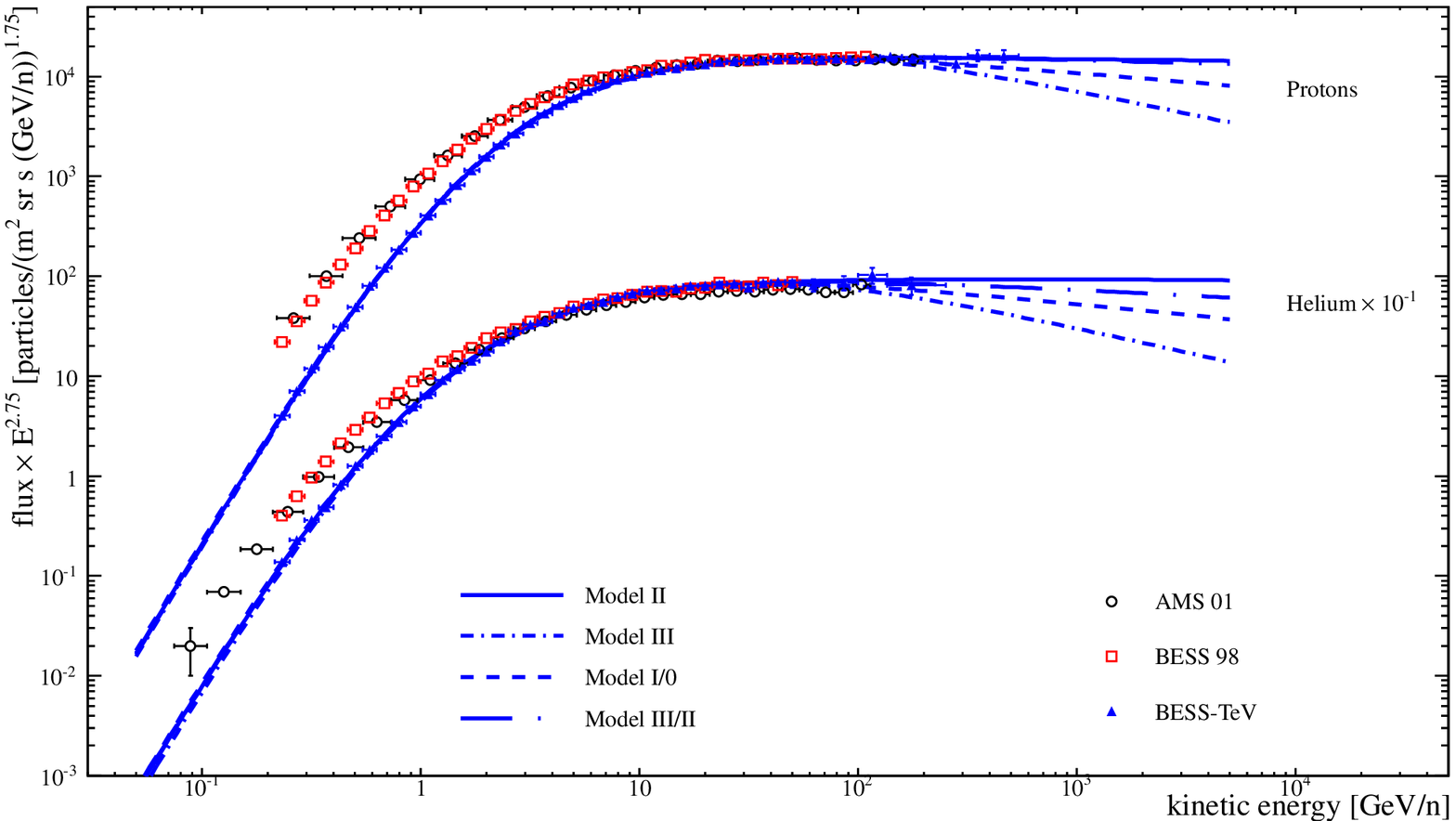}
\caption{TOA (modulated) fluxes (times $E_{k/n}^{2.75}$) as a function of the kinetic
energy per nucleon, for p and He. The symbols are black circles
for AMS-01, red squares for BESS-98, and blue triangles for BESS-TeV.
The curves correspond (for the four models of Table~\ref{tab:BC_models})
to the best-fit spectra obtained by a fit on the AMS-01 data 
(top panel), BESS98 data (middle panel) and BESS-TeV data 
(bottom panel).}
\label{fig:fluxes_fitAMSBESS}
\end{figure}
\begin{table}[t!]
\caption{Best-fit spectral index $\alpha$ for p and He fit
and associated $\chi^2_{\rm min/d.o.f.}$ (models/data correspond
to those shown in Fig.~\ref{fig:pHe_comparison}, and the number of data
for the fit is roughly the same for each species).}
\label{tab:pHe_BCfixed}
\centering
\begin{tabular}{lc|cc|cc|cc|c} \hline\hline
& \multicolumn{1}{c}{} \vspace{-0.30cm} \\ 
Model/Data &\multicolumn{2}{c}{AMS-01}&\multicolumn{2}{c}{BESS98}& \multicolumn{2}{c}{BESS-TeV} \\
           & \multicolumn{2}{c}{$\alpha^{\rm best}|\frac{\chi^2_{\rm min}}{d.o.f}$}
  & \multicolumn{2}{c}{\dots} & \multicolumn{2}{c}{\dots} \\\hline
  & \multicolumn{1}{c}{} \vspace{-0.25cm} \\ 
  \multicolumn{7}{c}{| Protons |} \vspace{0.10cm}\\ 
II	    &2.46&2.14 &2.45&0.70 &2.54&0.48 \vspace{0.05cm}\\
III    &2.30&3.72 &2.30&1.75 &2.36&2.07 \vspace{0.05cm}\\
III/II &2.24&2.05 &2.23&0.73 &2.30&0.93 \vspace{0.05cm}\\
I/0    &2.28&0.14 &2.28&0.18 &2.32&0.56 \vspace{0.05cm}\\
& \multicolumn{1}{c}{} \vspace{-0.25cm} \\ 
  \multicolumn{7}{c}{| Helium |} \vspace{0.10cm} \\ 
II	    &2.45&3.54 &2.41&0.69 &2.53&0.34 \vspace{0.05cm}\\
III    &2.30&2.85 &2.32&0.41 &2.39&0.40 \vspace{0.05cm}\\
III/II &2.27&2.02 &2.26&0.33 &2.36&0.25 \vspace{0.05cm}\\
I/0    &2.27&1.00 &2.30&0.21 &2.36&0.24 \vspace{0.10cm}\\
\hline
\end{tabular}
\end{table}
%
The data along with the best-fit spectra for all models are shown in
Fig.~\ref{fig:fluxes_fitAMSBESS}. An eye inspection shows a good match
to the data. More precisely, the $\chi^2_{\rm min/d.o.f.}$ values given in
Table~\ref{tab:pHe_BCfixed} tell us that the fit to the data is
very good for BESS98 and BESS-TeV, but not satisfactory for most of the
models with AMS-01 data (for which the first and last two bins are not well
reproduced given their small error bars). We remark that the spread in $\delta$ is larger
than the spread in $\alpha$ (see above). Hence, the smaller $\delta$, the
smaller $\gamma_{\rm asympt} (= \alpha+\delta)$. This is consistent with the
same ordering for all species of the propagated spectra|from larger to
smaller $\gamma_{\rm asympt}$|seen on Fig.~\ref{fig:fluxes_fitAMSBESS}. The
top curve is always Model II ($\delta=0.23$, solid lines), going down to
Model~III/II ($\delta=0.51$, long dash-dotted lines),  Model~I/0
($\delta=0.61$, dashed lines), and then Model~III ($\delta=0.86$, short
dash-dotted lines) for the bottom curve. This emphasises that more
accurate data in the high energy regime (TeV-PeV) are needed to
better constrain the asymptotic behaviour.

\begin{figure*}[t!]
\centering
\includegraphics[width = 0.32\textwidth]{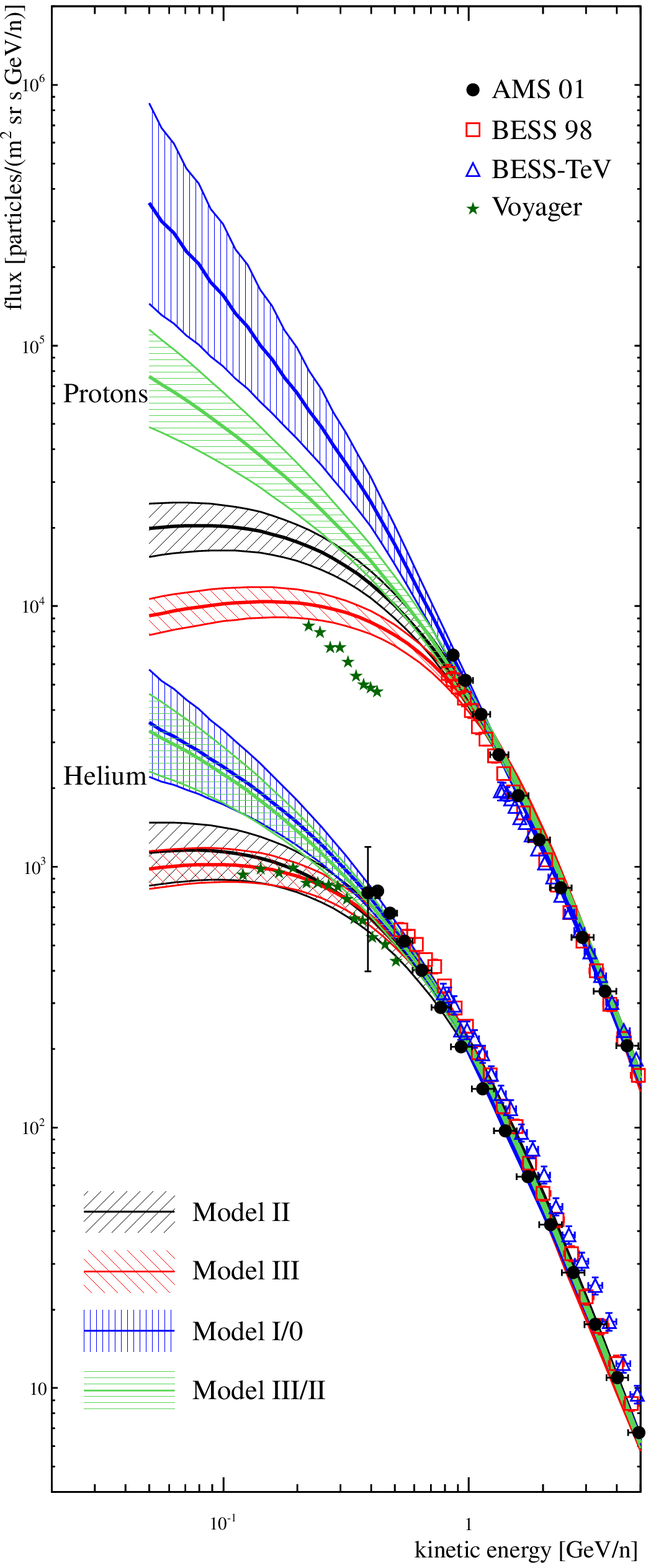}
\includegraphics[width = 0.32\textwidth]{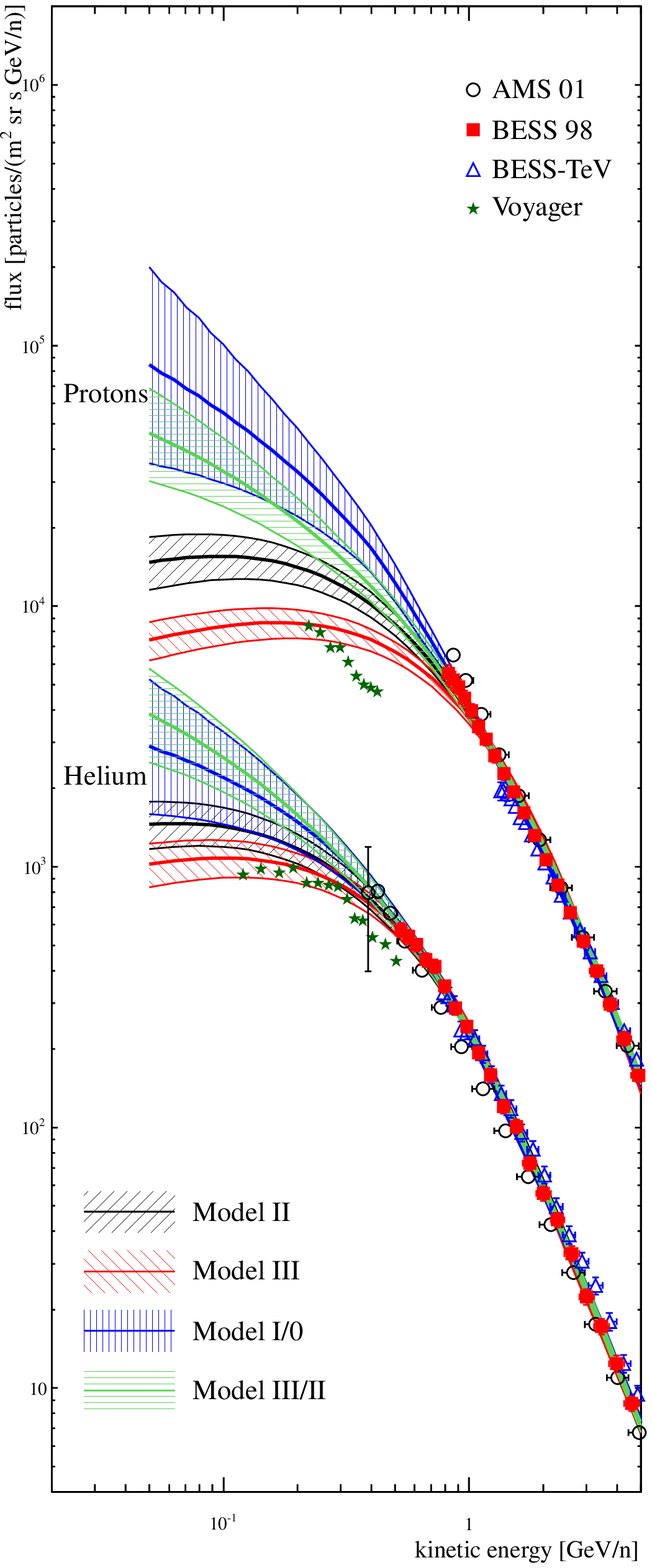}
\includegraphics[width = 0.32\textwidth]{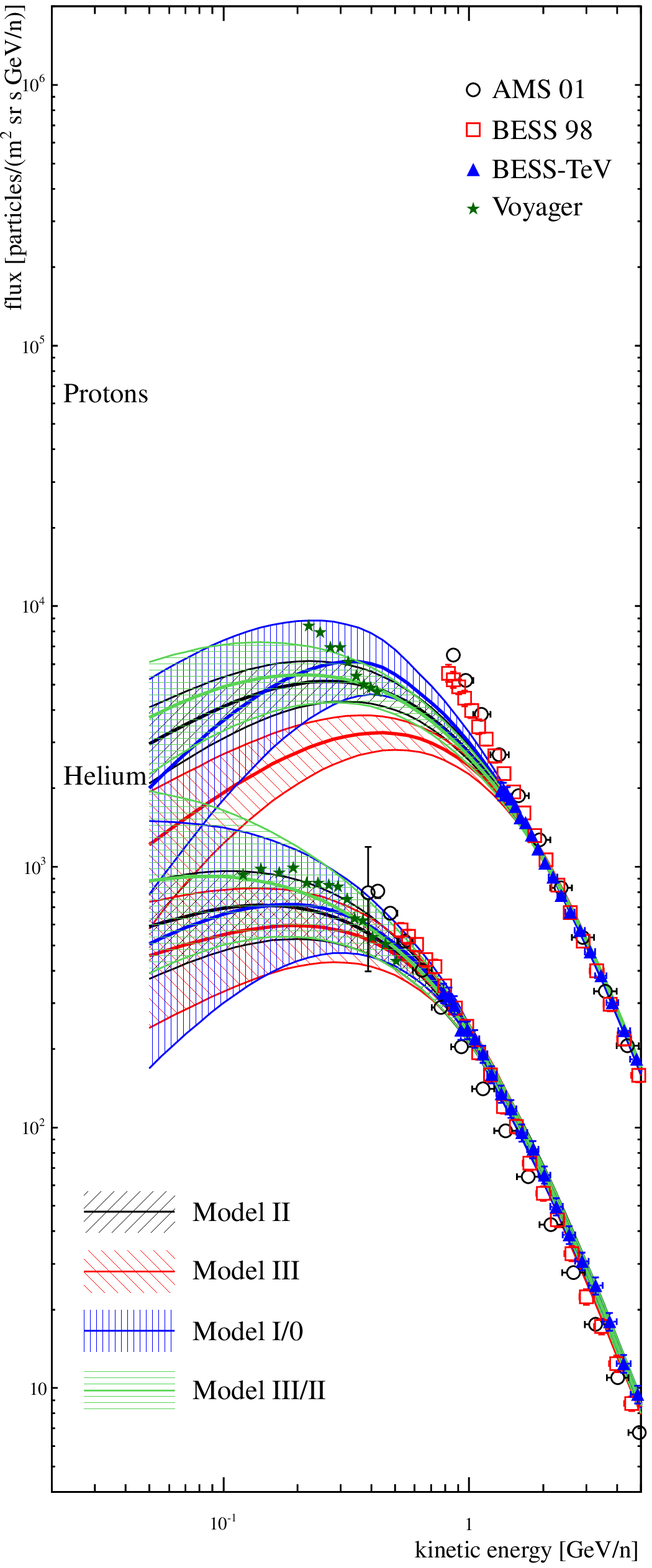}
\caption{95\% CL envelopes for the proton and helium fluxes for the four
propagation configurations of models of Table~\ref{tab:BC_models}.
The three panels correspond respectively to the
result of the MCMC analysis on AMS-01 data (left panel), BESS98 data
(middle panel), and BESS-TeV (right panel). For the sake of comparison,
the IS (demodulated) AMS-01 (black circles), BESS98 (red squares), BESS-TeV
(blue triangles), along with the Voyager data (stars).}
\label{fig:envelopesIS_LE}
\end{figure*}

\paragraph{Envelopes on IS fluxes and consistency with low-energy Voyager
data} Finally, the three panels of Fig.~\ref{fig:envelopesIS_LE} shows the
envelopes on the IS fluxes (obtained from the 95\% CIs on the parameters).
The flux is extrapolated down to an IS energy of $\sim 0.1$ GeV/n, where
the demodulated Voyager energies fall
\citep{2009JGRA..11402103W}\footnote{The estimated modulation parameter is
60 MV for these data \citep{2009JGRA..11402103W}.}. On the same plots are
shown the demodulated AMS-01, BESS98, BESS-TeV, and Voyager data. 
Showing here the demodulated envelopes underlines the spread of the different models at low energies and the possible constraints which can be obtained by low-energy data, such as from the Voyager spacecrafts.
The
low-energy envelopes obtained with AMS-01 and BESS98 data at Voyager
energies are hardly consistent with each other. BESS-TeV envelopes are more satisfactory 
in that respect. The Voyager data for helium is well reproduced.
This is possibly
related to the use of the force-field approximation which is known to fail
for the very low-energy protons \citep{1987A&A...184..119P}.
For the envelopes based on BESS-TeV data (right panel), almost all models are
allowed, except perhaps model III (standard diffusion with convection and
reacceleration). On the other hand, from the envelopes from AMS-01 and
BESS98 data, the modified diffusion scheme models (I/0 and III/II) are
disfavoured by Voyager data (that were not included in the fit).
Again, it is difficult to conclude given the inconsistencies between
the various data sets, but such plots clearly show the potential
of future analysis to access the low-energy source spectrum 
(using Voyager data closer to the IS state and/or more accurate
low-energy data from PAMELA and AMS-02).

    \subsection{Constraints on heavier primary species\label{sec:BCfixed_otherPrim}}

\subsubsection{Preamble}
Heavier primary species (from C to Fe) are less abundant and thus more
difficult to measure than p and He. As a result, the spread in their
measurements and their error bars are larger than those for the p and He
fluxes. But they still provide some useful information on cosmic-ray propagation and sources. Indeed, the heavier
the species, the larger its destructive rate. Hence, the universality of
the source spectrum can be checked against the above effect (which is
species dependent).

In this section, we repeat the analysis performed on p and He for the C, O, Ne, Mg,
Si, S, Ar, Ca and Fe elements. These elements are almost all completely
dominated by the primary contribution, except for S and Ar that receive a
$\sim 20\%$ secondary contribution. To speed up the calculation, but still
take into account this contribution, we separate the nuclei to propagate
in three families: $^{12}$C$-^{30}$Si, $^{32}$S$-^{48}$Ca,
and $^{54}$Fe$-^{64}$Ni. For instance, the nitrogen|a mixture in
comparable amount of primary and secondary contribution|is properly calculated
in this approach, and could be in principle used in this study. However,
we prefer to focus on the pure primary contribution to simplify the analysis
and the discussion (a full analysis of all nuclei is left to a future study).

Such a study complements and extends the analysis performed by the
HEAO-3 group \citep{1990A&A...233...96E}, the Ulysses group
\citep{1996ApJ...465..982D}, and the TRACER group
\citep{2009ApJ...697..106A}, in which only one propagation model, a
universal source spectral index $\alpha$ for all species, and a single
experiment was considered. Below, $\alpha_i$, $q_i$ and $\eta_S^i$ are
free parameters for each primary species, which allows us to i) test the
universality of the source spectra, ii) take into account the
correlations between the normalisation $q_i$ and the spectral index
$\alpha_i$ \citep{2009A&A...497..991P}, and iii) inspect the systematic
spread on the source parameters as several configurations of the
diffusion model are taken. The goal is to get more robust results
(as more potential sources of uncertainties are taken into account).
Besides, the MCMC technique is again helpful in providing a sound statistical
estimate of the error bars on the source parameters.

We restrict our analysis to the HEAO-3 \citep{1990A&A...233...96E},
TRACER \citep{2008ApJ...678..262A} and CREAM-II
\citep{2009ApJ...707..593A} data, to keep only the data covering as
large as possible an energy region, and also to avoid very low-energy
data which are more sensitive to solar modulation.

\subsubsection{Results} 

\begin{figure}[!t]
\centering
\includegraphics[width = 0.48\textwidth]{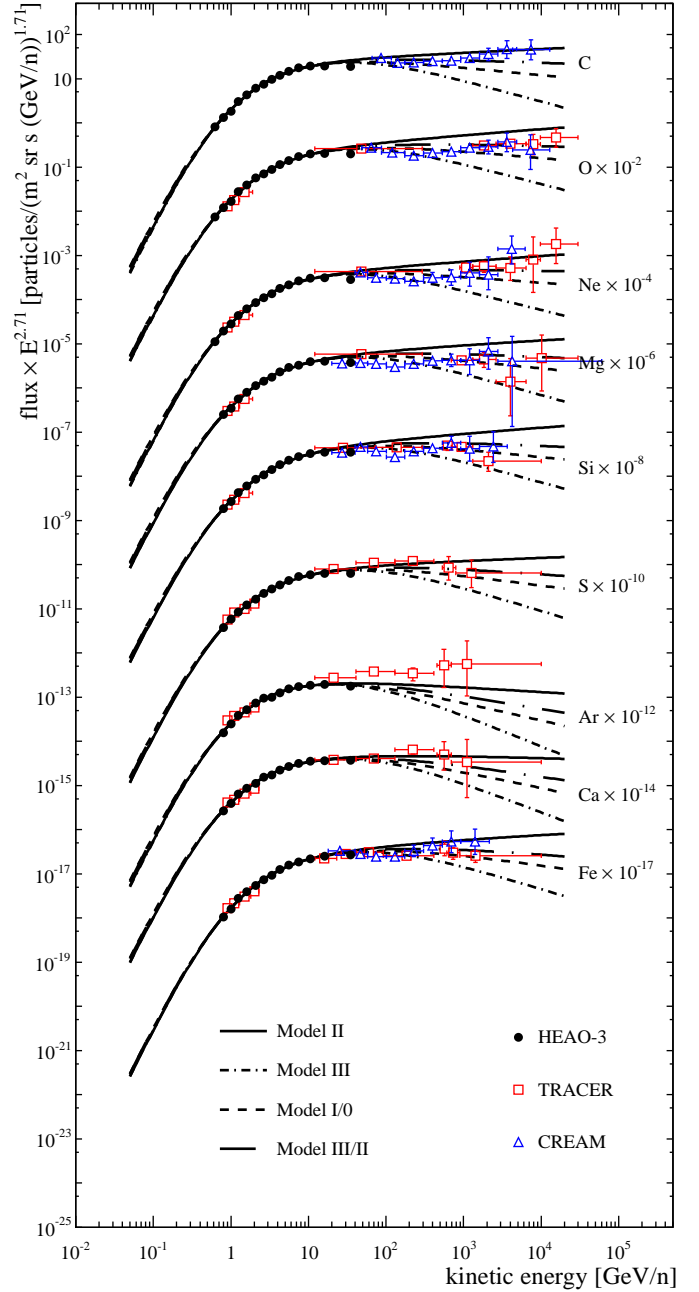}
\caption{TOA (modulated) fluxes (times $E_{k/n}^{2.71}$) as a function of the kinetic
energy per nucleon, for primary species from C to Fe. The symbols are
black circles for HEAO-3 \citep{1990A&A...233...96E}, red squares for
TRACER \citep{2008ApJ...678..262A} and blue triangles for CREAM-II
\citep{2009ApJ...707..593A}. The curves correspond (for the four models
of Table~\ref{tab:BC_models}) to the best-fit spectra obtained by a fit
on the HEAO-3 data only.}
\label{fig:fluxes_fitHEAO}
\end{figure}

\begin{figure}[!t]
\centering
\includegraphics[width = 0.48\textwidth]{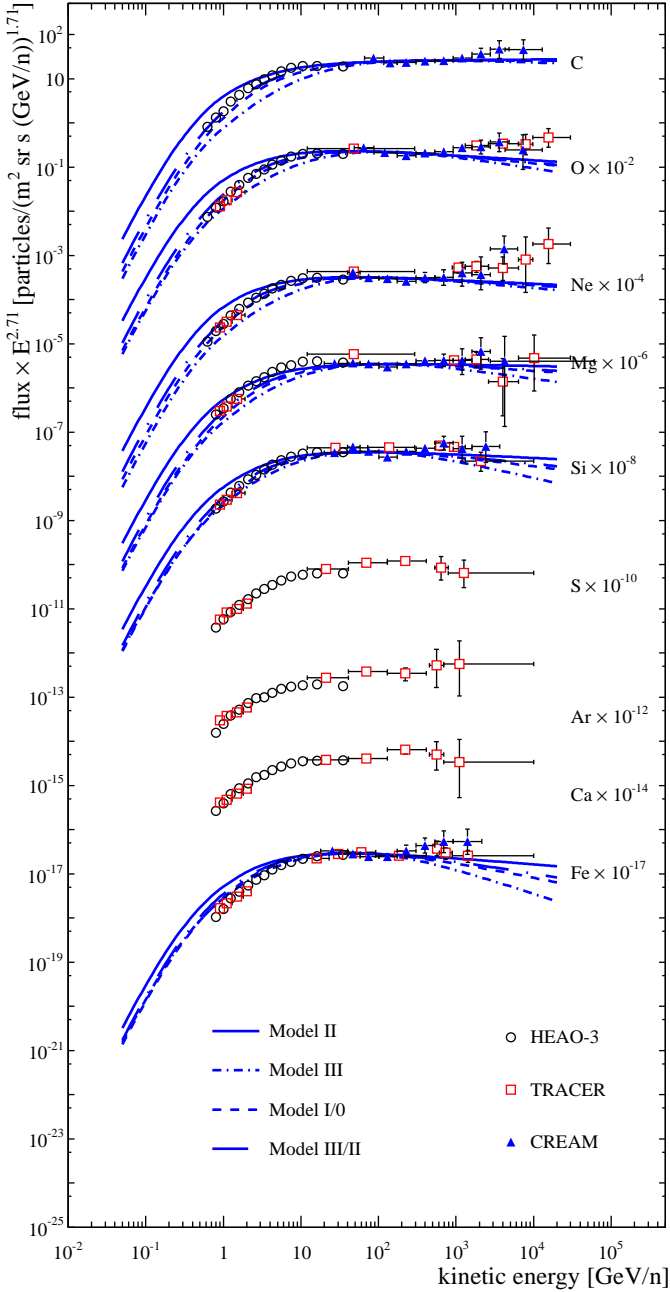}
\caption{Same as in Fig.~\ref{fig:fluxes_fitHEAO}, but the source spectra
are now fitted on the CREAM-II data only (no published data for
S, Ar, and Ca).}
\label{fig:fluxes_fitCREAM}
\end{figure}

\begin{figure}[!t]
\centering
\includegraphics[width = 0.48\textwidth]{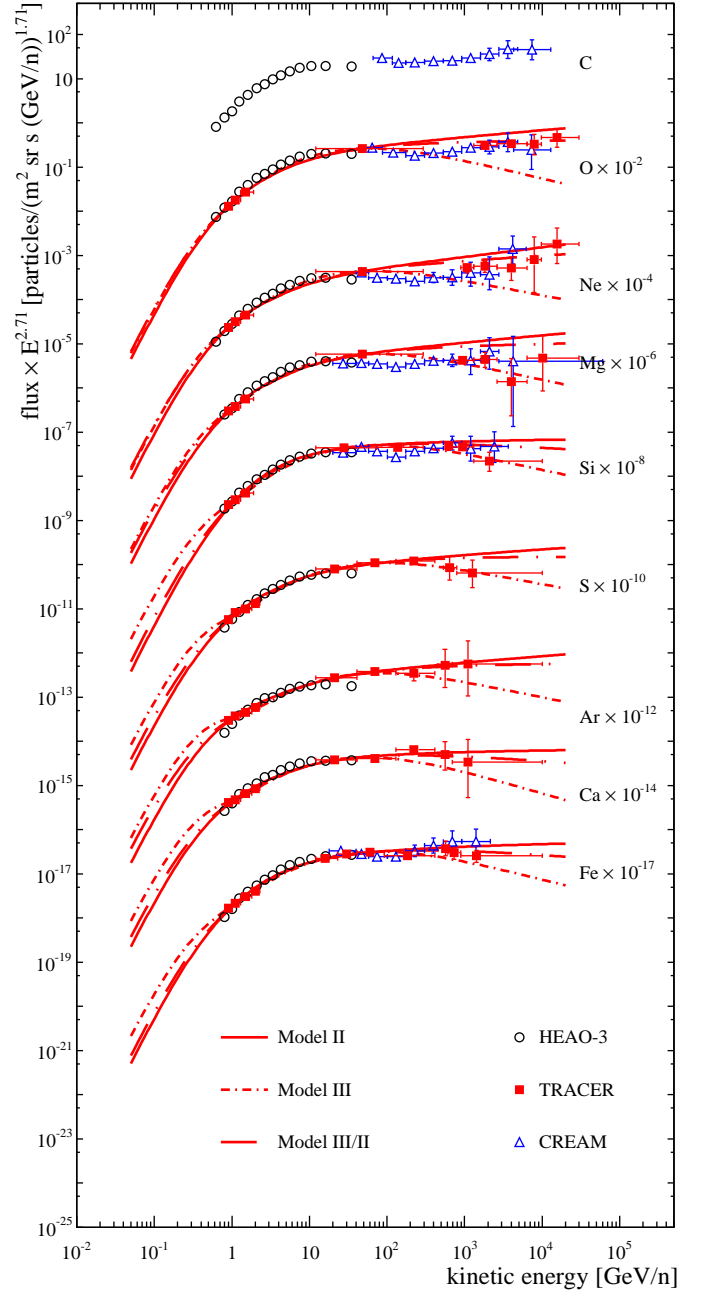}
\caption{Same as in Fig.~\ref{fig:fluxes_fitHEAO}, but the source spectra
are now fitted on the TRACER data only.}
\label{fig:fluxes_fitTRACER}
\end{figure}

We first show the fits to the data in Figs.~\ref{fig:fluxes_fitHEAO},
\ref{fig:fluxes_fitCREAM}, and \ref{fig:fluxes_fitTRACER}. They all
show the same sets of data, but each figure corresponds to a fit to a
single experiment (respectively, HEAO-3, CREAM-II and TRACER). Note 
that each species has three free parameters $\alpha$ (slope) , $q$
(normalisation), and $\eta_S$ (low-energy behaviour).

\paragraph{Fit to HEAO-3 data} In Fig.~\ref{fig:fluxes_fitHEAO} (i.e.
fit to HEAO-3), the four propagation configurations of
Table~\ref{tab:BC_models} lead to the same shape at low energy. This
is not surprising since this is where the bulk of HEAO-3 data lies. The
high-energy asymptotic behaviour is then influenced by the value of the
diffusion slope $\delta$, as is the case for p and He.
Similarly, the spread in $\delta$ is larger than the spread in $\alpha$ (see below).
It means that the smaller the value of $\delta$, the smaller $\gamma_{\rm asympt} =
\alpha+\delta$, so the same ordering according to the $\gamma_{\rm
asympt}$ value of the model is seen at high energy: Model II on top
(largest $\gamma_{\rm asympt}$), then Model III and I/0, and Model
III/II at bottom. An eye inspection shows that Model~I/0 and
Model~III/II are the ones in best agreement with the higher energy data
(CREAM-II and TRACER).

\paragraph{Fit to CREAM-II and TRACER data} 
Fig.~\ref{fig:fluxes_fitCREAM} shows the resulting best-fit spectra
for the same models, but now fitted to CREAM-II data only. The data
being at higher energy, the parameter $\eta_S$ is unconstrained and we
set it to -1 for this fit only. Not surprisingly, most models are
not able to match the lower-energy HEAO-3 data. For the lighter species,
$\alpha+\delta$ remains the same, regardless of the model. The CREAM-II
data for these species are $\gtrsim$ 100 GeV/n, in a regime where
the asymptotic slope $\gamma_{\rm asympt}$ is reached. For the heavier
species, where the data extend down to a few tens of GeV/n, a similar ordering
(though less clear) of the models with $\gamma_{\rm asympt}$ (as for the HEAO-3
fit) is seen for the high-energy asymptotic behaviour. 
The fits to TRACER data shown in Fig.~\ref{fig:fluxes_fitTRACER}
show an intermediate behaviour. Indeed, the energy range covers the same energy
range as CREAM-II, but a few data point at low energy give a turn-over in the spectrum. 
However, there is a gap between the two energy
regimes, where the curvature of the HEAO-3 data is not
reproduced\footnote{Model I/0 is not used on these data because it
leads to unphysical values. This is likely to be related to the unphysical
large negative value of $\eta_S$ required, that becomes an issue
for the numerical inversion of the energy losses
(no convection and reacceleration to smooth out the steep upturn in
the low-energy spectrum here, at variance with the three other models).}.

\paragraph{Goodness of fit} 
Table~\ref{tab:C_Fe_BCfixed} shows the best-fit spectral index
$\alpha_i$ and the associated $\chi^2_{\rm min/d.o.f.}$ value for
the various models, species, and data sets. Unsurprisingly, the
best-fit (smaller $\chi^2_{\rm min/d.o.f.}$ value) are for the
CREAM-II data that only cover the high energy range. It is indeed
more difficult to reproduce the low energy part, where data have
smaller error bars, but also where more effects (modulation,
continuous and catastrophic losses) shape the spectrum. For the
HEAO-3 case, the $\chi^2_{\rm min/d.o.f.}$ value is large for most
of the species because of the difficulty to fit the highest energy
point that has a very small error bar. For S, Ar and Ca, the fit
is better. Data from the next CREAM flights, or from the AMS-02
instrument should help clarify the situation, and confirm or otherwise
these discrepancies amongst the various data. Nevertheless, some
conclusions can still be drawn on the spectral indices (see
below), although they are less constraining than those derived from
the p and He data.
\begin{table}[t!]
	\caption{Best-fit spectral index $\alpha$ and associated
   $\chi^2_{\rm min/d.o.f.}$ for the fit of the source spectrum
   parameters. Each column correspond to a different set of data
   on which the fit is performed.}
\label{tab:C_Fe_BCfixed}
	\begin{tabular}{lc|cc|cc|c}
		\hline \hline
		Model & \multicolumn{2}{c}{HEAO-3} & \multicolumn{2}{c}{TRACER} & \multicolumn{2}{c}{CREAM}\\
		&  $\alpha_{\rm best}$ & $\frac{\chi^2_{\rm min}}{d.o.f.}$ &  \multicolumn{2}{c}{\dots} & \multicolumn{2}{c}{\dots} \\ \hline
		 \multicolumn{7}{c}{| Carbon |}\\
		 II     &  2.41 & 7.16 &  N/A & N/A  & 2.48 & 1.59 \\
		 III    &  2.33 & 6.14 &  N/A & N/A  & 1.90 & 2.15 \\
		 I/0    &  2.28 & 5.96 &  N/A & N/A  & 2.11 & 1.78 \\
		 III/II &  2.27 & 6.54 &  N/A & N/A  & 2.21 & 1.73 \\[1mm]
		 \multicolumn{7}{c}{| Oxygen |}\\
		 II     &  2.37 & 7.34 & 2.35 & 15.17& 2.61 & 3.17 \\
		 III    &  2.32 & 6.11 & 2.27 & 13.76& 2.18 & 4.59 \\
		 I/0    &  2.26 & 6.08 & N/A  & N/A  & 2.28 & 3.69 \\
		 III/II &  2.26 & 6.61 & 2.19 & 2.54 & 2.37 & 3.54 \\[1mm]
		 \multicolumn{7}{c}{| Neon |}\\
		 II     &  2.37 & 3.94 & 2.27 & 5.59 & 2.57 & 0.85 \\
		 III    &  2.30 & 2.85 & 2.19 & 3.06 & 2.01 & 1.09 \\
		 I/0    &  2.24 & 3.54 &  N/A &  N/A & 2.21 & 0.93 \\
		 III/II &  2.23 & 3.80 & 2.09 & 1.68 & 2.31 & 0.91 \\[1mm]
		 \multicolumn{7}{c}{| Magnesium |}\\
		 II     &  2.40 & 7.13 & 2.35 & 26.79& 2.54 & 0.69 \\
		 III    &  2.35 & 6.03 & 2.23 & 0.60 & 2.10 & 1.28 \\
		 I/0    &  2.29 & 6.37 &  N/A & N/A  & 2.22 & 0.90 \\
		 III/II &  2.29 & 6.79 & 2.17 & 11.47& 2.31 & 0.83 \\[1mm]
		 \multicolumn{7}{c}{| Silicon |}\\
		 II     &  2.38 & 3.95 & 2.49 & 53.88& 2.60 & 2.08 \\
		 III    &  2.34 & 3.15 & 2.24 &  3.89& 2.25 & 2.93 \\
		 I/0    &  2.29 & 3.35 & N/A  & N/A  & 2.33 & 2.41 \\
		 III/II &  2.29 & 3.63 & 2.31 & 35.14& 2.40 & 2.30 \\[1mm]
		 \multicolumn{7}{c}{| Sulfur |}\\
		 II     &  2.44 & 1.79 & 2.39 & 2.22 &  N/A  & N/A  \\
		 III    &  2.39 & 1.27 & 2.19 & 0.70 &  N/A  & N/A  \\
		 I/0    &  2.34 & 1.39 & N/A  & N/A  &  N/A  & N/A  \\
		 III/II &  2.33 & 1.58 & 2.21 & 1.35 &  N/A  & N/A  \\[1mm]
		 \multicolumn{7}{c}{| Argon |}\\
		 II     &  2.61 & 1.31 & 2.37 & 0.51 &  N/A  & N/A  \\
		 III    &  2.57 & 1.18 & 2.22 & 0.62 &  N/A  & N/A  \\
		 I/0    &  2.51 & 1.26 &  N/A &  N/A &  N/A  & N/A  \\
		 III/II &  2.50 & 1.31 & 2.19 & 0.29 &  N/A  & N/A  \\[1mm]
		 \multicolumn{7}{c}{| Calcium |}\\
		 II     &  2.56 & 1.94 & 2.49 & 0.83 &  N/A  & N/A  \\
		 III    &  2.52 & 1.76 & 2.36 & 1.08 &  N/A  & N/A  \\
		 I/0    &  2.48 & 1.84 &  N/A &  N/A &  N/A  & N/A  \\
		 III/II &  2.48 & 1.90 & 2.34 & 0.73 &  N/A  & N/A  \\[1mm]
		 \multicolumn{7}{c}{| Iron |}\\
		 II     &  2.43 & 4.05 & 2.48 & 19.60& 2.67 & 1.54 \\
		 III    &  2.39 & 3.77 & 2.29 & 2.48 & 2.42 & 2.37 \\
		 I/0    &  2.36 & 3.88 &  N/A & N/A  & 2.46 & 1.94 \\
		 III/II &  2.35 & 3.98 & 2.34 & 11.95& 2.52 & 1.80 \\ \hline
	\end{tabular}
\end{table}

\begin{figure*}[!t]
\centering
\hspace{0.36cm}\includegraphics[width = 0.94\textwidth]{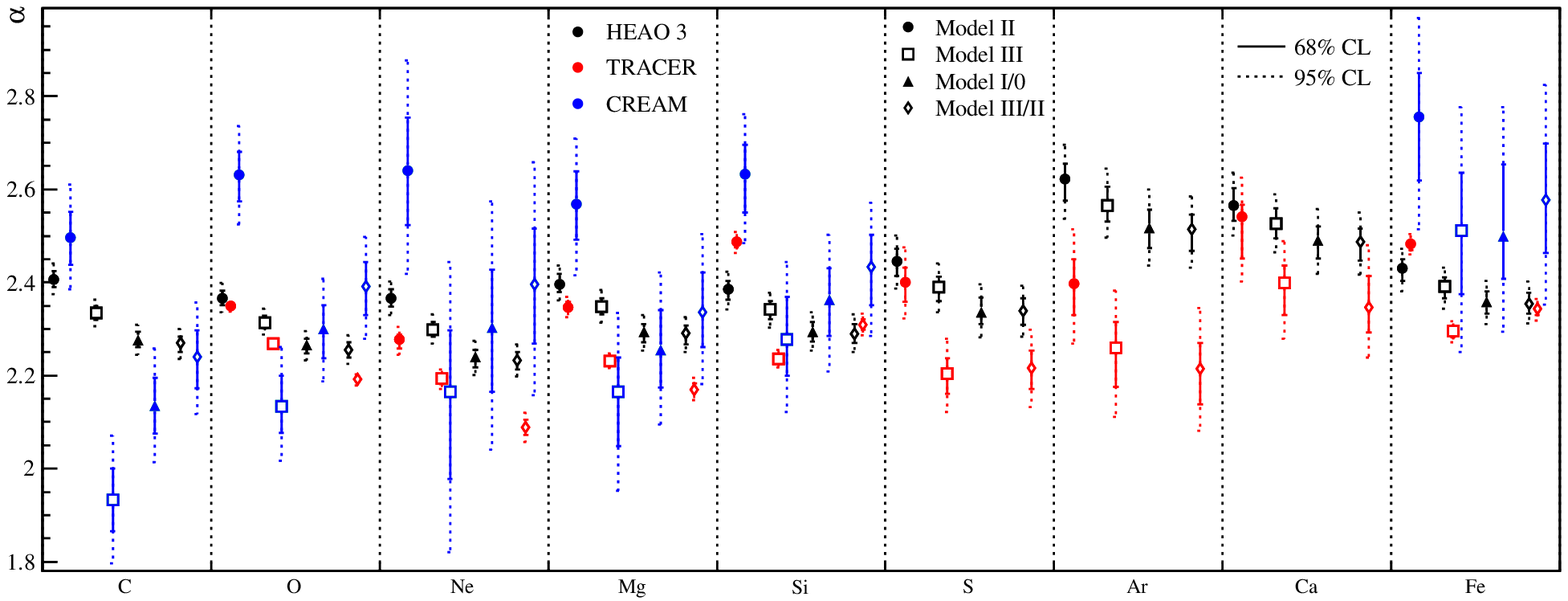}
\includegraphics[width = 0.96\textwidth]{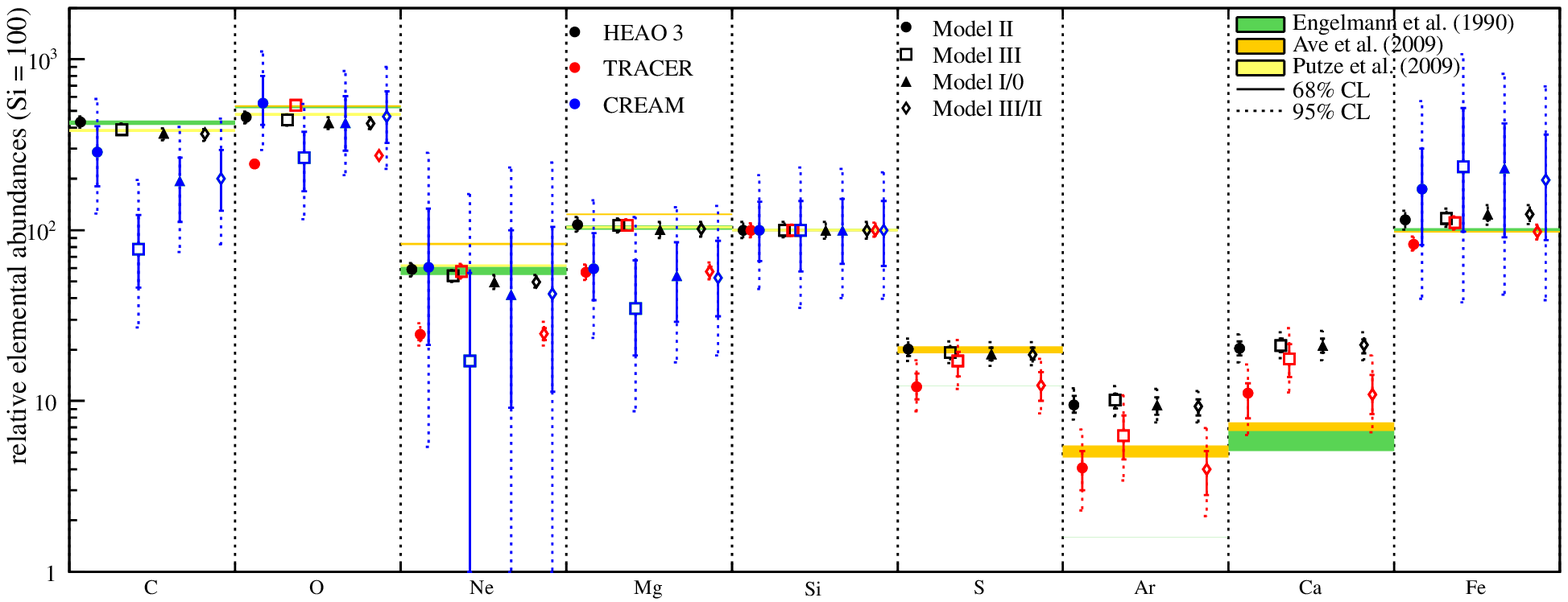}
\includegraphics[width = 0.96\textwidth]{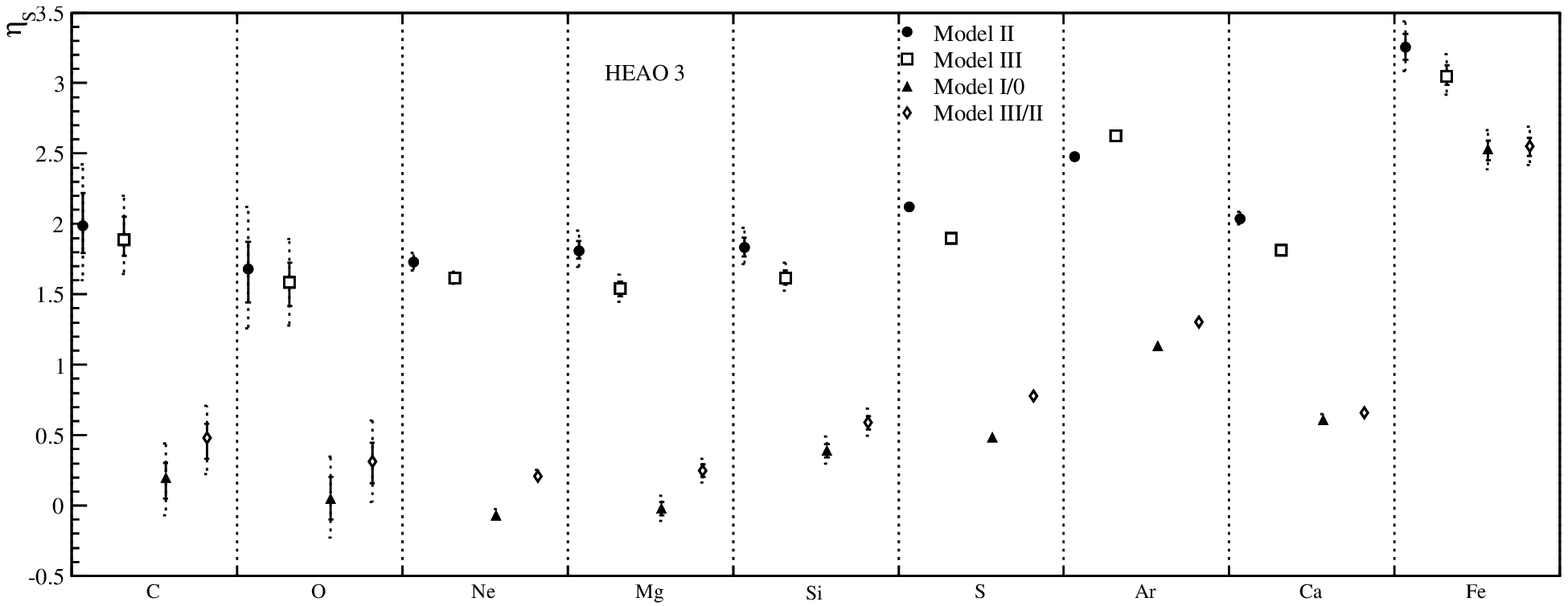}
\caption{Best-fit value (symbols), 68\% (dashed error bars) and
95\% (solid error bars) CIs for C to Fe source parameters from
the fit on CREAM-II, HEAO-3, and TRACER data.
Top panel: source spectral index $\alpha_i$. Middle panel:
relative abundances $q_i$. The results from the
analysis of the HEAO-3 group \citep{1990A&A...233...96E}
are shown as green boxes, and those from the TRACER group
\citep{2009ApJ...697..106A} as orange boxes. The yellow boxes
correspond to a leaky-box analysis of the abundances on HEAO-3
data performed in \citet{2009A&A...497..991P}. Bottom panel:
low-energy source parameter $\eta_S$ [see Eq.(\ref{eq:source_spec})].
Only the results from HEAO-3 data are plotted (for CREAM-II, $\eta_S$ is set
to -1, and for TRACER, the scatter is so large that the values are meaningless).}
\label{fig:alpha_abund_C_Fe}
\end{figure*}

\paragraph{Confidence intervals on $\alpha$, $q$ and $\eta_S$}
Figure~\ref{fig:alpha_abund_C_Fe} shows, along with the 68\% and 95\%
CIs, the best-fit values on the spectral indices $\alpha_i$ (top panel),
the relative source abundances $q_i$ (middle panel), and the source parameter
$\eta_S^i$, for all the primary species considered in this study. 
We first underline that the 95\% CL relative uncertainty for the parameters
ranges from $\lesssim 5\%$ on $\alpha_i$ and $\lesssim 20\%$ on $q_i$.
CREAM data cover too narrow an energy range to give stringent constraints,
so we do not comment on them further below.
The following trends are observed for the parameters:
\begin{itemize}
  \item $\alpha$ (top panel): as for the p and He data, Model II (reacceleration only, filled circle)
  always gives a larger value than the other models. Moreover, a similar range
  of slopes is found ($2.2-2.5$ for HEAO-3 data only, but more scatter when
  using TRACER data).
  \item $q_i$ (middle panel): the relative abundances from HEAO-3 data are
  quite insensitive to the propagation model used. We recover the values
  of \citet{1990A&A...233...96E} (green boxes), although with larger error
  bars. The \citet{1990A&A...233...96E} analysis is based on 
  the leaky-box model, so that their results are consistent with the
  \citet{2009A&A...497..991P} analysis (yellow boxes) performed in the same
  framework. Similar results are obtained in the diffusion model (this analysis),
  except for the discrepancy for S, Ar and Ca (our values are larger than
  those of the HEAO-3 analysis). The difference for Ar and Ca may be related
  to the fact these elements were assumed to be pure primary species in this
  analysis (in order to speed up the calculation). The lack of the secondary
  contribution translates in a higher primary flux required to match the
  data. This underlines the importance of taking properly into account
  all nuclei to derive the source abundances. Part of the discrepancy
  could also be related to the fact that, for
  the S, Ar, and Ca elements, the highest energy data point is better fitted than for the others
  (see above): this results in a larger value of $\eta_S$ and $\alpha$
  that may be responsible for the difference observed on the $q_i$. 
  The relative abundances obtained from the TRACER data are sensitive
  to the model chosen, presumably because of the lack of constraints
  in the intermediate energy range. 
  The relative abundances for Model~III (open squares)
  are consistent with those obtained from HEAO-3 data, whereas the obtained
  values for Model~II (filled circles) and III/II (open diamonds)
  systematically undershoot those from HEAO-3 data. If we look into
  Tab.~\ref{tab:C_Fe_BCfixed}, we remark that the fit to the Si data, on which
  all other abundances are normalised, is very poor for these models
  (large value of the $\chi^2_{\rm min/d.o.f.}$). The consistently low
  values for all elements for these two models can be simply explained
  in terms of a too large Si abundance obtained from the TRACER data for
  Models~II and III/II. The fact that Model~III, which gives a good fit
  to Si data, gives abundances in agreement with those derived from HEAO-3
  data, supports this explaination.
  \item $\eta_S$ (bottom panel): as for the p and He data, a trend is observed showing a
   dependence on the models. The pattern is the same, but with
  the value of $\eta_S$ larger by one than that for p and He. In terms of
  $\eta_S-\eta_T$, the C to Si primary species favour a value $\approx 1$,
  whereas p and He data favour $\approx 0$. There is more scatter
  in the S to Fe data, but we note that the HEAO-3 data are based
  on different use of sub-detectors for this heavier species.
\end{itemize}

\subsection{Summary for the source parameters}
An important result of the fixed transport parameter analysis
is that, independent of the model and data
considered, the source slope of p and He nuclei is constrained to fall in
the range $2.2-2.5$ (or $2.2-2.4$ if we discard Model~II). As the range
of $\delta$ covered by these models falls in the range $0.23-0.86$ (see
Table~\ref{tab:BC_models}), this is a robust prediction. It also means that
the asymptotic value for the propagated spectra ($\gamma_{\rm asymp.}
\equiv \alpha+\delta$), that falls in the range $2.7-3.0$, is not reached
in the GeV/n to TeV/n regime (as direct fits to the nuclei at 
$E_{k/n} > 100 $ GeV/n propagated spectra lead
to $\gamma_{\rm data} \approx 2.65$, \citealt{2008ApJ...678..262A}):
residual propagation effects (reacceleration, convection, spallations)
are still active. This is also supported by the results
on the heavier species.

Another important result is that regardless of the propagation model used,
the quantity $\eta_S-\eta_T$ is constrained  to be $\sim 0-1$ for all primary
nuclei considered. Hence, to
reproduce the data, if a pure power-law rigidity spectrum is assumed, a
{\em non-standard} low-energy diffusion coefficient (upturn at low energy)
is required. Conversely, if a {\em standard} diffusion coefficient is
assumed (i.e. $K(E)=K_0 \beta {\cal R}^\delta$), a flattening of the
low-energy source spectrum is required (i.e. $dQ/dR\propto \beta^{\eta_S+1}
{\cal R}^{-\alpha}$ with $\eta_S>-1$). The close-to-IS condition low-energy
Voyager data is a further piece of information to break the degeneracy
$\eta_S-\eta_T$, and would possibly provide the shape of the low-energy
source spectrum.

Finally, most of the derived relative source abundances are 
in agreement with those derived by earlier groups. Still, a
slight dependence on the propagation 
configuration is also observed. Due to the relevance of the
value of the source abundances in the context of acceleration
mechanisms \citep{1987ApJ...322..981D,1997ApJ...487..182M,1997ApJ...487..197E,
2009ApJ...695..666O},
this point deserves further investigation.

\section{Ratio of primary species\label{sec:ratios}}

\citet{1974Ap&SS..30..361W}, more than 30 years ago, recognised the importance
of looking at primary ratios. Such ratios may be, in principle, used to i) check
the consistency of spectral indices of various species, ii) inspect whether
source spectra are power-low in rigidity or power-law in kinetic energy, and
also iii) inspect whether solar modulation is a rigidity or total energy
effect. Below, we present several plots to illustrate some of these ideas,
but also underline the complications that arise due to the many degeneracies
between the source, transport, and modulation parameters (as underlined
in the previous sections).

\subsection{p/He ratio}
Concerning the p/He ratio\footnote{Note that because of the misidentification of
$^3$He and $^4$He, \citet{1974Ap&SS..30..361W} estimate a $\lesssim 2.5 \%$
effect in the data plotting the same measurement as rigidity spectra or kinetic
energy spectra, that is also not considered below.} on which
\citet{1974Ap&SS..30..361W} study mainly focused, the main conclusions were:
i) proton and helium source spectra are rigidity rather than energy/nucleon
spectra, ii) modulation effects dominate the shape of the p/He ratio for such
rigidity spectra when shown as a function of kinetic energy, and iii) modulation
effects is not a pure rigidity effect since it flattens the spectrum at low
energy relative to the interstellar flux.

\paragraph{p/He from the toy-model calculation}
Caution is in order when calculating the ratio p/He, whether we start
from the differential flux in energy or in rigidity. In an analogous
manner as for Eq.~(\ref{eq:E_p_R}),
\[
  \psi_{E_{k/n}}(E) \equiv \frac{d\psi}{dE_{k/n}}
\]
and
\[
  \psi_{\cal R}(E) \equiv \frac{d\psi}{d{\cal R}}
    = \frac{Z\beta}{A}\cdot \psi_{E_{k/n}}(E) \;,
\]
in order to define
\[
  \left.\frac{p}{He}\right|_{\cal R} 
    = \frac{\psi_{\cal R}^{\rm p}}{\psi_{\cal R}^{\rm He}}
    = \frac{2\beta_{\rm p}}{\beta_{\rm He}}\cdot\left.\frac{p}{He}\right|_{E_{k/n}} \;.
\]

In the 1D toy-model (energy gains and losses discarded),
assuming Eq.~(\ref{eq:source_spec}) for the source term|i.e.
$dQ/d{\cal R} =  q \beta^{\eta_S+1} {\cal R}^{- \alpha}$|,
and Eq.~(\ref{eq:eta_T}) for the diffusion coefficient|i.e.
$K({\cal R})= \beta^{\eta_T} K_0 {\cal R}^\delta$|, we have
the analog of Eq.~(\ref{eq:flux_spal}), but expressed in terms of
the rigidity:
\begin{eqnarray}
\left.\frac{p}{He}\right|_{\cal R} &=&
 \frac{2q_{\rm p}}{q_{\rm He}} \cdot \frac{\beta_{\rm p}^{\eta_S+2}({\cal R})}{\beta_{\rm He}^{\eta_S+2}({\cal R})}
 \cdot {\cal R}^{-(\alpha_{\rm p}-\alpha_{\rm He})}\nonumber\\
& \times &
\frac{\beta_{\rm He}^{\eta_T}({\cal R})}{\beta_{\rm p}^{\eta_T}({\cal R})}
\cdot
\frac{K_0 {\cal R}^\delta/(hL)
     + n c \cdot \beta_{\rm He}^{1-\eta_T}({\cal R})\cdot \sigma_{\rm He}}{
      K_0 {\cal R}^\delta/(hL) 
     + n c\cdot \beta_{\rm p}^{1-\eta_T}({\cal R})\cdot \sigma_{\rm p}}\;,
\label{eq:rig}
\end{eqnarray}
where we made explicit the rigidity dependence for all the terms
($c$ is the speed of light).
If the destruction rate is subdominant, we have
\begin{equation}
\left.\frac{p}{He}\right|_{\cal R} 
  \quad \stackrel{\sigma\rightarrow 0}{=}\quad
 \frac{2q_{\rm p}}{q_{\rm He}} 
   \cdot 
    \frac{\beta_{\rm p}^{\eta_S+2-\eta_T}({\cal R})}{
          \beta_{\rm He}^{\eta_S+2-\eta_T}({\cal R})}
    \cdot {\cal R}^{-(\alpha_{\rm p}-\alpha_{\rm He})}      
          \;.
\label{eq:rig_no_spall}
\end{equation}
In all the above formulae, we have 
\[
   \beta = \frac{{\cal R}}{\sqrt{{\cal R}^2 + m^2/Z^2}}
         \approx \frac{\sqrt{E_{k/n}(E_{k/n}+2)}}{E_{k/n}}
\]
For a proton $m^2/Z^2\approx 1$, whereas $\approx 4$ for a helium nucleus.
This is sufficient to distort the low energy p/He ratio whenever 
it is calculated from the differential fluxes in rigidity and
$\eta_S+2-\eta_T\neq 0$. However, if
the ratio is calculated from the differential fluxes in kinetic energy
per nucleon, for a given $E_{k/n}$, we have $\beta_{\rm p}(E_{k/n})\approx
\beta_{\rm He}(E_{k/n})\equiv \beta$,
and Eq.~(\ref{eq:rig}) reduces to
\begin{eqnarray}
\left.\frac{p}{He}\right|_{E_{k/n}} \!\!\!\!\!\!\!\!&\!\!\approx\!\!&\!\!
 \frac{q_{\rm p}}{q_{\rm He}} \cdot 
 \frac{[E_{k/n}(E_{k/n}+2)]^{-(\alpha_{\rm p}-\alpha_{\rm He})/2}}{2^{-\alpha_{\rm He}}}
 \nonumber\\
\!\!\!& \times &\!\!\!
\frac{K_0 [E_{k/n}(E_{k/n}+2)]^{\delta/2}2^\delta/(hL)
     + n c   \beta^{1-\eta_T}  \sigma_{\rm He}}{
      K_0 [E_{k/n}(E_{k/n}+2)]^{\delta/2}/(hL) 
     + n c \beta^{1-\eta_T}  \sigma_{\rm p}}\;.
\label{eq:ekn}
\end{eqnarray}
When the destruction rates are subdominant, we get
\begin{equation}
\left.\frac{p}{He}\right|_{E_{k/n}} 
   \stackrel{\sigma\rightarrow 0}{=}
 \frac{q_{\rm p}}{q_{\rm He}} \cdot 
 \frac{[E_{k/n}(E_{k/n}+2)]^{-(\alpha_{\rm p}-\alpha_{\rm He})/2}}{2^{-\alpha_{\rm He}-\delta}}
\;.
\label{eq:ekn_no_spall}
\end{equation}

\begin{figure*}[t!]
\includegraphics[width = \columnwidth]{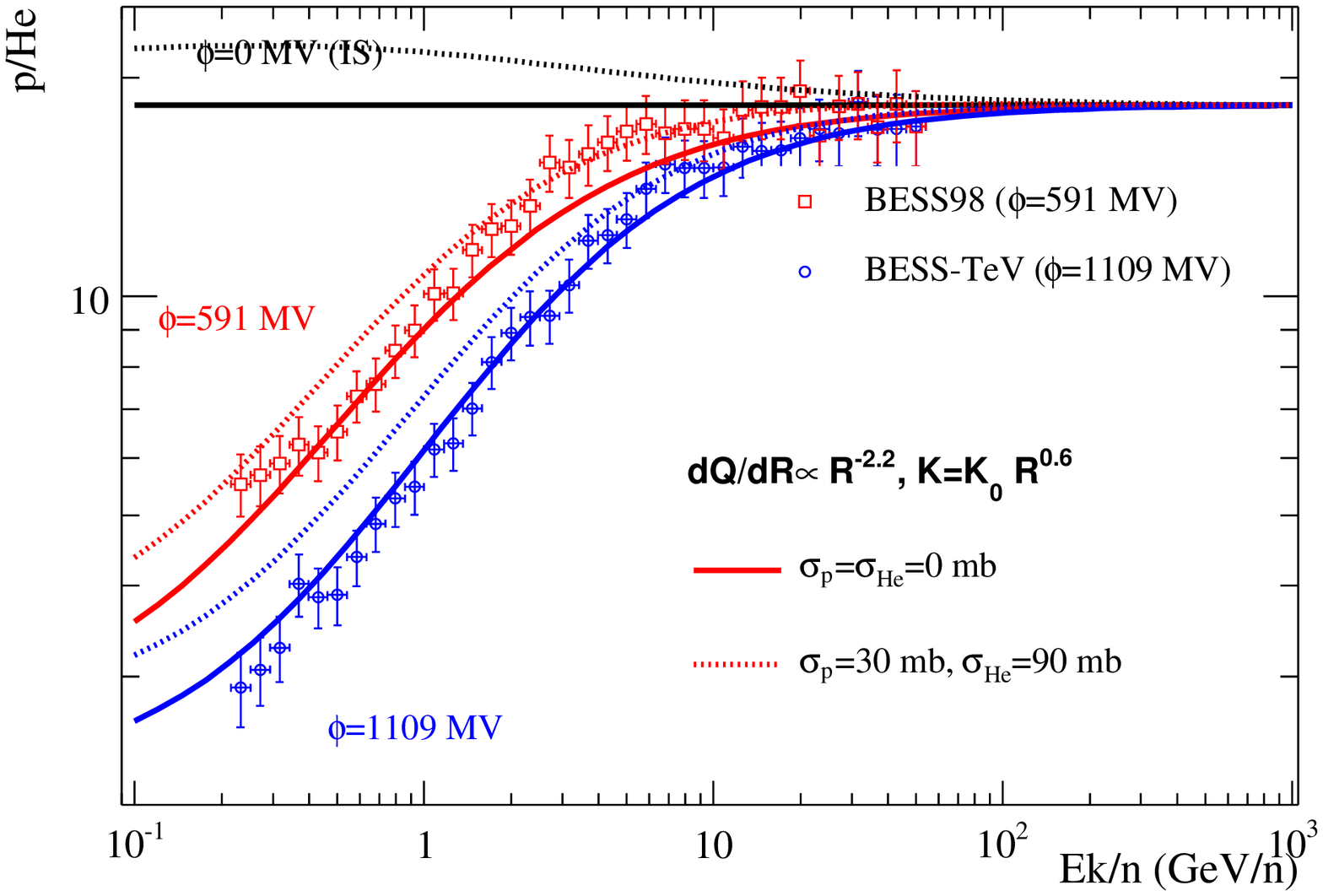}
\includegraphics[width = \columnwidth]{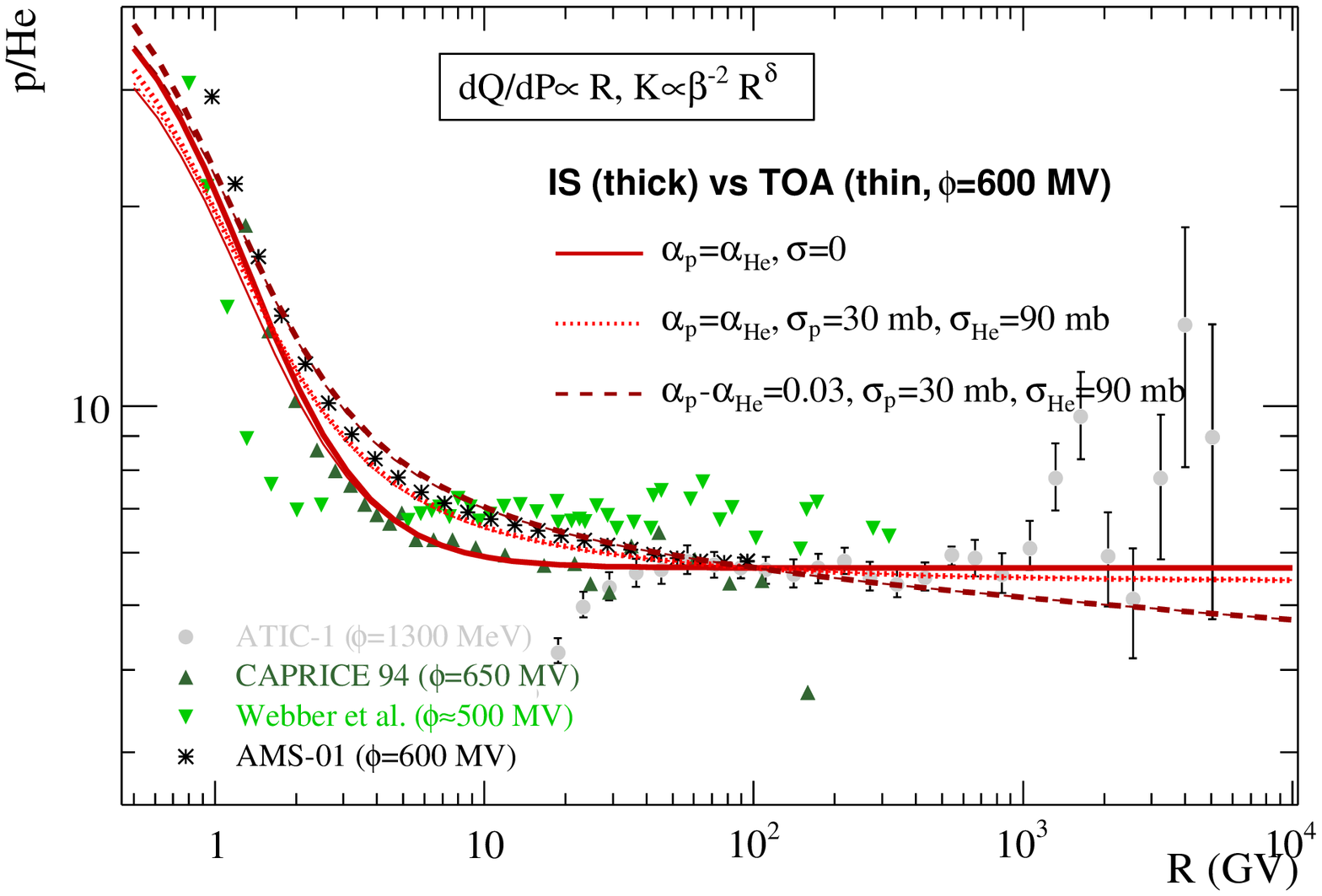}
\caption{Left panel: p/He ratio as a function of $E_{k/n}$, along with
BESS98 \citep{2000ApJ...545.1135S} and BESS-TeV \citep{2004PhLB..594...35H}
data. The lines show the toy-model calculation without the destruction term
(thick solid lines) and with it (thick dotted lines). The black, red and
blue lines are respectively modulated to $\Phi=0$ MV (IS),  $\Phi=591$~MV
and $\Phi=1109$. Left panel: same ratio, but as a function of the rigidity.
The data are AMS-01 \citep{2002PhR...366..331A}, ATIC 
\citep{2003ICRC....4.1829Z}, CAPRICE 94 \citep{1999ApJ...518..457B}, and
some balloon data \citep{1987ICRC....1..325W}. The three toy-model
calculations corresponds to the destruction rate set to zero (solid lines)
or to its required value (dotted lines), plus a difference in the spectral
index of p and He (dashed lines). In both plots, the normalisation is
arbitrarily set to match the data.} 
\label{fig:p_to_He}
\end{figure*}

\paragraph{Comparison to data}
The p/He ratio is displayed as a function of the kinetic energy per nucleon
in the left panel of Fig.~\ref{fig:p_to_He}, for the BESS98 (red squares)
and BESS-TeV (blue circles) data. The solid lines  (no inelastic reaction
terms) result from Eq.~(\ref{eq:ekn_no_spall}), with $\alpha_{\rm
p}=\alpha_{\rm He}$. The shape of the ratio as well as the differences
between the data taken at two different solar periods can be almost
completely ascribed to the modulation effect. The effect of the inelastic
reaction term is contained in Eq.~(\ref{eq:ekn}). A closer look to this
equation shows that the numerator and the denominator do not differ by a factor of more 
$\sim 3$, confirming the sub-dominant (though important)
role of this effect in determining the ratio (displayed versus kinetic energy
per nucleon).
The effects (not shown) of having different spectral indices for p and He,
having different values of $\alpha$, $\delta$, and $\eta_S-\eta_T$, when
varied within reasonable limits, is of the same amount as the effect of the destruction
rate. However, these effects are better seen when working with the rigidity.

The right panel of Fig.~\ref{fig:p_to_He} shows a few experiments that have
provided the p/He ratio as a function of the rigidity. Except for ATIC-1,
the error bars were not provided by the experiments, but are expected to be of the order of
the size of the symbols\footnote{The AMS-01 p and He flux given in
\citet{2002PhR...366..331A} are not calculated for the same rigidity
binning. We thus fitted both fluxes (expressed as a function of the nucleus
rigidity) with a simple polynomial function, and calculated the ratio from
these fits (shown as stars approximately along the same binning as the
original He data \citealt{2000PhLB..494..193A}). Calculating correctly the
associated error bars is not straightforward: as the data are just used for
eye comparison, we do not go into more details.
This difficulty in computing spectra binned in rigidity or kinetic energy
suggests that in future experimenters should provide both.}.  As underlined in the
previous sections, even though all experiments claim small error bars, not
many of them are consistent with each other for the p and He fluxes. On the
other hand, one would expect the ratio to have less systematics than fluxes.
Nevertheless, a large discrepancy remains, which cannot be explained by the different
level of solar modulation associated to each experiment. The various curves
show:
i) the effect of solar modulation is sub-dominant for p/He vs ${\cal R}$
(thick vs thin lines);
ii) the effect of inelastic interactions, which are switched off
(solid lines) or included (dotted lines);
iii) the distortion of the p/He ratio due to a possible difference in the
spectral indices of p and He (dashed lines). The shape of the ratio depends
mostly on the value of $\eta_S-\eta_T$. As found in
Eq.~(\ref{eq:rig_no_spall}), the ratio is constant if $\eta_S-\eta_T=-2$
(not shown on the figure).
The best-fit to the data is obtained for $\eta_S-\eta_T\approx 1$, in general
agreement with the results of the more complete analysis of Sect.~\ref{sec:BCfixed_pHe}.
A small difference $\alpha_{\rm p}-\alpha_{\rm He}\lesssim 0.05$ between p and He source indices
is not excluded. Note that the unperfect match to the data may result from the effect
of energy losses which is not implemented in the toy formula.

\subsection{Ratio relative to the Oxygen flux}
\begin{figure}[t!]
\includegraphics[width = \columnwidth]{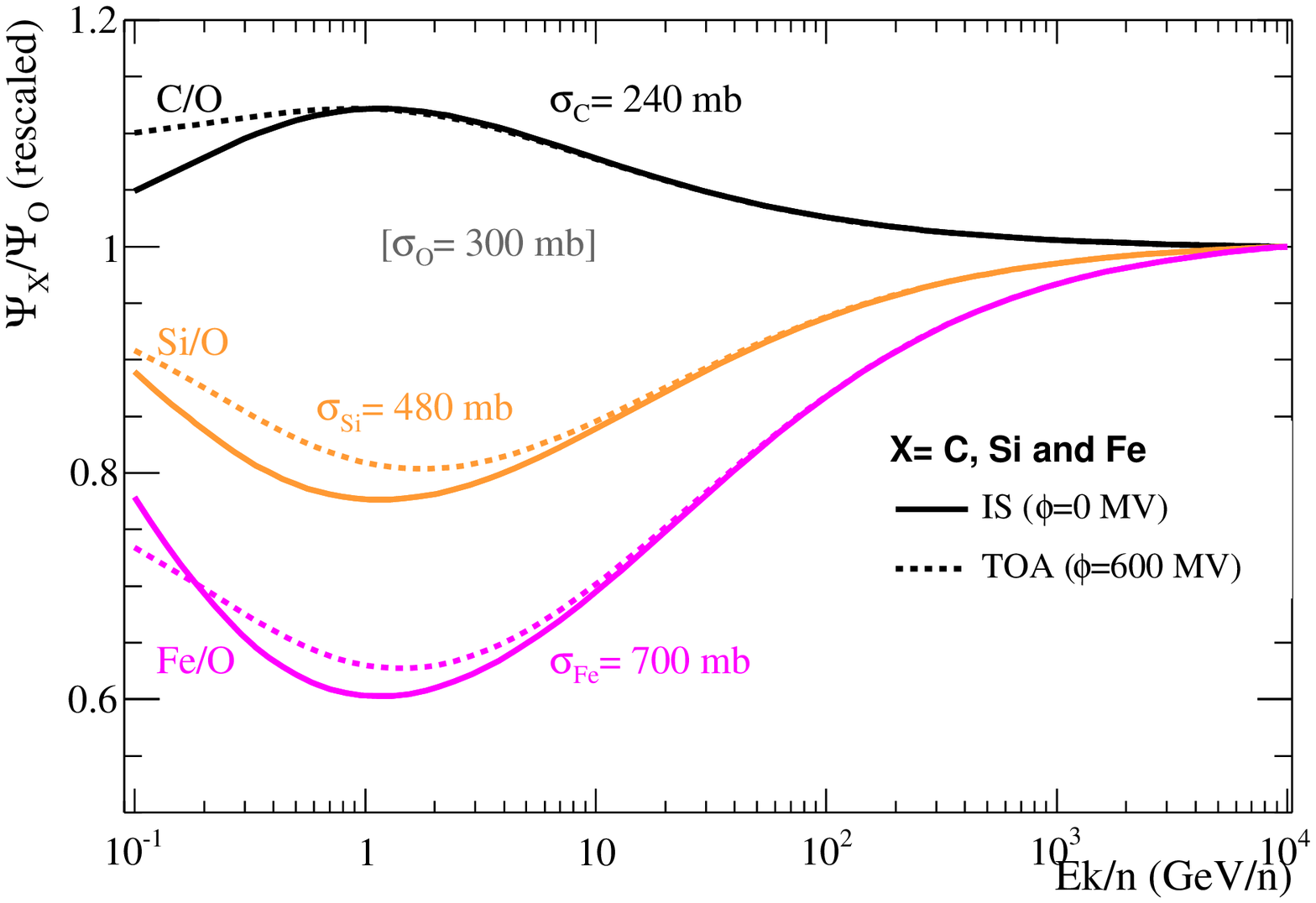}
\includegraphics[width = \columnwidth]{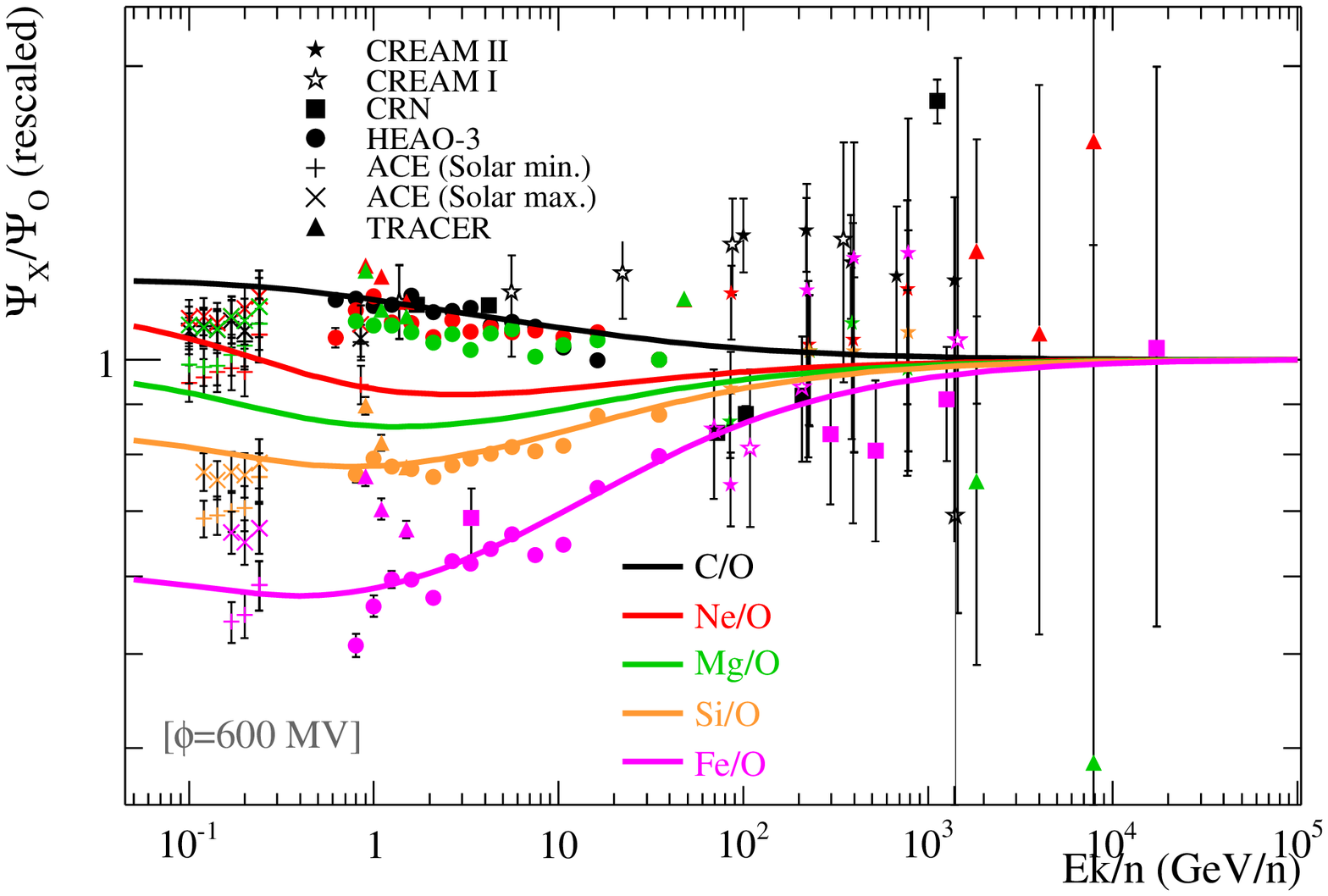}
\caption{Ratio of element to O as a function of the kinetic energy per
nucleon. Top panel: C/O, Si/O and Fe/O without (IS, solid lines) or with
(TOA, dashed lines) solar modulation. The destruction cross-section is
indicated for each element. Bottom panel: comparison of the simple toy-model
formula (thick solid lines) with the data (symbols): CREAM I and II
\citep{2008APh....30..133A,2009ApJ...707..593A,2010ApJ...714L..89A,2010ApJ...715.1400A},
CRN \citep{1990ApJ...349..625S,1991ApJ...374..356M}, HEAO-3
\citep{1990A&A...233...96E}, TRACER
\citep{2008ApJ...678..262A,2009ApJ...697..106A}, and low-energy ACE data
\citep{2009ApJ...698.1666G}.}
\label{fig:X_to_O}
\end{figure}
A similar analysis can be carried out for ratios of heavier species
($Z>2$). In that case, for any element, we have $A/Z\sim 2$, so that
for a given kinetic energy per nucleon, the rigidity or the $\beta$ is the
same for any element. The toy-model formulae is very similar to
those for p/He Eq.~(\ref{eq:ekn}): the parameters $\eta_S$ and
$\eta_T$, as well as the solar modulation effect are not expected to be
important. This is shown in the top panel of Fig.~\ref{fig:X_to_O}, for a
few elements: the main ingredient shaping the X/O ratios is the inelastic
scattering on the ISM.

There is a fair agreement with the data for C/O (black), Si/O (orange)
and Fe/O (magenta) as shown on the right panel of
Fig.~\ref{fig:X_to_O},  especially at low-energy with the ACE data
\citep{2009ApJ...698.1666G}. The discrepancy with Ne/O and Mg/O is only
at the level of $\sim 20\%$. This could be due to systematics in the data,
but this deserves further investigation, especially because some isotopic
anomalies in the Ne (and less likely for Mg) could be a signature for
a contribution of the Wolf-Rayet stars to the standard cosmic-ray abundances
\citep{1989ApJ...340.1124G,1997ApJ...476..766W,2005ApJ...634..351B,2008NewAR..52..427B},
or related to acceleration in super-bubbles \citep[e.g.,][]{2003ApJ...590..822H}.

\section{Conclusions\label{sec:conclusion}}

We have studied the spectra of proton, helium and other primary nuclei to
derive the possible contraints and degeneracies on the source spectrum and
the transport parameters. We have checked the compatibility of the primary
fluxes with the transport parameters derived from the B/C analysis, and
inspected  whether they add further constraints. We have then derived the
source parameters: slope of the power-law spectrum, abundance and
low-energy shape.

In Sect. 3, we have analysed the fluxes of  primary cosmic rays in diffusion models with 
particular attention to p and He.
The most recent data on p and He are well reproduced by a purely 
diffusive model, described by power-law source spectrum, isotropic 
diffusion coefficient, spallative destructions and electromagnetic energy losses. 
This conclusion holds for single data sets but it is not reproduced in a combined
data analysis (except for AMS01 and BESS98 proton data), due to the mean level of
consistency among the different data collections. 
The inspection of low energy ($\leq$ 100 GeV/n) p and He data
indicates that the purely diffusive regime is likely not reached due to the 
role of spallations and, to less extent, of energy losses.
The data are shown to be compatible with a wide 
class of purely diffusive models, but can also be accommodated by models with
convection and/or acceleration. In all scenarios, they do not put significant
constraints on the transport parameters and tend to favour values for the 
source and transport parameters outside reasonable physical limits. 

In Sect. 4 we have applied best-fit models previously selected from B/C data, to
the propagation of light and heavier primary nuclei, where we consider
the possibility of low-energy deviations both in the diffusion coefficient
$K(R)= K_0\beta^{\eta_T} {\cal R}^{\delta}$  
and in the acceleration spectrum $dQ/dR = q \beta^{\eta_S+1} {\cal R}^{-\alpha}$.
The main results on p and He are: 
 i) it is possible to accommodate these primary fluxes in diffusive models along 
 with B/C data (very good fit on BESS data, less satisfactory on AMS ones); 
 ii) $\alpha$ ranges between 2.2 and 2.5;
iii) for any given data set and species the spread in the source power index is 
$\alpha_p -\alpha_{\rm He}\lesssim 0.2$,  regardless of the model; 
iv) p and He point to very similar $\eta_S$, whose values depend on the model:
close to 1 for the reacceleration and convection/reacceleration models, 
whereas $\eta_S$ is close to -2 for Model I/0 and -1.5 for Model III/II, which
contain a low energy upturn in the diffusion coefficient by means of $\eta_T$. 
Indeed, the constraint which seems to emerge from any propagation model
is on their difference:  $\eta_S-\eta_T\approx 0-1$. 
We have demonstrated that a possible way to break the degeneracy  $\eta_S-\eta_T$ 
inherent the low-energy tail is by means of the Voyager data, taken at a
few hundreds of MeV/n, and in a quasi-IS regime. 

We have also studied heavier primary nuclei spectra, whose relevant destruction rate on the
ISM increases roughly with atomic number. Therefore, they can be tested against a
universal source spectral function. As for p and He, we have fitted data for C, O,
Ne, Mg, Si, S, Ar, Ca and Fe using the models selected based on B/C. 
As for light primaries, these nuclei point to $\alpha$ around 2.2-2.5 
regardless of the propagation configuration used.
The quantity $\eta_S-\eta_T$ is also constrained
from C to Fe primary data to be $\eta_S-\eta_T\approx 1$. Moreover, 
most of the derived relative source abundances are 
in agreement with those derived by earlier groups.
We have shown that it is possible to reproduce the primary
cosmic-ray spectra from protons to iron for all considered models (with or
without convection/reacceleration) without a need for an artificial break in
the injection slope at $10$~GV, as it is assumed, e.g., in \citet{2010arXiv1011.0037T}.
This is also consistent with the leaky-box analysis of
\citet{2009A&A...497..991P}.

In Sect. 5, we have studied the ratio of two primary species. The
shape of the p/He ratios, when plotted as a function of the kinetic energy
per nucleon, is driven by the modulation effect. This effect could be used
to monitor the modulation level at different periods. On the other hand, the
same ratio plotted as a function of the rigidity minimises the effect of the
modulation, and is well adapted to probe the values of $\eta_S-\eta_T$, and
$\alpha_{\rm p}-\alpha_{\rm He}$. A full analysis using energy gains and
losses is required to further develop such an approach, which is
complementary to the direct fit of the p and He fluxes, and which may suffer
less from systematics in the data. For X/O ratios, where X is an primary
element with $Z>2$, the behaviour as a function of $E{k/n}$ is driven by the
destruction rate on the ISM. Isotopic anomalies and/or non-universality of
the source slopes can be inspected by means of these ratios.

For the primary spectra and ratio of primary spectra, the
importance of the destructive terms on the measured spectra has not
previously been much considered. This is an innovation of the paper. We have
also detailed the degeneracy between the source and propagation parameters,
and provided probability density functions for the spectral index, abundance
and low-energy shape of the source spectra for all species, as a means
to address the question of the universality of the source spectra and
their low-energy shape.
Our analysis reinforces the need of more accurate data on light primary nuclei 
not only in the low-energy regime but also in the TeV/n-PeV/n range, as well as 
accurate measurement of primary--to--primary ratio. With such data it would be
possible to significantly constrain the low-energy shape of the diffusion 
coefficient and the source spectrum, and fix the asymptotic behaviour of propagated
nuclei.


\begin{acknowledgements}
We thank Alexander Panov for providing us an ASCII file of the ATIC data
and Makoto Hareyama for providing us the RUNJOB and other experiment data
points. D.~M. thanks Laurent Derome and William Gillard for useful
comments and discussions on the paper.
\end{acknowledgements}

\bibliographystyle{aa}
\bibliography{pHe}

\begin{thebibliography}{77}
\expandafter\ifx\csname natexlab\endcsname\relax\def\natexlab#1{#1}\fi

\bibitem[{{Ahn} {et~al.}(2009){Ahn}, {Allison}, {Bagliesi}, {Barbier},
  {Beatty}, {Bigongiari}, {Brandt}, {Childers}, {Conklin}, {Coutu}, {Du
  Vernois}, {Ganel}, {Han}, {Jeon}, {Kim}, {Lee}, {Maestro}, {Malinine},
  {Marrocchesi}, {Minnick}, {Mognet}, {Nam}, {Nutter}, {Park}, {Park}, {Seo},
  {Sina}, {Walpole}, {Wu}, {Yang}, {Yoon}, {Zei}, \&
  {Zinn}}]{2009ApJ...707..593A}
{Ahn}, H.~S., {Allison}, P., {Bagliesi}, M.~G., {et~al.} 2009, \apj, 707, 593

\bibitem[{{Ahn} {et~al.}(2010{\natexlab{a}}){Ahn}, {Allison}, {Bagliesi},
  {Beatty}, {Bigongiari}, {Childers}, {Conklin}, {Coutu}, {DuVernois}, {Ganel},
  {Han}, {Jeon}, {Kim}, {Lee}, {Lutz}, {Maestro}, {Malinin}, {Marrocchesi},
  {Minnick}, {Mognet}, {Nam}, {Nam}, {Nutter}, {Park}, {Park}, {Seo}, {Sina},
  {Wu}, {Yang}, {Yoon}, {Zei}, \& {Zinn}}]{2010ApJ...714L..89A}
{Ahn}, H.~S., {Allison}, P., {Bagliesi}, M.~G., {et~al.} 2010{\natexlab{a}},
  \apjl, 714, L89

\bibitem[{{Ahn} {et~al.}(2010{\natexlab{b}}){Ahn}, {Allison}, {Bagliesi},
  {Barbier}, {Beatty}, {Bigongiari}, {Brandt}, {Childers}, {Conklin}, {Coutu},
  {DuVernois}, {Ganel}, {Han}, {Jeon}, {Kim}, {Lee}, {Lee}, {Maestro},
  {Malinin}, {Marrocchesi}, {Minnick}, {Mognet}, {Na}, {Nam}, {Nam}, {Nutter},
  {Park}, {Park}, {Seo}, {Sina}, {Walpole}, {Wu}, {Yang}, {Yoon}, {Zei}, \&
  {Zinn}}]{2010ApJ...715.1400A}
{Ahn}, H.~S., {Allison}, P.~S., {Bagliesi}, M.~G., {et~al.} 2010{\natexlab{b}},
  \apj, 715, 1400

\bibitem[{{Ahn} {et~al.}(2008){Ahn}, {Allison}, {Bagliesi}, {Beatty},
  {Bigongiari}, {Boyle}, {Brandt}, {Childers}, {Conklin}, {Coutu}, {Duvernois},
  {Ganel}, {Han}, {Hyun}, {Jeon}, {Kim}, {Lee}, {Lee}, {Lutz}, {Maestro},
  {Malinin}, {Marrocchesi}, {Minnick}, {Mognet}, {Nam}, {Nutter}, {Park},
  {Park}, {Seo}, {Sina}, {Swordy}, {Wakely}, {Wu}, {Yang}, {Yoon}, {Zei}, \&
  {Zinn}}]{2008APh....30..133A}
{Ahn}, H.~S., {Allison}, P.~S., {Bagliesi}, M.~G., {et~al.} 2008, Astroparticle
  Physics, 30, 133

\bibitem[{{Alcaraz} {et~al.}(2000){Alcaraz}, {Alpat}, {Ambrosi}, {Anderhub},
  {Ao}, {Arefiev}, {Azzarello}, {Babucci}, {Baldini}, {Basile}, {Barancourt},
  {Barao}, {Barbier}, {Barreira}, {Battiston}, {Becker}, {Becker},
  {Bellagamba}, {B{\'e}n{\'e}}, {Berdugo}, {Berges}, {Bertucci}, {Biland},
  {Bizzaglia}, {Blasko}, {Boella}, {Boschini}, {Bourquin}, {Brocco}, {Bruni},
  {Buenerd}, {Burger}, {Burger}, {Cai}, {Camps}, {Cannarsa}, {Capell},
  {Casadei}, {Casaus}, {Castellini}, {Cecchi}, {Chang}, {Chen}, {Chen}, {Chen},
  {Chernoplekov}, {Chiueh}, {Chuang}, {Cindolo}, {Commichau}, {Contin},
  {Crespo}, {Cristinziani}, {da Cunha}, {Dai}, {Deus}, {Dinu}, {Djambazov},
  {D'Antone}, {Dong}, {Emonet}, {Engelberg}, {Eppling}, {Eronen}, {Esposito},
  {Extermann}, {Favier}, {Fiandrini}, {Fisher}, {Fluegge}, {Fouque},
  {Galaktionov}, {Gervasi}, {Giusti}, {Grandi}, {Grimm}, {Gu}, {Hangarter},
  {Hasan}, {Hermel}, {Hofer}, {Huang}, {Hungerford}, {Ionica}, {Ionica},
  {Jongmanns}, {Karlamaa}, {Karpinski}, {Kenney}, {Kenny}, {Kim}, {Klimentov},
  {Kossakowski}, {Koutsenko}, {Kraeber}, {Laborie}, {Laitinen}, {Lamanna},
  {Laurenti}, {Lebedev}, {Lee}, {Levi}, {Levtchenko}, {Liu}, {Liu}, {Lopes},
  {Lu}, {Lu}, {L{\"u}belsmeyer}, {Luckey}, {Lustermann}, {Ma{\~n}a},
  {Margotti}, {Mayet}, {McNeil}, {Meillon}, {Menichelli}, {Mihul}, {Mourao},
  {Mujunen}, {Palmonari}, {Papi}, {Park}, {Pauluzzi}, {Pauss}, {Perrin},
  {Pesci}, {Pevsner}, {Pimenta}, {Plyaskin}, {Pojidaev}, {Pohl}, {Postolache},
  {Produit}, {Rancoita}, {Rapin}, {Raupach}, {Ren}, {Ren}, {Ribordy},
  {Richeux}, {Riihonen}, {Ritakari}, {Roeser}, {Roissin}, {Sagdeev},
  {Sartorelli}, {Schultz von Dratzig}, {Schwering}, {Scolieri}, {Seo},
  {Shoutko}, {Shoumilov}, {Siedling}, {Son}, {Song}, {Steuer}, {Sun}, {Suter},
  {Tang}, {Ting}, {Ting}, {Tornikoski}, {Torsti}, {Tr{\"u}mper}, {Ulbricht},
  {Urpo}, {Usoskin}, {Valtonen}, {Vandenhirtz}, {Velcea}, {Velikhov},
  {Verlaat}, {Vetlitsky}, {Vezzu}, {Vialle}, {Viertel}, {Vit{\'e}}, {Von
  Gunten}, {Waldmeier Wicki}, {Wallraff}, {Wang}, {Wang}, {Wang}, {Wiik},
  {Williams}, {Wu}, {Xia}, {Yan}, {Yan}, {Yang}, {Yang}, {Ye}, {Yeh}, {Xu},
  {Zhang}, {Zhang}, {Zhao}, {Zhu}, {Zhu}, {Zhuang}, {Zichichi}, \&
  {Zimmermann}}]{2000PhLB..490...27A}
{Alcaraz}, J., {Alpat}, B., {Ambrosi}, G., {et~al.} 2000, Physics Letters B,
  490, 27

\bibitem[{{AMS Collaboration} {et~al.}(2002){AMS Collaboration}, {Aguilar},
  {Alcaraz}, {Allaby}, {Alpat}, {Ambrosi}, {Anderhub}, {Ao}, {Arefiev},
  {Azzarello}, {Babucci}, {Baldini}, {Basile}, {Barancourt}, {Barao},
  {Barbier}, {Barreira}, {Battiston}, {Becker}, {Becker}, {Bellagamba},
  {B{\'e}n{\'e}}, {Berdugo}, {Berges}, {Bertucci}, {Biland}, {Bizzaglia},
  {Blasko}, {Boella}, {Boschini}, {Bourquin}, {Brocco}, {Bruni}, {Bu{\'e}nerd},
  {Burger}, {Burger}, {Cai}, {Camps}, {Cannarsa}, {Capell}, {Casadei},
  {Casaus}, {Castellini}, {Cecchi}, {Chang}, {Chen}, {Chen}, {Chen},
  {Chernoplekov}, {Chiueh}, {Cho}, {Choi}, {Choi}, {Chuang}, {Cindolo},
  {Commichau}, {Contin}, {Cortina-Gil}, {Cristinziani}, {da Cunha}, {Dai},
  {Delgado}, {Deus}, {Dinu}, {Djambazov}, {D'Antone}, {Dong}, {Emonet},
  {Engelberg}, {Eppling}, {Eronen}, {Esposito}, {Extermann}, {Favier},
  {Fiandrini}, {Fisher}, {Fluegge}, {Fouque}, {Galaktionov}, {Gervasi},
  {Giusti}, {Grandi}, {Grimms}, {Gu}, {Hangarter}, {Hasan}, {Hermel}, {Hofer},
  {Huang}, {Hungerford}, {Ionica}, {Ionica}, {Jongmanns}, {Karlamaa},
  {Karpinski}, {Kenney}, {Kenny}, {Kim}, {Kim}, {Kim}, {Kim}, {Klimentov},
  {Kossakowski}, {Koutsenko}, {Kraeber}, {Laborie}, {Laitinen}, {Lamanna},
  {Lanciotti}, {Laurenti}, {Lebedev}, {Lechanoine-Leluc}, {Lee}, {Lee}, {Levi},
  {Levtchenko}, {Liu}, {Liu}, {Lopes}, {Lu}, {Lu}, {L{\"u}belsmeyer}, {Luckey},
  {Lustermann}, {Ma{\~n}a}, {Margotti}, {Mayet}, {McNeil}, {Meillon},
  {Menichelli}, {Mihul}, {Mourao}, {Mujunen}, {Palmonari}, {Papi}, {Park},
  {Park}, {Pauluzzi}, {Pauss}, {Perrin}, {Pesci}, {Pevsner}, {Pimenta},
  {Plyaskin}, {Pojidaev}, {Pohl}, {Postolache}, {Produit}, {Rancoita}, {Rapin},
  {Raupach}, {Ren}, {Ren}, {Ribordy}, {Richeux}, {Riihonen}, {Ritakari}, {Ro},
  {Roeser}, {Rossin}, {Sagdeev}, {Santos}, {Sartorelli}, {Sbarra}, {Schael},
  {Schultz von Dratzig}, {Schwering}, {Scolieri}, {Seo}, {Shin}, {Shoutko},
  {Shoumilov}, {Siedling}, {Son}, {Song}, {Steuer}, {Sun}, {Suter}, {Tang},
  {Ting}, {Ting}, {Tornikoski}, {Torsti}, {Tr{\"u}mper}, {Ulbricht}, {Urpo},
  {Valtonen}, {Vandenhirtz}, {Velcea}, {Velikhov}, {Verlaat}, {Vetlitsky},
  {Vezzu}, {Vialle}, {Viertel}, {Vit{\'e}}, {Gunten}, {Wicki}, {Wallraff},
  {Wang}, {Wang}, {Wang}, {Wiik}, {Williams}, {Wu}, {Xia}, {Yan}, {Yan},
  {Yang}, {Yang}, {Yang}, {Ye}, {Yeh}, {Xu}, {Zhang}, {Zhang}, {Zhao}, {Zhu},
  {Zhu}, {Zhuang}, {Zichichi}, {Zimmermann}, \& {Zuccon}}]{2002PhR...366..331A}
{AMS Collaboration}, {Aguilar}, M., {Alcaraz}, J., {et~al.} 2002, \physrep,
  366, 331

\bibitem[{{AMS Collaboration} {et~al.}(2000){AMS Collaboration}, {Alcaraz},
  {Alpat}, {Ambrosi}, {Anderhub}, {Ao}, {Arefiev}, {Azzarello}, {Babucci},
  {Baldini}, {Basile}, {Barancourt}, {Barao}, {Barbier}, {Barreira},
  {Battiston}, {Becker}, {Becker}, {Bellagamba}, {B{\'e}n{\'e}}, {Berdugo},
  {Berges}, {Bertucci}, {Biland}, {Bizzaglia}, {Blasko}, {Boella}, {Boschini},
  {Bourquin}, {Brocco}, {Bruni}, {Buenerd}, {Burger}, {Burger}, {Cai}, {Camps},
  {Cannarsa}, {Capell}, {Casadei}, {Casaus}, {Castellini}, {Cecchi}, {Chang},
  {Chen}, {Chen}, {Chen}, {Chernoplekov}, {Chiueh}, {Chuang}, {Cindolo},
  {Commichau}, {Contin}, {Cristinziani}, {da Cunha}, {Dai}, {Deus}, {Dinu},
  {Djambazov}, {D'Antone}, {Dong}, {Emonet}, {Engelberg}, {Eppling}, {Eronen},
  {Esposito}, {Extermann}, {Favier}, {Fiandrini}, {Fisher}, {Fluegge},
  {Fouque}, {Galaktionov}, {Gervasi}, {Giusti}, {Grandi}, {Grimm}, {Gu},
  {Hangarter}, {Hasan}, {Hermel}, {Hofer}, {Huang}, {Hungerford}, {Ionica},
  {Ionica}, {Jongmanns}, {Karlamaa}, {Karpinski}, {Kenney}, {Kenny}, {Kim},
  {Klimentov}, {Kossakowski}, {Koutsenko}, {Kraeber}, {Laborie}, {Laitinen},
  {Lamanna}, {Laurenti}, {Lebedev}, {Lee}, {Levi}, {Levtchenko}, {Liu}, {Liu},
  {Lopes}, {Lu}, {Lu}, {L{\"u}belsmeyer}, {Luckey}, {Lustermann}, {Ma{\~n}a},
  {Margotti}, {Mayet}, {McNeil}, {Meillon}, {Menichelli}, {Mihul}, {Mourao},
  {Mujunen}, {Palmonari}, {Papi}, {Park}, {Pauluzzi}, {Pauss}, {Perrin},
  {Pesci}, {Pevsner}, {Pimenta}, {Plyaskin}, {Pojidaev}, {Pohl}, {Postolache},
  {Produit}, {Rancoita}, {Rapin}, {Raupach}, {Ren}, {Ren}, {Ribordy},
  {Richeux}, {Riihonen}, {Ritakari}, {Roeser}, {Roissin}, {Sagdeev},
  {Sartorelli}, {Schultz von Dratzig}, {Schwering}, {Scolieri}, {Seo},
  {Shoutko}, {Shoumilov}, {Siedling}, {Son}, {Song}, {Steuer}, {Sun}, {Suter},
  {Tang}, {Ting}, {Ting}, {Tornikoski}, {Torsti}, {Tr{\"u}mper}, {Ulbricht},
  {Urpo}, {Usoskin}, {Valtonen}, {Vandenhirtz}, {Velcea}, {Velikhov},
  {Verlaat}, {Vetlitsky}, {Vezzu}, {Vialle}, {Viertel}, {Vit{\'e}}, {Von
  Gunten}, {Waldmeier Wicki}, {Wallraff}, {Wang}, {Wang}, {Wang}, {Wiik},
  {Williams}, {Wu}, {Xia}, {Yan}, {Yan}, {Yang}, {Yang}, {Ye}, {Yeh}, {Xu},
  {Zhang}, {Zhang}, {Zhao}, {Zhu}, {Zhu}, {Zhuang}, {Zichichi}, {Zimmermann},
  \& {Zuccon}}]{2000PhLB..494..193A}
{AMS Collaboration}, {Alcaraz}, J., {Alpat}, B., {et~al.} 2000, Physics Letters
  B, 494, 193

\bibitem[{{Asakimori} {et~al.}(1998){Asakimori}, {Burnett}, {Cherry}, {Chevli},
  {Christ}, {Dake}, {Derrickson}, {Fountain}, {Fuki}, {Gregory}, {Hayashi},
  {Holynski}, {Iwai}, {Iyono}, {Johnson}, {Kobayashi}, {Lord}, {Miyamura},
  {Moon}, {Nilsen}, {Oda}, {Ogata}, {Olson}, {Parnell}, {Roberts}, {Sengupta},
  {Shiina}, {Strausz}, {Sugitate}, {Takahashi}, {Tominaga}, {Watts}, {Wefel},
  {Wilczynska}, {Wilczynski}, {Wilkes}, {Wolter}, {Yokomi}, \&
  {Zager}}]{1998ApJ...502..278A}
{Asakimori}, K., {Burnett}, T.~H., {Cherry}, M.~L., {et~al.} 1998, \apj, 502,
  278

\bibitem[{{Ave} {et~al.}(2008){Ave}, {Boyle}, {Gahbauer}, {H{\"o}ppner},
  {H{\"o}randel}, {Ichimura}, {M{\"u}ller}, \&
  {Romero-Wolf}}]{2008ApJ...678..262A}
{Ave}, M., {Boyle}, P.~J., {Gahbauer}, F., {et~al.} 2008, \apj, 678, 262

\bibitem[{{Ave} {et~al.}(2009){Ave}, {Boyle}, {H{\"o}ppner}, {Marshall}, \&
  {M{\"u}ller}}]{2009ApJ...697..106A}
{Ave}, M., {Boyle}, P.~J., {H{\"o}ppner}, C., {Marshall}, J., \& {M{\"u}ller},
  D. 2009, \apj, 697, 106

\bibitem[{{Binns} {et~al.}(2008){Binns}, {Wiedenbeck}, {Arnould}, {Cummings},
  {de Nolfo}, {Goriely}, {Israel}, {Leske}, {Mewaldt}, {Stone}, \& {von
  Rosenvinge}}]{2008NewAR..52..427B}
{Binns}, W.~R., {Wiedenbeck}, M.~E., {Arnould}, M., {et~al.} 2008, \nar, 52,
  427

\bibitem[{{Binns} {et~al.}(2005){Binns}, {Wiedenbeck}, {Arnould}, {Cummings},
  {George}, {Goriely}, {Israel}, {Leske}, {Mewaldt}, {Meynet}, {Scott},
  {Stone}, \& {von Rosenvinge}}]{2005ApJ...634..351B}
{Binns}, W.~R., {Wiedenbeck}, M.~E., {Arnould}, M., {et~al.} 2005, \apj, 634,
  351

\bibitem[{{Boezio} {et~al.}(2003){Boezio}, {Bonvicini}, {Schiavon}, {Vacchi},
  {Zampa}, {Bergstr{\"o}m}, {Carlson}, {Francke}, {Hansen}, {Mocchiutti},
  {Suffert}, {Hof}, {Kremer}, {Menn}, {Simon}, {Ambriola}, {Bellotti},
  {Cafagna}, {Ciacio}, {Circella}, {de Marzo}, {Finetti}, {Papini}, {Piccardi},
  {Spillantini}, {Vannuccini}, {Bartalucci}, {Ricci}, {Casolino}, {de Pascale},
  {Morselli}, {Picozza}, {Sparvoli}, {Mitchell}, {Ormes}, {Stephens},
  {Streitmatter}, {Bravar}, \& {Stochaj}}]{2003APh....19..583B}
{Boezio}, M., {Bonvicini}, V., {Schiavon}, P., {et~al.} 2003, Astroparticle
  Physics, 19, 583

\bibitem[{{Boezio} {et~al.}(1999){Boezio}, {Carlson}, {Francke}, {Weber},
  {Suffert}, {Hof}, {Menn}, {Simon}, {Stephens}, {Bellotti}, {Cafagna},
  {Castellano}, {Circella}, {de Marzo}, {Finetti}, {Papini}, {Piccardi},
  {Spillantini}, {Ricci}, {Casolino}, {de Pascale}, {Morselli}, {Picozza},
  {Sparvoli}, {Barbiellini}, {Bravar}, {Schiavon}, {Vacchi}, {Zampa},
  {Mitchell}, {Ormes}, {Streitmatter}, {Golden}, \&
  {Stochaj}}]{1999ApJ...518..457B}
{Boezio}, M., {Carlson}, P., {Francke}, T., {et~al.} 1999, \apj, 518, 457

\bibitem[{{Boulares}(1989)}]{1989ApJ...342..807B}
{Boulares}, A. 1989, \apj, 342, 807

\bibitem[{{Caprioli} {et~al.}(2010){Caprioli}, {Amato}, \&
  {Blasi}}]{2010APh....33..160C}
{Caprioli}, D., {Amato}, E., \& {Blasi}, P. 2010, Astroparticle Physics, 33,
  160

\bibitem[{{Delahaye} {et~al.}(2010){Delahaye}, {Lavalle}, {Lineros}, {Donato},
  \& {Fornengo}}]{2010arXiv1002.1910D}
{Delahaye}, T., {Lavalle}, J., {Lineros}, R., {Donato}, F., \& {Fornengo}, N.
  2010, ArXiv e-prints

\bibitem[{{Derbina} {et~al.}(2005){Derbina}, {Galkin}, {Hareyama}, {Hirakawa},
  {Horiuchi}, {Ichimura}, {Inoue}, {Kamioka}, {Kobayashi}, {Kopenkin},
  {Kuramata}, {Managadze}, {Matsutani}, {Misnikova}, {Mukhamedshin},
  {Nagasawa}, {Nakano}, {Namiki}, {Nakazawa}, {Nanjo}, {Nazarov}, {Ohata},
  {Ohtomo}, {Osedlo}, {Oshuev}, {Publichenko}, {Rakobolskaya}, {Roganova},
  {Saito}, {Sazhina}, {Semba}, {Shibata}, {Shuto}, {Sugimoto}, {Suzuki},
  {Sveshnikova}, {Taran}, {Yajima}, {Yamagami}, {Yashin}, {Zamchalova},
  {Zatsepin}, \& {Zayarnaya}}]{2005ApJ...628L..41D}
{Derbina}, V.~A., {Galkin}, V.~I., {Hareyama}, M., {et~al.} 2005, \apjl, 628,
  L41

\bibitem[{{Diehl} {et~al.}(2003){Diehl}, {Ellithorpe}, {M{\"u}ller}, \&
  {Swordy}}]{2003APh....18..487D}
{Diehl}, E., {Ellithorpe}, D., {M{\"u}ller}, D., \& {Swordy}, S.~P. 2003,
  Astroparticle Physics, 18, 487

\bibitem[{{Donato} {et~al.}(2008){Donato}, {Fornengo}, \&
  {Maurin}}]{2008PhRvD..78d3506D}
{Donato}, F., {Fornengo}, N., \& {Maurin}, D. 2008, \prd, 78, 043506

\bibitem[{{Donato} {et~al.}(2009){Donato}, {Maurin}, {Brun}, {Delahaye}, \&
  {Salati}}]{2009PhRvL.102g1301D}
{Donato}, F., {Maurin}, D., {Brun}, P., {Delahaye}, T., \& {Salati}, P. 2009,
  Physical Review Letters, 102, 071301

\bibitem[{{Donato} {et~al.}(2001){Donato}, {Maurin}, {Salati}, {Barrau},
  {Boudoul}, \& {Taillet}}]{2001ApJ...563..172D}
{Donato}, F., {Maurin}, D., {Salati}, P., {et~al.} 2001, \apj, 563, 172

\bibitem[{{Donato} {et~al.}(2002){Donato}, {Maurin}, \&
  {Taillet}}]{2002A&A...381..539D}
{Donato}, F., {Maurin}, D., \& {Taillet}, R. 2002, Astronomy and Astrophys.,
  381, 539

\bibitem[{{Donato} \& {Serpico}(2010)}]{2010arXiv1010.5679D}
{Donato}, F. \& {Serpico}, P.~D. 2010, ArXiv e-prints

\bibitem[{{Drury}(1983)}]{1983RPPh...46..973D}
{Drury}, L.~O. 1983, Reports on Progress in Physics, 46, 973

\bibitem[{{Duvernois} \& {Thayer}(1996)}]{1996ApJ...465..982D}
{Duvernois}, M.~A. \& {Thayer}, M.~R. 1996, \apj, 465, 982

\bibitem[{{Dwyer} \& {Meyer}(1987)}]{1987ApJ...322..981D}
{Dwyer}, R. \& {Meyer}, P. 1987, \apj, 322, 981

\bibitem[{{Ellison} {et~al.}(1997){Ellison}, {Drury}, \&
  {Meyer}}]{1997ApJ...487..197E}
{Ellison}, D.~C., {Drury}, L.~O., \& {Meyer}, J.-P. 1997, \apj, 487, 197

\bibitem[{{Engelmann} {et~al.}(1990){Engelmann}, {Ferrando}, {Soutoul},
  {Goret}, \& {Juliusson}}]{1990A&A...233...96E}
{Engelmann}, J.~J., {Ferrando}, P., {Soutoul}, A., {Goret}, P., \& {Juliusson},
  E. 1990, \aap, 233, 96

\bibitem[{{Ferrand} {et~al.}(2008){Ferrand}, {Downes}, \&
  {Marcowith}}]{2008MNRAS.383...41F}
{Ferrand}, G., {Downes}, T., \& {Marcowith}, A. 2008, \mnras, 383, 41

\bibitem[{{Ferrand} \& {Marcowith}(2010)}]{2010A&A...510A.101F}
{Ferrand}, G. \& {Marcowith}, A. 2010, \aap, 510, A101+

\bibitem[{{George} {et~al.}(2009){George}, {Lave}, {Wiedenbeck}, {Binns},
  {Cummings}, {Davis}, {de Nolfo}, {Hink}, {Israel}, {Leske}, {Mewaldt},
  {Scott}, {Stone}, {von Rosenvinge}, \& {Yanasak}}]{2009ApJ...698.1666G}
{George}, J.~S., {Lave}, K.~A., {Wiedenbeck}, M.~E., {et~al.} 2009, \apj, 698,
  1666

\bibitem[{{Gilmore} {et~al.}(1992){Gilmore}, {Gustafsson}, {Edvardsson}, \&
  {Nissen}}]{1992Natur.357..379G}
{Gilmore}, G., {Gustafsson}, B., {Edvardsson}, B., \& {Nissen}, P.~E. 1992,
  \nat, 357, 379

\bibitem[{{Gupta} \& {Webber}(1989)}]{1989ApJ...340.1124G}
{Gupta}, M. \& {Webber}, W.~R. 1989, \apj, 340, 1124

\bibitem[{{Haino} {et~al.}(2004){Haino}, {Sanuki}, {Abe}, {Anraku}, {Asaoka},
  {Fuke}, {Imori}, {Itasaki}, {Maeno}, {Makida}, {Matsuda}, {Matsui},
  {Matsumoto}, {Mitchell}, {Moiseev}, {Nishimura}, {Nozaki}, {Orito}, {Ormes},
  {Sasaki}, {Seo}, {Shikaze}, {Streitmatter}, {Suzuki}, {Takasugi}, {Tanaka},
  {Tanizaki}, {Yamagami}, {Yamamoto}, {Yamamoto}, {Yamato}, {Yoshida}, \&
  {Yoshimura}}]{2004PhLB..594...35H}
{Haino}, S., {Sanuki}, T., {Abe}, K., {et~al.} 2004, Physics Letters B, 594, 35

\bibitem[{{Herbst} {et~al.}(2010){Herbst}, {Kopp}, {Heber}, {Steinhilber},
  {Fichtner}, {Scherer}, \& {Matthi{\"a}}}]{2010JGRA..11500I20H}
{Herbst}, K., {Kopp}, A., {Heber}, B., {et~al.} 2010, Journal of Geophysical
  Research (Space Physics), 115

\bibitem[{{Higdon} \& {Lingenfelter}(2003)}]{2003ApJ...590..822H}
{Higdon}, J.~C. \& {Lingenfelter}, R.~E. 2003, \apj, 590, 822

\bibitem[{{Ichimura} {et~al.}(1993){Ichimura}, {Kogawa}, {Kuramata}, {Mito},
  {Murabayashi}, {Nanjo}, {Nakamura}, {Ohba}, {Ohuchi}, {Ozawa}, {Yamada},
  {Matsutani}, {Watanabe}, {Kamioka}, {Kirii}, {Kitazawa}, {Kobayashi},
  {Mihashi}, {Shibata}, {Shibuta}, {Sugimoto}, \&
  {Nakazawa}}]{1993PhRvD..48.1949I}
{Ichimura}, M., {Kogawa}, M., {Kuramata}, S., {et~al.} 1993, \prd, 48, 1949

\bibitem[{{Ivanenko} {et~al.}(1993){Ivanenko}, {Shestoperov}, \& {et
  al.}}]{1993ICRC....2...17I}
{Ivanenko}, I.~P., {Shestoperov}, V.~Y., \& {et al.} 1993, ICRC, 2, 17

\bibitem[{{Jones}(1994)}]{1994ApJS...90..561J}
{Jones}, F.~C. 1994, \apjs, 90, 561

\bibitem[{{Jones} {et~al.}(2001){Jones}, {Lukasiak}, {Ptuskin}, \&
  {Webber}}]{2001ApJ...547..264J}
{Jones}, F.~C., {Lukasiak}, A., {Ptuskin}, V., \& {Webber}, W. 2001, \apj, 547,
  264

\bibitem[{{Lavalle}(2010)}]{2010arXiv1011.3063L}
{Lavalle}, J. 2010, ArXiv e-prints

\bibitem[{{Lemoine} {et~al.}(1998){Lemoine}, {Vangioni-Flam}, \&
  {Casse}}]{1998ApJ...499..735L}
{Lemoine}, M., {Vangioni-Flam}, E., \& {Casse}, M. 1998, \apj, 499, 735

\bibitem[{{Lionetto} {et~al.}(2005){Lionetto}, {Morselli}, \&
  {Zdravkovic}}]{2005JCAP...09..010L}
{Lionetto}, A.~M., {Morselli}, A., \& {Zdravkovic}, V. 2005, Journal of
  Cosmology and Astro-Particle Physics, 9, 10

\bibitem[{{Maurin} {et~al.}(2001){Maurin}, {Donato}, {Taillet}, \&
  {Salati}}]{2001ApJ...555..585M}
{Maurin}, D., {Donato}, F., {Taillet}, R., \& {Salati}, P. 2001, \apj, 555, 585

\bibitem[{{Maurin} {et~al.}(2010){Maurin}, {Putze}, \&
  {Derome}}]{2010A&A...516A..67M}
{Maurin}, D., {Putze}, A., \& {Derome}, L. 2010, \aap, 516, A67

\bibitem[{{Maurin} {et~al.}(2002){Maurin}, {Taillet}, \&
  {Donato}}]{2002A&A...394.1039M}
{Maurin}, D., {Taillet}, R., \& {Donato}, F. 2002, Astronomy and Astrophys.,
  394, 1039

\bibitem[{{Menn} {et~al.}(2000){Menn}, {Hof}, {Reimer}, {Simon}, {Davis},
  {Labrador}, {Mewaldt}, {Schindler}, {Barbier}, {Christian}, {Krombel},
  {Krizmanic}, {Mitchell}, {Ormes}, {Streitmatter}, {Golden}, {Stochaj},
  {Webber}, \& {Rasmussen}}]{2000ApJ...533..281M}
{Menn}, W., {Hof}, M., {Reimer}, O., {et~al.} 2000, \apj, 533, 281

\bibitem[{{Meyer} {et~al.}(1997){Meyer}, {Drury}, \&
  {Ellison}}]{1997ApJ...487..182M}
{Meyer}, J.-P., {Drury}, L.~O., \& {Ellison}, D.~C. 1997, \apj, 487, 182

\bibitem[{{Mueller} {et~al.}(1991){Mueller}, {Swordy}, {Meyer}, {L'Heureux}, \&
  {Grunsfeld}}]{1991ApJ...374..356M}
{Mueller}, D., {Swordy}, S.~P., {Meyer}, P., {L'Heureux}, J., \& {Grunsfeld},
  J.~M. 1991, \apj, 374, 356

\bibitem[{{Nath} \& {Biermann}(1994)}]{1994MNRAS.267..447N}
{Nath}, B.~B. \& {Biermann}, P.~L. 1994, \mnras, 267, 447

\bibitem[{{Ogliore} {et~al.}(2009){Ogliore}, {Stone}, {Leske}, {Mewaldt},
  {Wiedenbeck}, {Binns}, {Israel}, {von Rosenvinge}, {de Nolfo}, \&
  {Moskalenko}}]{2009ApJ...695..666O}
{Ogliore}, R.~C., {Stone}, E.~C., {Leske}, R.~A., {et~al.} 2009, \apj, 695, 666

\bibitem[{{Osborne} \& {Ptuskin}(1988)}]{1988SvAL...14..132O}
{Osborne}, J.~L. \& {Ptuskin}, V.~S. 1988, Soviet Astronomy Letters, 14, 132

\bibitem[{{Padovani} {et~al.}(2009){Padovani}, {Galli}, \&
  {Glassgold}}]{2009A&A...501..619P}
{Padovani}, M., {Galli}, D., \& {Glassgold}, A.~E. 2009, \aap, 501, 619

\bibitem[{{Panov} {et~al.}(2009){Panov}, {Adams Jr.}, {Ahn}, {Bashinzhagyan},
  {Watts}, {Wefel}, {Wu}, {Ganel}, {Guzik}, {Zatsepin}, {Isbert}, {Kim},
  {Christl}, {Kouznetsov}, {Panasyuk}, {Seo}, {Sokolskaya}, {Chang}, {Schmidt},
  \& {Fazely}}]{2009APh....30..133A}
{Panov}, A.~D., {Adams Jr.}, J.~H., {Ahn}, A.~S., {et~al.} 2009, Bulletin of
  the Russian Academy of Sciences: Physics, 73, 564

\bibitem[{{Perko}(1987)}]{1987A&A...184..119P}
{Perko}, J.~S. 1987, \aap, 184, 119

\bibitem[{{Ptuskin} {et~al.}(2006){Ptuskin}, {Moskalenko}, {Jones}, {Strong},
  \& {Zirakashvili}}]{2006ApJ...642..902P}
{Ptuskin}, V.~S., {Moskalenko}, I.~V., {Jones}, F.~C., {Strong}, A.~W., \&
  {Zirakashvili}, V.~N. 2006, \apj, 642, 902

\bibitem[{{Putze} {et~al.}(2010){Putze}, {Derome}, \&
  {Maurin}}]{2010A&A...516A..66P}
{Putze}, A., {Derome}, L., \& {Maurin}, D. 2010, \aap, 516, A66

\bibitem[{{Putze} {et~al.}(2009){Putze}, {Derome}, {Maurin}, {Perotto}, \&
  {Taillet}}]{2009A&A...497..991P}
{Putze}, A., {Derome}, L., {Maurin}, D., {Perotto}, L., \& {Taillet}, R. 2009,
  \aap, 497, 991

\bibitem[{{Sanuki} {et~al.}(2000){Sanuki}, {Motoki}, {Matsumoto}, {Seo},
  {Wang}, {Abe}, {Anraku}, {Asaoka}, {Fujikawa}, {Imori}, {Maeno}, {Makida},
  {Matsui}, {Matsunaga}, {Mitchell}, {Mitsui}, {Moiseev}, {Nishimura},
  {Nozaki}, {Orito}, {Ormes}, {Saeki}, {Sasaki}, {Shikaze}, {Sonoda},
  {Streitmatter}, {Suzuki}, {Tanaka}, {Ueda}, {Yajima}, {Yamagami}, {Yamamoto},
  {Yoshida}, \& {Yoshimura}}]{2000ApJ...545.1135S}
{Sanuki}, T., {Motoki}, M., {Matsumoto}, H., {et~al.} 2000, \apj, 545, 1135

\bibitem[{{Scherer} {et~al.}(2008){Scherer}, {Fichtner}, {Ferreira},
  {B{\"u}sching}, \& {Potgieter}}]{2008ApJ...680L.105S}
{Scherer}, K., {Fichtner}, H., {Ferreira}, S.~E.~S., {B{\"u}sching}, I., \&
  {Potgieter}, M.~S. 2008, \apjl, 680, L105

\bibitem[{{Seo} \& {Ptuskin}(1994)}]{1994ApJ...431..705S}
{Seo}, E.~S. \& {Ptuskin}, V.~S. 1994, \apj, 431, 705

\bibitem[{{Shikaze} {et~al.}(2007){Shikaze}, {Haino}, {Abe}, {Fuke}, {Hams},
  {Kim}, {Makida}, {Matsuda}, {Mitchell}, {Moiseev}, {Nishimura}, {Nozaki},
  {Orito}, {Ormes}, {Sanuki}, {Sasaki}, {Seo}, {Streitmatter}, {Suzuki},
  {Tanaka}, {Yamagami}, {Yamamoto}, {Yoshida}, \&
  {Yoshimura}}]{2007APh....28..154S}
{Shikaze}, Y., {Haino}, S., {Abe}, K., {et~al.} 2007, Astroparticle Physics,
  28, 154

\bibitem[{{Strong} {et~al.}(2007){Strong}, {Moskalenko}, \&
  {Ptuskin}}]{2007ARNPS..57..285S}
{Strong}, A.~W., {Moskalenko}, I.~V., \& {Ptuskin}, V.~S. 2007, Annual Review
  of Nuclear and Particle Science, 57, 285

\bibitem[{{Swordy} {et~al.}(1990){Swordy}, {Mueller}, {Meyer}, {L'Heureux}, \&
  {Grunsfeld}}]{1990ApJ...349..625S}
{Swordy}, S.~P., {Mueller}, D., {Meyer}, P., {L'Heureux}, J., \& {Grunsfeld},
  J.~M. 1990, \apj, 349, 625

\bibitem[{{Trotta} {et~al.}(2010){Trotta}, {Johannesson}, {Moskalenko},
  {Porter}, {Ruiz de Austri}, \& {Strong}}]{2010arXiv1011.0037T}
{Trotta}, R., {Johannesson}, G., {Moskalenko}, I.~V., {et~al.} 2010, ArXiv
  e-prints

\bibitem[{{Wang} {et~al.}(2002){Wang}, {Seo}, {Anraku}, {Fujikawa}, {Imori},
  {Maeno}, {Matsui}, {Matsunaga}, {Motoki}, {Orito}, {Saeki}, {Sanuki}, {Ueda},
  {Yoshimura}, {Makida}, {Suzuki}, {Tanaka}, {Yamamoto}, {Yoshida}, {Mitsui},
  {Matsumoto}, {Nozaki}, {Sasaki}, {Mitchell}, {Moiseev}, {Ormes},
  {Streitmatter}, {Nishimura}, {Yajima}, \& {Yamagami}}]{2002ApJ...564..244W}
{Wang}, J.~Z., {Seo}, E.~S., {Anraku}, K., {et~al.} 2002, \apj, 564, 244

\bibitem[{{Webber}(1987)}]{1987A&A...179..277W}
{Webber}, W.~R. 1987, \aap, 179, 277

\bibitem[{{Webber}(1998)}]{1998ApJ...506..329W}
{Webber}, W.~R. 1998, \apj, 506, 329

\bibitem[{{Webber} {et~al.}(2008){Webber}, {Cummings}, {McDonald}, {Stone},
  {Heikkila}, \& {Lal}}]{2008JGRA..11310108W}
{Webber}, W.~R., {Cummings}, A.~C., {McDonald}, F.~B., {et~al.} 2008, Journal
  of Geophysical Research (Space Physics), 113, 10108

\bibitem[{{Webber} {et~al.}(1987){Webber}, {Golden}, \&
  {Stephens}}]{1987ICRC....1..325W}
{Webber}, W.~R., {Golden}, R.~L., \& {Stephens}, S.~A. 1987, ICRC, 1, 325

\bibitem[{{Webber} \& {Higbie}(2009)}]{2009JGRA..11402103W}
{Webber}, W.~R. \& {Higbie}, P.~R. 2009, Journal of Geophysical Research (Space
  Physics), 114, 2103

\bibitem[{{Webber} \& {Lezniak}(1974)}]{1974Ap&SS..30..361W}
{Webber}, W.~R. \& {Lezniak}, J.~A. 1974, \apss, 30, 361

\bibitem[{{Webber} {et~al.}(1997){Webber}, {Lukasiak}, \&
  {McDonald}}]{1997ApJ...476..766W}
{Webber}, W.~R., {Lukasiak}, A., \& {McDonald}, F.~B. 1997, \apj, 476, 766

\bibitem[{{Webber} {et~al.}(2003){Webber}, {Soutoul}, {Kish}, \&
  {Rockstroh}}]{2003ApJS..144..153W}
{Webber}, W.~R., {Soutoul}, A., {Kish}, J.~C., \& {Rockstroh}, J.~M. 2003,
  \apjs, 144, 153

\bibitem[{{Zatsepin} {et~al.}(2003){Zatsepin}, {Adams}, {Ahn},
  {Bashindzhangyan}, {Batkov}, {Chang}, {Christl}, {Fazely}, {Ganel},
  {Gunasingha}, {Guzik}, {Isbert}, {Kim}, {Kouznetsov}, {Panasyuk}, {Panov},
  {Schmidt}, {Seo}, {Sokolskaya}, {Wang}, {Wefel}, \&
  {Wu}}]{2003ICRC....4.1829Z}
{Zatsepin}, V.~I., {Adams}, J.~H., {Ahn}, H.~S., {et~al.} 2003, ICRC, 4, 1829

\bibitem[{{Zatsepin} {et~al.}(1993){Zatsepin}, {Zamchalova}, \& {et
  al.}}]{1993ICRC....2...13Z}
{Zatsepin}, V.~I., {Zamchalova}, E.~A., \& {et al.} 1993, ICRC, 2, 13

\end{thebibliography}

\Online

\begin{appendix} 
\onecolumn
\section{Source slopes and relative elemental abundances}

\begin{table}[!ht]
	\centering
  \caption{Most probable values and 68\% CIs on the spectral index $\alpha$ and
the relative source abundance $q_Z/q_{\rm Si}$ for the fit of the source
spectrum parameters. Each column corresponds to a different set of data on
which the fit is performed.}
\label{tab:C_Fe_BCfixed_MPV}
	\begin{tabular}{lc|cc|cc|c}
		\hline \hline
		Model & \multicolumn{2}{c}{HEAO-3} & \multicolumn{2}{c}{TRACER} & \multicolumn{2}{c}{CREAM}\\
		&  $\alpha$ & $q_Z/q_{\rm Si}$ &  \multicolumn{2}{c}{\dots} & \multicolumn{2}{c}{\dots} \\ \hline
		 \multicolumn{7}{c}{| Carbon |}\\
		 II     &  $2.41^{+0.02}_{-0.02}$ & $428^{+19}_{-16}$ &  N/A & N/A  & $2.50^{+0.06}_{-0.06}$ & $285^{+121}_{-106}$ \\
		 III    &  $2.34^{+0.01}_{-0.02}$ & $388^{+17}_{-12}$&  N/A & N/A  & $1.93^{+0.07}_{-0.07}$ & $77^{+46}_{-31}$ \\
		 I/0    &  $2.28^{+0.02}_{-0.02}$ &$368^{+12}_{-18}$ &  N/A & N/A  & $2.13^{+0.06}_{-0.06}$ & $194^{+73}_{-82}$ \\
		 III/II &  $2.27^{+0.01}_{-0.02}$ & $365^{+11}_{-17}$ &  N/A & N/A  & $2.24^{+0.06}_{-0.07}$ & $200^{+95}_{-70}$ \\[1mm]
		 \multicolumn{7}{c}{| Oxygen |}\\
		 II     &  $2.37^{+0.02}_{-0.02}$ & $458^{+16}_{-19}$ & $2.35^{+0.01}_{-0.01}$ & $244^{+6}_{-7}$ & $2.63^{+0.05}_{-0.06}$ & $552^{+249}_{-138}$ \\
		 III    &  $2.31^{+0.02}_{-0.01}$ & $443^{+14}_{-17}$ & $2.27^{+0.01}_{-0.01}$ & $540^{+11}_{-10}$ & $2.13^{+0.06}_{-0.06}$ & $266^{+111}_{-97}$ \\
		 I/0    &  $2.27^{+0.01}_{-0.02}$ & $424^{+18}_{-18}$ & N/A  & N/A  & $2.30^{+0.05}_{-0.06}$ & $424^{+183}_{-131}$\\
		 III/II &  $2.25^{+0.02}_{-0.02}$ & $423^{+16}_{-18}$ & $2.19^{+0.01}_{-0.01}$ & $274^{+6}_{-10}$ & $2.40^{+0.06}_{-0.07}$ & $462^{+190}_{-139}$ \\[1mm]
		 \multicolumn{7}{c}{| Neon |}\\
		 II     &  $2.37^{+0.02}_{-0.02}$ & $59^{+3}_{-3}$ & $2.28^{+0.01}_{-0.02}$ & $25^{+2}_{-2}$ & $2.64^{+0.11}_{-0.11}$ & $61^{+73}_{-39}$ \\
		 III    &  $2.30^{+0.02}_{-0.02}$ & $54^{+2}_{-2}$ & $2.19^{+0.01}_{-0.01}$ & $57^{+3}_{-3}$ & $2.16^{+0.07}_{-0.12}$ & $17^{+39}_{-17}$ \\
		 I/0    &  $2.24^{+0.02}_{-0.02}$ & $50^{+3}_{-2}$ &  N/A &  N/A & $2.30^{+0.12}_{-0.14}$ & $42^{+57}_{-33}$ \\
		 III/II &  $2.23^{+0.02}_{-0.02}$ & $50^{+3}_{-2}$ & $2.09^{+0.02}_{-0.02}$ & $25^{+2}_{-2}$ & $2.40^{+0.12}_{-0.13}$ & $42^{+63}_{-31}$ \\[1mm]
		 \multicolumn{7}{c}{| Magnesium |}\\
		 II     &  $2.40^{+0.02}_{-0.02}$ & $108^{+6}_{-5}$ & $2.35^{+0.01}_{-0.01}$ & $57^{+3}_{-3}$ & $2.57^{+0.07}_{-0.08}$ & $60^{+37}_{-21}$ \\
		 III    &  $2.35^{+0.02}_{-0.02}$ & $107^{+5}_{-5}$ & $2.23^{+0.01}_{-0.01}$ & $107^{+4}_{-4}$ & $2.16^{+0.07}_{-0.12}$ & $35^{+32}_{-16}$ \\
		 I/0    &  $2.29^{+0.02}_{-0.02}$ & $101^{+6}_{-6}$ &  N/A & N/A  & $2.26^{+0.08}_{-0.08}$ & $54^{+31}_{-25}$ \\
		 III/II &  $2.29^{+0.02}_{-0.02}$ & $101^{+5}_{-5}$ & $2.17^{+0.01}_{-0.01}$ & $57^{+4}_{-3}$ & $2.34^{+0.09}_{-0.07}$ & $52^{+34}_{-21}$  \\[1mm]
		 \multicolumn{7}{c}{| Silicon |}\\
		 II     &  $2.39^{+0.02}_{-0.02}$ & $100^{+6}_{-5}$ & $2.49^{+0.01}_{-0.01}$ & $100^{+4}_{-5}$ & $2.63^{+0.06}_{-0.08}$ & $100^{+47}_{-34}$ \\
		 III    &  $2.34^{+0.02}_{-0.02}$ & $100^{+6}_{-5}$ & $2.24^{+0.01}_{-0.01}$ & $100^{+4}_{-4}$ & $2.28^{+0.09}_{-0.08}$ & $100^{+48}_{-43}$ \\
		 I/0    &  $2.29^{+0.02}_{-0.02}$ & $100^{+5}_{-6}$ & N/A  & N/A  & $2.36^{+0.07}_{-0.08}$ & $100^{+53}_{-36}$ \\
		 III/II &  $2.29^{+0.02}_{-0.02}$ & $100^{+5}_{-6}$ & $2.31^{+0.01}_{-0.01}$ & $100^{+5}_{-4}$ & $2.43^{+0.07}_{-0.08}$ & $100^{+48}_{-38}$ \\[1mm]
		 \multicolumn{7}{c}{| Sulfur |}\\
		 II     &  $2.45^{+0.03}_{-0.03}$ & $20^{+1}_{-2}$ & $2.40^{+0.03}_{-0.04}$ & $12^{+2}_{-2}$ &  N/A  & N/A  \\
		 III    &  $2.39^{+0.02}_{-0.03}$ & $19^{+2}_{-1}$ & $2.20^{+0.03}_{-0.04}$ & $17^{+2}_{-3}$ &  N/A  & N/A  \\
		 I/0    &  $2.34^{+0.03}_{-0.03}$ & $19^{+2}_{-1}$ & N/A  & N/A  &  N/A  & N/A  \\
		 III/II &  $2.34^{+0.03}_{-0.03}$ & $19^{+2}_{-1}$ & $2.22^{+0.04}_{-0.04}$ & $12^{+2}_{-2}$ &  N/A  & N/A  \\[1mm]
		 \multicolumn{7}{c}{| Argon |}\\
		 II     &  $2.62^{+0.03}_{-0.03}$ & $9^{+1}_{-1}$ & $2.40^{+0.05}_{-0.07}$ & $4^{+1}_{-1}$ &  N/A  & N/A  \\
		 III    &  $2.57^{+0.04}_{-0.03}$ & $10^{+1}_{-1}$ & $2.26^{+0.05}_{-0.08}$ & $6^{+2}_{-2}$ &  N/A  & N/A  \\
		 I/0    &  $2.52^{+0.04}_{-0.04}$ & $10^{+1}_{-1}$ &  N/A &  N/A &  N/A  & N/A  \\
		 III/II &  $2.51^{+0.03}_{-0.05}$ & $9^{+1}_{-1}$ & $2.21^{+0.06}_{-0.08}$ & $4^{+1}_{-1}$ &  N/A  & N/A  \\[1mm]
		 \multicolumn{7}{c}{| Calcium |}\\
		 II     &  $2.57^{+0.04}_{-0.04}$ & $20^{+2}_{-2}$ & $2.54^{+0.03}_{-0.09}$ & $11^{+2}_{-3}$ &  N/A  & N/A  \\
		 III    &  $2.53^{+0.04}_{-0.03}$ & $21^{+2}_{-2}$ & $2.40^{+0.04}_{-0.07}$ & $18^{+4}_{-4}$ &  N/A  & N/A  \\
		 I/0    &  $2.49^{+0.04}_{-0.04}$ & $21^{+2}_{-2}$ &  N/A &  N/A &  N/A  & N/A  \\
		 III/II &  $2.49^{+0.03}_{-0.04}$ & $21^{+2}_{-2}$ & $2.35^{+0.07}_{-0.05}$ & $11^{+3}_{-3}$ &  N/A  & N/A  \\[1mm]
		 \multicolumn{7}{c}{| Iron |}\\
		 II     &  $2.43^{+0.02}_{-0.03}$ & $115^{+7}_{-7}$ & $2.48^{+0.01}_{-0.01}$ & $83^{+4}_{-3}$ & $2.76^{+0.09}_{-0.14}$ & $175^{+126}_{-93}$ \\
		 III    &  $2.39^{+0.02}_{-0.02}$ & $117^{+8}_{-6}$ & $2.29^{+0.01}_{-0.01}$ & $110^{+5}_{-6}$ & $2.51^{+0.12}_{-0.14}$ & $235^{+283}_{-137}$ \\
		 I/0    &  $2.36^{+0.02}_{-0.02}$ & $124^{+7}_{-9}$ &  N/A & N/A  & $2.50^{+0.15}_{-0.09}$ & $230^{+190}_{-139}$ \\
		 III/II &  $2.35^{+0.02}_{-0.02}$ & $124^{+7}_{-8}$ & $2.34^{+0.01}_{-0.01}$ & $98^{+5}_{-5}$ & $2.58^{+0.12}_{-0.11}$ & $196^{+167}_{-109}$ \\ \hline
	\end{tabular}
\end{table}
\end{appendix}
\end{document}